\documentclass[superscriptaddress,aps,preprintnumbers,amsmath,amssymb,prd,nofootinbib,preprint,longbibliography]{revtex4-1}
\pdfoutput=1

\usepackage{cancel}
\usepackage{graphicx,array}
\usepackage{color}
\usepackage{latexsym}
\usepackage{amsthm}
\usepackage{amsmath}
\usepackage{amssymb}
\usepackage{hyperref} 
\usepackage{bbold}
\usepackage{mathtools}
\usepackage{enumitem}
\usepackage{makecell}
\usepackage[table]{xcolor}
\usepackage{multirow}
\usepackage{booktabs}
\usepackage{setspace}
\usepackage{stackengine}
\usepackage{comment}
\usepackage{tikz}
\usepackage{verbatim}
\usepackage[normalem]{ulem}
\usepackage{cancel}

\numberwithin{equation}{section}

\makeatletter
\renewcommand{\p@subsection}{}
\renewcommand{\p@subsubsection}{}
\makeatother

\def\simgt{\mathrel{\lower2.5pt\vbox{\lineskip=0pt\baselineskip=0pt
           \hbox{$>$}\hbox{$\sim$}}}}
\def\simlt{\mathrel{\lower2.5pt\vbox{\lineskip=0pt\baselineskip=0pt
           \hbox{$<$}\hbox{$\sim$}}}}

\newcommand{\be}{\begin{equation}}
\newcommand{\ee}{\end{equation}}
\newcommand{\bea}{\begin{eqnarray}}
\newcommand{\eea}{\end{eqnarray}}

\newcommand{\GeV}{\textrm{ GeV}}
\newcommand{\TeV}{\textrm{ TeV}}
\newcommand{\gsim}{\lower.7ex\hbox{$\;\stackrel{\textstyle>}{\sim}\;$}}
\newcommand{\lsim}{\lower.7ex\hbox{$\;\stackrel{\textstyle<}{\sim}\;$}}

\definecolor{nicered}{rgb}{0.7,0.1,0.1}
\definecolor{nicegreen}{rgb}{0.1,0.5,0.1}
\definecolor{PatColor}{rgb}{0,.8,0}
\hypersetup{colorlinks,citecolor= blue,linkcolor= black}

\begin{document}

\title{Non-Gaussianity from explicit $U(1)$-breaking interactions}

\author{Raymond T. Co}
\email{rco@iu.edu}
\affiliation{Physics Department, Indiana University, Bloomington, IN, 47405, USA}
\author{Taegyu Lee}
\email{taeglee@iu.edu}
\affiliation{Physics Department, Indiana University, Bloomington, IN, 47405, USA}
\author{Sai Chaitanya Tadepalli}
\email{saictade@iu.edu}
\affiliation{Physics Department, Indiana University, Bloomington, IN, 47405, USA}

\begin{abstract}
We investigate primordial non-Gaussianity (NG) arising from the explicit $U(1)$ symmetry-breaking interactions during inflation involving a nearly massless axial component of a complex scalar field $P$. 
We analyze the induced NG parameter $f_{\mathrm{NL}}$ under scenarios where the axial field functions as either a curvaton or cold dark matter (CDM). 
In the curvaton framework, there is a conventional contribution to the local NG of $f_{\rm NL} \simeq -O(1)$.
Additional positive local NG can result from either the self-interactions of axial field fluctuations, their interactions with a light radial partner, or kinetic mixing with the inflaton via $U(1)$ symmetry-breaking terms. We identify parameter regions where the interactions lead to cancellations, suppressing the overall local NG to $|f^{\rm loc}_{\mathrm{NL}}| \lesssim O(0.1)$, while leaving the trispectrum largely unaffected. In the CDM scenario, these interactions enhance the NG in the isocurvature fluctuations. Moreover, interactions between the axial field and another light scalar, such as a curvaton, can generate $O(1)$ curvature NG signals and significant mixed curvature-isocurvature NGs that are within the reach of future experiments with $\sigma(f^{\rm loc}_{{\rm NL}})\sim1$. We also explore the role of a heavy radial field in generating oscillating correlation signals, noting that such signals can dominate the shape of the mixed adiabatic-isocurvature bispectrum. In certain cases, an oscillatory isocurvature bispectrum signal may be observable in the future, aiding in distinguishing between certain types of the $U(1)$-breaking self-interactions of the axial field.

\end{abstract}

\maketitle

\vspace{1cm}

\begingroup
\hypersetup{linkcolor=black}
\renewcommand{\baselinestretch}{1.26}\normalsize
\tableofcontents
\renewcommand{\baselinestretch}{2}\normalsize
\endgroup

\newpage

\section{Introduction}

Non-Gaussianities (NG) in the primordial density perturbations provide a crucial probe into the physics of the early universe, particularly the dynamics of cosmic inflation. In the simplest inflationary models, quantum fluctuations of the inflaton field lead to Gaussian-distributed density perturbations. For reviews on inflationary cosmology, refer to Refs.~\cite{Riotto:2002yw,Linde:2007fr}. However, a wide range of inflationary scenarios predict deviations from Gaussianity due to nonlinear interactions during inflation \cite{Bartolo:2004if}. Detecting NG will offer insights into the mechanisms that generated the initial fluctuations. One of the primary statistical tools for studying NG is the \textit{bispectrum}, which is the Fourier transform of the real-space three-point correlation function of cosmological perturbations. In most scenarios, the bispectrum can effectively capture the amplitude and shape of a non-Gaussian signal. Observational efforts have focused extensively on measuring the bispectrum using data from the Cosmic Microwave Background (CMB)~\cite{Planck:2019kim, Abazajian:2019eic} and Large-Scale Structure (LSS) surveys \cite{eBOSS:2021jbt,Cabass:2022wjy,DAmico:2022gki}. Experiments like the Planck satellite have provided high-precision measurements of the CMB anisotropies, placing stringent constraints on the level of primordial non-Gaussianity (PNG). Currently, the amplitude of local PNG from Planck is $f_{\rm NL}^{\rm loc}=-0.9\pm5.1$~($68\%$~C.L.)~\cite{Planck:2019kim}. So far, observations have been consistent with Gaussian initial conditions, but the tight bounds on non-Gaussian parameters have significantly informed and constrained early universe models. Upcoming 3D LSS experiments such as SPHEREX \cite{SPHEREx:2014bgr}, MegaMapper \cite{Schlegel:2019eqc}, LSST \cite{LSSTScience:2009jmu} are estimated to reach a precision of $\sigma(f_{\rm NL})\sim 1$, while 21-cm experiments in future will be able to probe PNG with sensitivities reaching $\sigma(f_{\rm NL})\sim 10^{-3}$ \cite{Meerburg:2016zdz}. CMB-based future surveys such as CMB-S4 \cite{CMB-S4:2016ple} and SO \cite{SimonsObservatory:2018koc} can also reach $\sigma(f_{\rm NL})\sim 2$.

In this work, we consider the presence of a spectator complex scalar field $P$ during inflation and assume that the $U(1)$ symmetry is spontaneously broken during inflation such that the radial component of $P$ is displaced at a large field value. Specifically, we study interactions arising from higher-dimensional terms that explicitly break the $U(1)$ symmetry. These terms generate nonlinear interactions for the nearly massless axial component of $P$. We investigate the NG generated from such nonlinear interactions during inflation. Since we focus on interactions between nearly massless fields, we find that the NG has a \textit{local}~\cite{Komatsu:2001rj} shape and is produced by the superhorizon evolution of the quantum mode functions. 

\newpage

After inflation, the axial field can function as either cold dark matter (CDM) or a curvaton. In the curvaton framework, the dominant contribution to the curvature perturbation originates from the decay of the oscillating curvaton field after inflation. NG arises primarily through the nonlinear mapping of curvaton fluctuations onto curvature perturbations. However, a finite contribution to NG also emerges when curvaton fluctuations develop non-Gaussian features during inflation via interactions, meaning that they are not purely Gaussian after inflation. Previous studies \cite{Enqvist:2005pg,Enqvist:2008gk,Huang:2008zj,Enqvist:2009zf,Enqvist:2009eq,Enqvist:2009ww,Enqvist:2010dt,Fonseca:2011aa} have primarily focused on NG contributions from the self-interactions of the background curvaton field, considering that the contribution from the field fluctuations is negligible. In this work, we discuss parametric regions where $U(1)$-breaking self-interactions of the axial field fluctuations during inflation can lead to significant NG. Additionally, we also consider scenarios where the radial partner of the axial field is heavy and analyze the amplitude of the oscillatory power spectrum and bispectrum signals generated during the classical oscillations of a massive background radial field. 

If the axial field functions as CDM after inflation, its self-interactions can lead to sizable isocurvature NGs. Previous studies on axion isocurvature perturbations, such as Refs.~\cite{Kawasaki:2008sn,Hikage:2012be,Geller:2024upd}\footnote{The axion is the axial component of the Peccei-Quinn (PQ) complex scalar field and a pseudo-Nambu-Goldstone boson associated with the spontaneous breaking of the global $U(1)_{\rm{PQ}}$ symmetry \cite{Peccei:1977hh,Peccei:1977ur,Weinberg:1977ma,Wilczek:1977pj}. See Refs.~\cite{Marsh:2015xka,DiLuzio:2020wdo} for some recent reviews. } and on oscillatory signals from cosmological collider (CC) physics \cite{Lu:2021gso,Chen:2023txq} have not considered scenarios involving $U(1)$-breaking self-interactions of the axial field fluctuations or $U(1)$-breaking couplings between the axial field and a light radial partner during inflation. In scenarios where the complex scalar field $P$ couples via explicit $U(1)$-breaking to another light scalar $\phi$, such as an inflaton or a curvaton, direct couplings between $\phi$ and the axial field can generate various enhanced bispectrum signals with a local shape, including mixed curvature-isocurvature modes. Additionally, they can produce bispectrum signals with oscillatory templates, offering distinctive features that could be probed observationally.

The order of the presentation is as follows. In Sec.~\ref{sec:1} we introduce an explicit $U(1)$-breaking potential and discuss the NG arising from the self-interactions of the axial field fluctuations in the context of scenarios where the axial field serves as CDM, an oscillating curvaton, or a curvaton rotating in complex field space. In Subsecs.~\ref{sec:classical-effects-radial} and \ref{sec:light-radial-field} we analyze cases where the radial partner of the axial field is either massive or light. Sec.~\ref{Couplings with a light scalar} focuses on $U(1)$-breaking couplings of the complex scalar \(P\) with another light scalar \(\phi\) (inflaton/curvaton), leading to interactions between the axial field and \(\phi\). In Subsecs.~\ref{subsec:modelA} and \ref{Sec:KineticMixing}, we consider two models distinguished by the nature of the couplings and study the bispectrum signals generated from these interactions, both with and without radial oscillations. We conclude in Sec.~\ref{sec:conclusion}. Several appendices follow in Apps.~\ref{app:Evolution-of-radial}-\ref{App:pwrbispec}
that supplement the technical details and discussion in the main body of the paper. 

\section{Explicit $U(1)$-breaking potential}
\label{sec:1}

In this section, we explore the
NG generated during inflation in the stochastic fluctuations of a
nearly massless axial component of a complex scalar field $P$ for
scenarios where its radial partner is displaced at a large field or vacuum expectation value (vev). A
large radial displacement allows higher-dimensional terms to make
dominant contributions. We primarily evaluate non-Gaussian contributions
from a higher-dimensional $U(1)$-breaking term such as that utilized in the Affleck-Dine
mechanism \cite{Affleck:1984fy}. To this end, we consider a $U(1)$-charged
complex scalar field $P$ governed by the Lagrangian $\mathcal{L}=-g^{\mu\nu}\partial_{\mu}P^{*}\partial_{\nu}P-V$
where $g_{\mu\nu}$ is an externally sourced background metric.\footnote{The background metric and the Hubble scale $H$ are set by the inflaton
dynamics. Hence, $P$ is a spectator field in this scenario.} We consider that the potential $V$ consists of a Hubble-induced
mass term $-cH^{2}|P|^{2}$ and a vacuum mass term $m_{s}^{2}|P|^{2}$
such that during inflation $cH^{2}\gg m_{s}^{2}$.
We also assume that the self-interactions of $P$ are dominated by
some higher-dimensional terms such as $V\propto O\left(|P|^{2m}/\Lambda^{2m}\right)$,
and a $U(1)$-breaking term $V_{\cancel{U(1)}} = - g^{(4)} P^{n}/\Lambda^{n}+{\rm h.c.}$,
where $\Lambda \le M_{\rm Pl}$ is a cutoff scale of the theory with $M_{\rm Pl}$ the Planck scale, and $g^{(4)}$ is a dimension-four quantity. The complete form
for these higher-dimensional terms depends on the relevant UV theory such as supergravity \cite{Nilles:1983ge}. For instance, the $U(1)$-conserving and -breaking terms mentioned above can be derived from the $F$- and $A$-terms in supersymmetry~\cite{Mazumdar:2010sa}. Ref.~\cite{Co:2022qpr} discusses a specific implementation of such a model. 

We parameterize the complex scalar field $P$ as 
\begin{equation}
P=\frac{S}{\sqrt{2}}e^{i\theta},
\end{equation}
where $S$ is the radial scalar partner of the axial field $A \equiv S\theta$. A large displacement of the background radial field $S_{0}$ from
its true minimum at low temperatures $S_{\min}$ can be achieved if $c>0$.\footnote{Henceforth, we will refer to all background homogeneous and isotropic
fields with a subscript 0.} The
minimum is then defined by the balance between the negative Hubble-induced mass term and the higher-dimensional $U(1)$-symmetric term $\propto O\left(|P|^{2m}/\Lambda^{2m}\right)$. Here we assume that the $U(1)$-preserving quartic term $|P|^4$ is sufficiently suppressed, which is naturally the case in supersymmetric models with dimension transmutation~\cite{Moxhay:1984am} or a moduli space of two fields~\cite{Kasuya:1996ns}. 
We will denote this local minimum of the radial field as $S_I$.
In terms of the radial and axial fields, the explicit $U(1)$-breaking term can be written as \begin{equation}
V_{\cancel{U(1)}}=- g^{(4)} \left(\frac{S}{\Lambda\sqrt{2}}\right)^{n}2\cos(n\theta)\label{Vtheta} .
\end{equation}
For simplicity, we assume
that $g^{(4)}>0$ without any loss of generality. While Eq.~(\ref{Vtheta}) is the dominant potential at $S_I$. One could consider a vacuum potential at $S_{\min}$ for the axial direction as $V = m_{A,\rm vac}^2S^2_{\rm min}(1- \cos(N A/S_{\min}))$ for $N \geq 1$.

For a stationary radial field at its minimum, the linear fluctuations
in the radial and axial fields are given by $\delta S$ and $\delta A=S_I\delta\theta$
respectively. At the local minimum $S_I$, the radial and axial
mass-squared terms are 
\begin{eqnarray}
m_{S}^{2} & =\left.\frac{\partial^{2}V}{\partial S^{2}}\right|_{S_I} & \propto cH^{2},\label{mSsq}\\
m_{A}^{2} & =\left.\frac{\partial^{2}V}{\partial A^{2}}\right|_{S_I} & =\frac{n^{2}}{S_I^{2}} g^{(4)} \left(\frac{S_I}{\Lambda\sqrt{2}}\right)^{n}2\cos(n\theta_{0})\equiv m_{A,I}^{2}\cos(n\theta_{0}),
\label{mAsq}
\end{eqnarray}
where $m_{A,I}$ is the mass of the axial field at $S=S_I,~\theta_0=0$.
To ensure that the axial mass is positive, we choose $n\theta_{0} \in (-\pi/2,\pi/2)$ such that $g^{(4)} \cos(n\theta_{0})>0$.
Outside this range, the axial field is tachyonic, leading to
fluctuations that could grow as the field rolls down the potential.
We do not address this scenario in the current work. Ref.~\cite{Kawasaki:2008mc} discusses the motion of the background axial field, $A_0(t)$, along the hilltop trajectory after inflation.

For the cosine potential along the axial direction as given in Eq.~(\ref{Vtheta}),
the mass of the axial field depends on the angular displacement $n\theta_{0}(t)$
as derived in Eq.~(\ref{mAsq}). In the slow-roll limit, the background
angular velocity can be approximated by 
\begin{equation}
H\partial_{t}(n\theta)\approx\frac{-1}{3}m_{A,I}^{2}\sin\left(n\theta\right).\label{angular velocity}
\end{equation}
During inflation, the angular field evolves from an initial displacement
$\theta_{0}(t_{i})$ to $\theta_{0}(t_{e})$ at the end of inflation,
given by the solution to the above differential equation as 
\begin{equation}
\tan\left(\frac{n\theta_{0}(t)}{2}\right)=\tan\left(\frac{n\theta_{0}(t_{i})}{2}\right)e^{-N(t)\frac{m_{A,I}^{2}/H^{2}}{3}}, \label{eq:theta-soln}
\end{equation}
where $N(t)$ is the number of inflationary e-folds elapsed at time
$t$ starting from $t_{i}$, such that $N(t_{e})=N_{\rm inf}$ is
the total number of inflationary e-folds. In this work, we are interested
in a light axial scalar field with $m_{A} \ll H$. As the axial field slowly rolls down the potential
during inflation, the dynamical axial mass increases over time saturating
to a constant value of $m_{A,I}$ as $n\theta\rightarrow0$. Requiring
the mass of the axial field during its motion to remain much smaller than $H$ implies that we only consider $m_{A,I}\ll H$.

Following the above discussion, we will restrict ourselves to a light axial
field such that the axial slow-roll parameter, $\eta_{AA}=m_{A}^{2}/H^{2}$, is
$\lesssim O(0.02)$. Consequently,
the power spectrum for the axial field fluctuations, 
$\delta A$, is nearly scale-invariant, and its amplitude is given
by~\cite{Linde:2007fr}
\begin{equation}
\mathcal{A}_{\delta A}(k)\approx\frac{k^{3}}{2\pi^{2}}\left\langle \delta A(\vec{k},t)\delta A(\vec{p},t)\right\rangle '=\left(\frac{H}{2\pi}\right)^{2}, \label{eq:pwrspec-scaleinvariant}
\end{equation}where $\delta A(\vec{k},t)$ is the Fourier transform of the real-space fluctuation $\delta A(\vec{x},t)$. Here, the $(')$ denotes that the momentum-conserving
Dirac-delta function has been factored out. Throughout this discussion, we refer to the power spectrum of a field $X$ as $P_{X} = \left\langle \delta X(\vec{k},t)\delta X(\vec{p},t)\right\rangle'$.

After inflation, the field value of $P$ may evolve in different manners. As we will elaborate in Sec.~\ref{subsec:bispectrum}, the dynamics depend on the size of $m_{A,\rm vac}$ and the time (in)dependence of $g^{(4)}$. The field $P$ may receive a kick in the angular direction by $m_A^2$, the dynamics of which is discussed in App.~\ref{app:Evolution-of-radial}. Alternatively, $P$ may oscillate in the radial direction in the case where $g^{(4)}$ decreases with Hubble after inflation, and the axial field subsequently oscillates because of $m_{A,\rm vac}$. These different scenarios allow the axial field to become cold dark matter or a rotating/oscillating curvaton. As the radial field evolves toward $S_{\min}$ after inflation, the specific form of the radial potential can lead to a significant amplification of the superhorizon relative perturbative quantity \( \delta S/S \). This amplification, due to the coupling between \( S \) and \( \theta \), can subsequently influence \( \delta \theta/\theta \). However, for a massive radial field during inflation, the associated radial fluctuations are negligibly small, ensuring that \( \delta \theta/\theta \) remains unaffected, even if \( \delta S/S \) increases substantially post-inflation. In contrast, for a light radial scalar field as considered in Sec.~\ref{sec:light-radial-field}, we assume the radial potential to be nearly quadratic to ensure the conservation of axial fluctuations. Accordingly, this work assumes that the superhorizon Goldstone fluctuations remain conserved after inflation. We defer the investigation of scenarios in which the post-inflationary rolling of the radial field has a significant impact on the evolution of axial fluctuations to a future work.

\subsection{Axial self-interactions}\label{subsec:selfintr}
The self-interactions of the axial field are obtained from the $U(1)$-breaking term in Eq.~(\ref{Vtheta}). The leading cubic order interaction
Lagrangian is given as 
\begin{equation}
\mathcal{L}_{3}\supset a^3\frac{n^{3}}{3S_I^{3}} g^{(4)} \left(\frac{S_I}{\Lambda\sqrt{2}}\right)^{n}\sin(n\theta_{0})\,(\delta A)^{3}\label{A3},
\end{equation}
where $a\equiv a(t)$ is the scale factor.
In terms of the axial mass-squared quantity $m_{A,I}^{2}$, we rewrite
the cubic term as 
\begin{equation}
\mathcal{L}_{3}^{(\delta A)^{3}}=a^3\frac{n}{6S_I}m_{A,I}^{2}\sin(n\theta_{0})\,(\delta A)^{3}.\label{A3-mA}
\end{equation}
Hence, we find that the cubic interaction term is suppressed by the
axial slow-roll parameter, $\eta_{AA}=m_{A}^{2}/H^{2}$, and a factor
of $H/S_I$. This suppression is true for all higher-order self-interactions
of the axial field with increasing powers of $H/S_I$. 
\begin{figure}
\centering \begin{tikzpicture}  
\node at (4.3,0) {$t$};
\draw[thick] (0,0) -- (4,0);  
\node at (2,-3.4) {$\left(\delta A\right)^{3}(t')$};       
\draw[thick] (1,0) -- (2,-3);     
\draw[thick] (2,0) -- (2,-3);          
\draw[thick] (3,0) -- (2,-3);          
 
\end{tikzpicture} \caption{Feynman diagram representation for the cubic-interaction contribution
to the axial field bispectrum, where $\delta A$ is the axial field fluctuation.}
\label{fig1}
\end{figure}
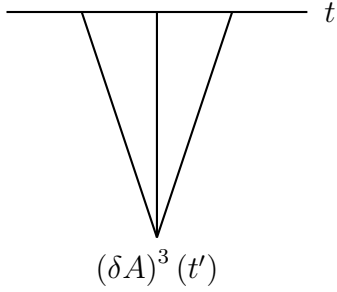

We will now evaluate the perturbative contribution to the three-point
correlation function from the $U(1)$-breaking term $\propto\left(\delta A\right)^{3}$.
This interaction is represented diagrammatically in Fig.~\ref{fig1}.
Using the in-in formalism \cite{JS1960,Bakshi:1962dv,Bakshi:1963bn,Keldysh:1964ud}, we write the first-order cubic correlation
function for the operator $W(t)=\delta A_{k_{1}}(t)\delta A_{k_{2}}(t)\delta A_{k_{3}}(t)$
as
\begin{eqnarray}
\left\langle W(t)\right\rangle ^{(1)} & = & i\left(\int_{-\infty^{-}}^{t}dt'\left\langle 0\right|H(t')W(t)\left|0\right\rangle -\int_{-\infty^{+}}^{t}dt'\left\langle 0\right|W(t)H(t')\left|0\right\rangle \right)\\
 & = & -2{\rm Im}\left[\int_{-\infty}^{t}dt'\left\langle 0\right|H(t')W(t)\left|0\right\rangle \right]\label{A3-corr} \nonumber ,
\end{eqnarray}
where all fields are in interaction-picture and the cubic interaction
Hamiltonian $H(t)$ is given by the expression 
\begin{eqnarray}
H(t) & = & \int d^{3}\vec{x}\mathcal{H}_{3}^{I}=a^{3}c_3\int d^{3}\vec{x}\left(\delta A\right)^{3}
 =  a^{3}c_3\int_{p}\int_{q}\delta A_{p}\delta A_{q}\delta A_{-p-q}~, \label{eq:cubic-intr-Hamiltonian}
\end{eqnarray}
where 
\begin{equation}
c_3=\frac{1}{6S_I}\left(-m_{A,I}^{2}n\sin\left(n\theta_{0}\right)\right)<0,\label{eq:c3-coeff}
\end{equation}
$\int_{k}\equiv\int d^{3}k/(2\pi)^{3}$ and $\delta A_p \equiv \delta A(\vec{p},t)$. To evaluate the correlation
function, we consider that the axial field in the interaction picture
is represented by 
\begin{equation}
\delta A(x,t)=\int\frac{d^{3}p}{\left(2\pi\right)^{3}}\left[b_{\vec{p}}u_{\vec{p}}(t)+b_{-\vec{p}}^{\dagger}u_{-\vec{p}}^{*}(t)\right]e^{i\vec{p}\cdot\vec{x}},
\end{equation}
where we utilize the mode function for a massless scalar field 
\begin{equation}
u_{k}(\tau)=\frac{H}{\sqrt{2k^{3}}}\left(1+ik\tau\right)e^{-ik\tau}\label{massless_modefn}
\end{equation}to evaluate the correlations of the axial field. Here, $\tau=-1/(aH)$ is the conformal time and the ladder operators
follow the usual commutation relation $\left[b_{\vec{k}},b_{-\vec{p}}^{\dagger}\right]=(2\pi)^{3}\delta^{3}\left(\vec{k}+\vec{p}\right)$
and zero for all other combinations.

The calculation of the correlation function from the cubic self-interaction in Eq.~(\ref{A3-corr})
for massless mode functions is a standard result \cite{Bartolo:2004if}.
The final expression for the bispectrum of the axial field is given
as 
\begin{equation}
B_{A}\left(\vec{k}_{1},\vec{k}_{2},\vec{k}_{3}\right)=\left\langle \delta A_{k_{1}}\delta A_{k_{2}}\delta A_{k_{3}}\right\rangle '\approx c_3\frac{H^{2}}{2k_{1}^{3}k_{2}^{3}k_{3}^{3}}\left(\sum_{i=1}^{3}k_{i}^{3}\left(-N_{k}\right)+O(1)f(k_{i})\right), \label{BA3}
\end{equation}
where $N_{k}$ is the number of e-foldings elapsed from
the horizon exit of the mode $k$ to the end of inflation, and $f(k_{i})$
is a scale-dependent cubic polynomial function. For CMB modes, $N_{k}$
can be as large as $\approx50-60$, dominating over the remaining
$O(1)$ terms. Thus, the bispectrum for the cubic self-interaction
is dominated by the classical evolution of the field after it exits
the horizon. 

The bispectrum signal derived above is directly proportional to the mass-squared, $m^2_{A}$,
of the axial field, through the coefficient $c_3$ defined in Eq.~(\ref{eq:c3-coeff}).
In the limit $m_{A,I}\rightarrow0$, the cubic interaction term in Eq.~(\ref{A3-mA})
vanishes, leading to the disappearance of the bispectrum signal in
Eq.~(\ref{BA3}). Conversely, as $m_{A}/H$ increases, the perturbations decay over time after horizon exit during inflation. Quantitatively, for $m_{A}>0$,
the mode function of the axial field follows the standard massive mode function, as described in Eq.~(\ref{massive-mode-fn}). These massive mode functions undergo dilution by a suppression factor $\gamma$, relative to the massless
case. For instance, when $m_{A}^{2}/H^{2}=0.02$, the factor $\gamma\sim0.7$
for CMB modes that exit the horizon $\sim50$ e-folds before the end
of inflationary de-Sitter phase. Consequently, correlation functions such as power spectrum and bispectrum, evaluated using massive mode functions,  are suppressed
by factors of $\gamma$, as illustrated in Eqs.~(\ref{eq:massive_pwr_spec})
and (\ref{eq:bispec-light-scalar}), respectively. 

In Fig.~\ref{fig:bispec-light-scalar} in Appendix~\ref{App:light-scalar}, we present the bispectrum signal for a self-interaction term proportional to the mass (analogous to Eq.~(\ref{A3-mA})), computed using massive mode functions for $\delta A$ rather than massless ones. The \textit{normalized} bispectrum in Fig.~\ref{fig:bispec-light-scalar}
exhibits a peak in the mass range $0.01\lesssim m_{A}^{2}/H^{2}\lesssim0.03$, with a sharp decline outside this interval. As described earlier, the suppression of the signal at higher masses is due to the redshift of perturbations. For masses smaller than $0.1H$, the bispectrum approaches zero as the coefficient $c_3\rightarrow0$.  Accordingly, we will focus our
analysis on light axial scalar fields within this mass range. 

In Appendix~\ref{App:light-scalar},
we also show that for these light scalar fields with $m_{A}^{2}/H^{2}\lesssim0.03$, calculations performed
using massless mode functions provide a reasonable approximation while significantly simplifying the evaluations.
Therefore, for the remainder of our discussion, we will employ massless
mode functions to analyze the bispectrum and power spectrum for these light
scalar fields.

\begin{figure}
\begin{centering}
\includegraphics[width=0.5\columnwidth]{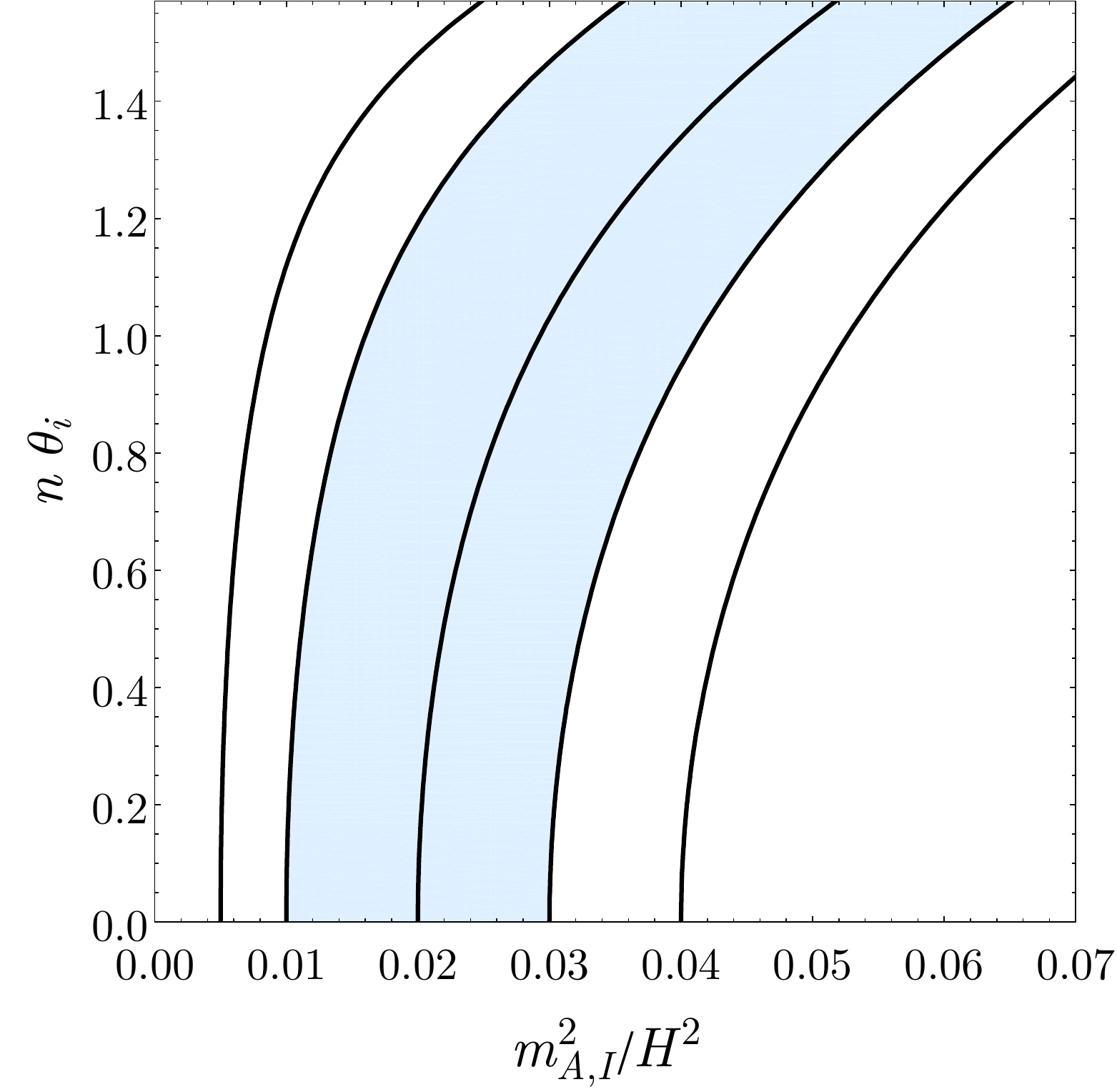}
\par\end{centering}
\caption{\label{fig:mass-range} Contours showing the mean mass-squared
$m_{A}^{2}/H^{2}$ as a function of $m_{A,I}^{2}/H^{2}$ and initial
angular displacement $n\theta_{i}$. The mass of the axial field as
it rolls down potential is time-dependent and a function of $\cos(n\theta(t))$.
Using the solution for $\theta(t)$ given in Eq.~(\ref{eq:theta-soln}),
we show the time-average mass-squared of the axial field for $N_{\rm inf}=50$.
The region where $0.01\protect\leq m_{A}^{2}/H^{2}\protect\leq0.03$
is shaded in light blue.}
\end{figure}

The mass of the axial field is time-dependent yet remains nearly constant due to
the slow-roll dynamics along the axial potential. Starting from $\theta_{i}$
at $t_{i}$, the time-average mass of the axial field can be analytically
determined using the expressions provided in Eqs.~(\ref{mAsq}) and
(\ref{eq:theta-soln}). The average mass of the axial field depends
upon the initial angle $\theta_{i}$, $m_{A,I}$ and total number
of inflationary e-folds $N_{\rm inf}$. In Fig.~\ref{fig:mass-range},
we present contours of the dimensionless average axial field
mass-squared $m_{A}^{2}/H^{2}$ quantity in $\{m_{A,I}^{2}/H^{2},n\theta_{i}\}$
parametric space for a fixed $N_{\rm inf}=50$. The area where $0.01\leq m_{A}^{2}/H^{2}\leq0.03$
is shaded in light blue, highlighting the parametric region where
our results are most relevant. For example, if $n=6$, the axial mass-squared $m_{A,I}^{2}$ during inflation has the fiducial value 
\begin{align}
\frac{m_{A,I}^{2}}{H^{2}}\approx 0.03\times \left( \frac{g^{(4)}}{0.1\Lambda^{2} H^{2}}\right) \left(\frac{S_I}{0.3\Lambda\sqrt{2}}\right)^{4}.
\label{eq:mAI-example}
\end{align}

We note in passing that the bispectrum signal in Eq.~(\ref{BA3}) for axial
field fluctuations arising from the interactions during inflation can also be evaluated using an equation-of-motion (EoM) approach \cite{Huang:2008zj}. In this case, the field fluctuation $\delta A$ is expressed perturbatively up to
second order, where the first-order fluctuations $\delta A^{(1)}$
are Gaussian and the cubic interactions $\propto\partial_{A}^{3}V_{{\rm \cancel{U(1)}}}$
generate a non-Gaussian second-order contribution $\delta A^{(2)}\propto N_k \partial_{A}^{3}V_{{\rm \cancel{U(1)}}}\left(\delta A^{(1)}\right)^{2}$.

\subsubsection{Local bispectrum}\label{subsec:bispectrum}
It is often customary to write the bispectrum of curvature $\zeta$
or isocurvature $\mathcal{S}$ field in terms of the local-shape template
defined as 
\begin{eqnarray}
B_{i}(\vec{k_{1}},\vec{k_{2}},\vec{k_{3}}) & = & f_{{\rm NL}}^{i}\left(P_{i}(k_{1})P_{i}(k_{2})+P_{i}(k_{1})P_{i}(k_{3})+P_{i}(k_{2})P_{i}(k_{3})\right), \label{fNLi}
\end{eqnarray}
where $P_{i}(k)$ is the power spectrum of the linear fluctuations,
and $f_{{\rm NL}}^{i}$ is the non-linearity parameter associated
with the perturbation. Here, $i$ denotes either $\zeta$ (curvature) or $\mathcal{S}$ (isocurvature) \cite{Komatsu:2001rj}. If the curvature (isocurvature)
is related at linear order with the scalar field fluctuations $\delta A$
as $\zeta = C_{\zeta}\delta A$ ($\mathcal{S}=C_{\mathcal{S}} \delta A$) for a linear coefficient $C$, then the above expression can be written as 
\begin{eqnarray}
B_{A}(\vec{k_{1}},\vec{k_{2}},\vec{k_{3}}) & = & f_{{\rm NL}}^{i}\times C_i\left(P_{A}(k_{1})P_{A}(k_{2})+P_{A}(k_{1})P_{A}(k_{3})+P_{A}(k_{2})P_{A}(k_{3})\right).\label{eq:fNL_general_expr}
\end{eqnarray}
Since the power spectrum for light scalar fields is approximated
as $P_{A}(k)\approx H^{2}/(2k^{3})$, substituting this in the above
expression reveals that the bispectrum from the cubic interaction
in Eq.~(\ref{BA3}) has a local-shape. We will now estimate the $f_{{\rm NL}}$ parameter from the cubic interaction of the axial field fluctuations
for the following scenarios. 
\begin{enumerate}
\item \textbf{CDM}: \\
 If the axial field acts as CDM after inflation, then the homogeneous
quantum fluctuations of $\delta A$ can be mapped to the isocurvature
fluctuations of the CDM. If the fraction of the axial field's energy
density relative to the total CDM is $\omega_{A}$, then the conserved superhorizon CDM isocurvature fluctuations $\mathcal{S}$ are given as 
\begin{equation}
\mathcal{S}=\omega_{A}\mathcal{S}_{A}\approx\omega_{A}\left(\frac{2\delta A}{A_{0}}+\left(\frac{\delta A}{A_{0}}\right)^{2}\right).\label{eq:AtoS}
\end{equation}
The background field value $A_{0}$ evolves under the combined effects
of a classical slow-roll and quantum fluctuation $\delta A$ \cite{Dimopoulos:2003az}. If
\begin{equation}
m_{A}^{2}A_{0}\gg H^{3}\implies\theta_{0} \frac{S_I}{H} \gg \frac{H^{2}}{m_{A}^{2}} ,
\end{equation}
the dynamics are dominated by classical rolling, and $A_{0}\approx\theta_{0}S_I$.
This is true for light scalars when $S_I\gg H$ where $H\equiv H_{\rm inf}$
is the Hubble parameter during inflation.\footnote{For the rest of our discussion, the inflationary Hubble scale will be referred to as $H$, unless explicitly specified otherwise.} In Ref.~\cite{Kawasaki:2008sn},
the authors evaluated the NG generated through a post-inflationary
local mapping of the axial field fluctuations to isocurvature perturbations
up to quadratic order in $\delta A$ as given in Eq.~(\ref{eq:AtoS}).
For $\delta A<A$, the isocurvature non-linearity parameter $f_{{\rm NL}}^{\mathcal{S}}$
is approximately given as $\omega_{A}^{-1}/2$. Below we give the
contribution to $f_{{\rm NL}}^{\mathcal{S}}$ from the self-interaction of axial
field fluctuations during inflation
\begin{equation}
f_{{\rm NL}}^{\mathcal{S}}=C_{\mathcal{S}}^{-1}\frac{B_{A}(\vec{k_{1}},\vec{k_{2}},\vec{k_{3}})}{P_{A}(k_{1})P_{A}(k_{2})+P_{A}(k_{1})P_{A}(k_{3})+P_{A}(k_{2})P_{A}(k_{3})} ,
\end{equation}
where the linear coefficient is $C_{\mathcal{S}}=2\omega_{A}/(\theta_{0}S_I)$.
Substituting $B_{A}(\vec{k_{1}},\vec{k_{2}},\vec{k_{3}})$ from Eq.~(\ref{BA3}),
we obtain 
\begin{eqnarray}
\Delta f_{{\rm NL}}^{\mathcal{S}} & = & C_{\mathcal{S}}^{-1}2c_3H^{-2}\left(-N_{k}\right),\nonumber \\
 & = & \frac{1}{6\omega_{A}}n\theta_{0}\sin\left(n\theta_{0}\right)\frac{m_{A,I}^{2}N_{k}}{H^{2}},\nonumber \\
 & \approx & \frac{1}{4\omega_{A}}\left(\frac{n\theta_{0}}{1}\frac{\sin\left(n\theta_{0}\right)}{0.85}\right)\left(\frac{m_{A,I}^{2}/H^{2}}{0.035}\frac{N_{k}}{50}\right).\label{CDM_fNL_1}
\end{eqnarray}
In the last expression above, we have chosen the following benchmarks
(fiducial) values lying within the blue-shaded region in Fig.~\ref{fig:mass-range}: $N_{k}\sim50-60$, $m_{A,I}^{2}/H^{2}\sim0.035$,
$n\theta_{0}\sim O(\pm1)$, thus yielding $f_{{\rm NL}}^{\mathcal{S}}\lesssim\omega_{A}^{-1}/4$
which implies that the isocurvature NG from cubic self-interactions of the axial field provides a significant
contribution that increases the overall level of NG. During the matter-dominated era, both curvature and
isocurvature fluctuations grow at the same rate. Since CDM carries
the isocurvature component, the curvature fluctuations receive a contribution
equal to $\mathcal{S}/3$ from isocurvature\footnote{This is true at all orders in perturbation theory.}.
Consequently, a fraction of NG is transferred from isocurvature to
curvature perturbations. This type of NG is suppressed by a square
of $P_{\mathcal{S}}/P_{\zeta}\lesssim0.04$, the ratio of isocurvature power
to curvature power on CMB scales. Hence, the CDM isocurvature-induced
local curvature NG is negligible unless $f_{{\rm NL}}^{\mathcal{S}}\gg1$ for
$\omega_{A}\ll1$. A lower bound on $\omega_{A}$ is obtained by setting
$\mathcal{S}\sim\omega_{A}$, which yields an upper bound of $f_{{\rm NL}}\lesssim20$.
For more details see discussion in Ref.~\cite{Kawasaki:2008sn}.

We now comment on how to estimate the dark matter fraction $\omega_A$ for a given model. For simplicity, we discuss the dynamics when $g^{(4)}$ is a constant. Given that we assume non-vanishing $m_{A,I}/H$ during inflation, we expect that when the radial field starts oscillating, the axial field gets a kick from the mass term $m_{A,I}$. Such a mass term from the higher-dimensional operator quickly becomes suppressed after the kick, and the complex field undergoes a rotational motion similar to that in the Affleck-Dine mechanism. This rotation may result in a contribution to the dark matter abundance via the conventional misalignment~\cite{Preskill:1982cy, Dine:1982ah,Abbott:1982af} or kinetic misalignment~\cite{Co:2019jts} mechanisms depending on the vacuum mass $m_{A,{\rm vac}}$ of the axial field at $S_{\min}$. Since the radial field value decreases by redshift, the rotation will eventually get to the minimum of the wine-bottle potential. Subsequently, the angular speed redshifts according to $\partial_t\theta \equiv \partial_t A/S_{\min} \propto a^3$. If $\partial_t\theta \simeq m_{A,{\rm vac}}$ occurs when $H \gg m_{A,{\rm vac}}$, the rotation stops at an essentially fixed value of $\theta$. The oscillations towards the final minimum of $A$ start when $H \simeq m_{A,{\rm vac}}$, and conventional misalignment determines the final abundance.  On the other hand, if $\partial_t\theta \simeq m_{A,{\rm vac}}$ occurs when $H \ll m_{A,{\rm vac}}$, then the kinetic misalignment applies. Thus, with the freedom of choosing $m_{A,{\rm vac}}$, one can realize any value of $\omega_A$.

\item \textbf{Curvaton:} \\
In the curvaton mechanism \cite{Lyth:2001nq,Bartolo:2003jx,Sasaki:2006kq},
the primordial curvature perturbations evolve outside the horizon.
In this paradigm, $\zeta$ receives significant contribution from
the fluctuations of another field called the \textit{curvaton}. This happens
after inflation, when the curvaton is assumed to begin oscillation.
Eventually, the curvaton must decay before the Big-Bang Nucleosynthesis (BBN)
and give rise to standard radiation dominated era. If the axial field
$A$ acts as a curvaton, its fluctuations $\delta A$ will determine
the primordial curvature perturbation $\zeta$  \cite{Dimopoulos:2003az,Chun:2004gx,Bartolo:2004if,Dimopoulos:2005bp,Lyth:2005fi,Kawasaki:2008mc,Chingangbam:2009xi}.
\begin{enumerate}
\item \textbf{Oscillating curvaton}: \\
We assume that after inflation, the axial field begins to oscillate
only when $H\sim m_{A,{\rm vac}}$.%
\footnote{The dynamics described above about the complex field rotation initiated by the higher-dimensional operator can be avoided if $g^{(4)}$ is varying. For example, $g^{(4)}$ can be proportional to $H^2$ from a coupling to the inflaton potential energy. This implies that the axial mass decreases with Hubble and the oscillations would not begin until $H \lesssim m_{A,{\rm vac}}$.}
If the inflaton fluctuations are subdominant
when the curvaton decays and the curvaton's potential is purely quadratic,
we can approximate the localized generation of curvature perturbations
from axial fluctuations (in the absence of any contribution from the
interactions during inflation) as
\begin{equation}
\zeta\approx\frac{2r}{3}\left(\frac{\delta A}{A_{0}}\right)-\left(\frac{3-2r(r+2)}{9r}\right)\left(\frac{\delta A}{A_{0}}\right)^{2}, \label{curvaton_zeta}
\end{equation}
which remains conserved on superhorizon scales and where $r$ is the fraction of curvaton's energy density to the
total energy budget at the time of its decay. The amplitude of the
linear-order curvature power spectrum is given as $\mathcal{A}_{\zeta}\approx(H/(\pi\theta_{0}S_I))^{2}/9$
and the spectral tilt is given by the expression $n_{s}-1=-2\epsilon+2\eta_{AA}/3$
where $\epsilon=-\partial_t{H}/H^{2}$ during inflation. Since $\eta_{AA}\approx0.02$
in this work, we require large field inflationary models to satisfy
the current $n_{s}$ constraints \cite{Lyth:1996im}. From the
non-linear mapping of the superhorizon curvaton fluctuations $\delta A$
to the curvature perturbations at the quadratic order as given by
Eq.~(\ref{curvaton_zeta}), the local-type NG parameter evaluates
to\footnote{$\zeta=\zeta_{g}+3f_{{\rm NL}}^{{\rm loc}}\zeta_{g}^{2}/5$ where
$\zeta_{g}$ is the Gaussian stochastic fluctuation} 
\begin{equation}
f_{{\rm NL}}^{{\rm loc}}=\frac{5}{4r}-\frac{5}{3}-\frac{5r}{6}.\label{fNL_curv_loc}
\end{equation}
When the curvaton dominates the energy density of the universe during its decay, $r\approx1$ and hence
\begin{equation}
f_{{\rm NL}}^{{\rm loc}}=-\frac{5}{4}.
\end{equation}
This value is much larger than the $f_{{\rm NL}}^{{\rm loc}}$ in the squeezed limit from single-field
inflationary models \cite{Creminelli:2004yq}. We note that the above
formula is only applicable to a curvaton with quadratic potential.
When the potential deviates from a quadratic form, the $f_{{\rm NL}}$
receives additional contribution 
\begin{equation}
\Delta f_{{\rm NL}}=\frac{5}{4r}h,
\end{equation}
where the dimensionless coefficient $h$ encodes contributions due
to the interactions of the field fluctuations $\delta A$ during inflation \cite{Huang:2008zj}
and from the non-linear mapping of the background curvaton field $A_{0}$
from the horizon exit to the onset of the oscillation \cite{Lyth:2003dt,Enqvist:2005pg,Sasaki:2006kq}.
For $|n\theta_{0}|\sim 1$, the deviation of the background cosine
potential $\cos(n\theta)$ from a quadratic form is $\sim10\%$.
For such small deviations, the resulting contribution to $h$ from
the background nonlinear mapping is $\lesssim5\%$ \cite{Huang:2008bg}.
However, the contribution from the self-interaction $\propto(\delta A)^{3}$
during inflation can be significantly large within the region $0.01\lesssim m_{A}^{2}/H^{2}\lesssim0.03$ and $n\theta_{0}\sim1$.
Using the linear relation $\zeta=2r\delta A/(3A_{0})$, we convert
the axial bispectrum from Eq.~(\ref{BA3}) into the curvature
bispectrum. Hence, the NG arising from the explicit $U(1)$-breaking cubic interaction
term, will transform into local NG in curvature perturbations. Utilizing
Eq.~(\ref{eq:fNL_general_expr}), we evaluate the non-linearity parameter
as
\begin{eqnarray}
\Delta f_{{\rm NL}}^{{\rm loc}} & = & \frac{5}{6}C_\zeta^{-1}\frac{B_{A}(\vec{k_{1}},\vec{k_{2}},\vec{k_{3}})}{P_{A}(k_{1})P_{A}(k_{2})+P_{A}(k_{1})P_{A}(k_{3})+P_{A}(k_{2})P_{A}(k_{3})}\nonumber \\
 & = & \frac{5}{4r}\times\frac{n\theta_{0}}{3}\sin(n\theta_{0})\frac{m_{A,I}^{2}N_{k}}{H^{2}}\nonumber \\
\Delta f_{{\rm NL}}^{{\rm loc}} & \approx & \frac{5}{4r}\times\frac{1}{2}\left(\frac{n\theta_{0}}{1}\frac{\sin\left(n\theta_{0}\right)}{0.85}\right)\left(\frac{m_{A,I}^{2}/H^{2}}{0.035}\frac{N_{k}}{50}\right),\label{fNL_NRcurv}
\end{eqnarray}
where we have set the linear coefficient $C_\zeta=2r/(3\theta_{0}S_I)$
and the factor of $5/6$ in the right-hand side (RHS) comes from the standard definition
of the local-type NG for the Bardeen potential, $f_{{\rm NL}}^{{\rm loc}}=\frac{5}{6}f_{{\rm NL}}^{\zeta}$.

\end{enumerate}
From the expression derived in Eq.~(\ref{fNL_NRcurv}), and using
the same benchmark values given below Eq.~(\ref{CDM_fNL_1}), we
find that $\Delta f_{{\rm NL}}^{{\rm loc}}\approx5/(8r)$. Since the
$f_{{\rm NL}}$ in Eq.~(\ref{fNL_NRcurv}) represents a positive
contribution, the contribution from the self-interaction of the axial
field during inflation results in a partial cancellation of the NG
between the contributions from Eqs.~(\ref{fNL_curv_loc}) and (\ref{fNL_NRcurv}).
This could result in an effective $\left|f_{{\rm NL}}^{{\rm loc}}\right|\lesssim O(0.5)$ for $r\sim1$.
\item \textbf{Rotating curvaton}: \\
 Recently, the authors in Ref.~\cite{Co:2022qpr} explored an interesting
scenario where the post-inflationary axial rotations of a complex scalar field associated
with an approximate $U(1)$ symmetry act as a curvaton, due to the lightness
of the angular direction and the longevity of the rotation.\footnote{Large axial rotations of a spectator PQ field \textit{during} inflation have been recently studied in Ref.~\cite{Chung:2024ctx} and give rise to blue-tilted primordial isocurvature fluctuations.} The present-day
entropy of the Universe is created by the (partial) washout of the
$U(1)$ charge, and the fluctuations of the non-zero $U(1)$ charge create
curvature perturbations given by the following expression in Ref.~\cite{Co:2022qpr}
\begin{equation}
\zeta=\frac{\partial_{\theta_{i}}n_{i}}{3n_{i}}\delta\theta(x)+\frac{n_{i}\partial_{\theta_{i}}^{2}n_{i}-(\partial_{\theta_{i}}n_{i})^{2}}{6n_{i}^{2}}\delta\theta^{2}(x)+...~,
\end{equation}
where $n_{i}\equiv n(\theta_{i})$ and $\delta\theta(x)$ are the initial $U(1)$ charge and fluctuations generated at the time of the kick for an angle $\theta_{i}$. Thus, the linear
coefficient $C_\zeta=\partial_{\theta_{i}}n_{i}/(3n_{i}S_I)$. As
the angular field is approximately massless, the amplitude of the
curvature power spectrum is 
\begin{equation}
\mathcal{A}_{\zeta}\approx\left(\frac{\partial_{\theta_{i}}n_{i}}{3n_{i}S_I}\frac{H}{2\pi}\right)^{2}.
\end{equation}
Since the rotation is initiated after inflation, we generalize by
assuming that the $U(1)$-breaking potential responsible for the rotations
can be different from the one in Eq.~(\ref{Vtheta}). Hence, we consider
that $V_{{\rm rot}}\propto\cos(p\theta)$ where $p\leq n$. In the
weak-kick limit, the $U(1)$ charge can be approximated as $n_{i}\sim\partial_{\theta_{i}}V_{{\rm rot}}\propto p\sin(p\theta_{i})$, and hence 
\begin{equation}
\frac{\partial_{\theta_{i}}n_{i}}{3n_{i}}\approx\frac{p}{3\tan(p\theta_{i})}.
\end{equation}
Similar to the oscillating curvaton scenario, the term quadratic
in $\delta\theta$ determines the local NG parameter given as 
\begin{equation}
f_{{\rm NL}}^{{\rm loc}}\approx-\frac{5}{2}\left(1-\frac{n_{i}\partial_{\theta_{i}}^{2}n_{i}}{(\partial_{\theta_{i}}n_{i})^{2}}\right)\underrightarrow{\mbox{\ weak-kick \ }}-\frac{5}{2\cos^{2}(p\theta_{i})}.\label{fNL_Rotcurv_loc}
\end{equation}
We now estimate the contribution to the above NG from the cubic self-interaction
during inflation. In the weak-kick limit, 
\begin{align}
\Delta f_{{\rm NL}}^{{\rm loc}} & \approx\frac{5}{2}\times0.76\left(\frac{n/p}{1}\right)\left(\frac{\tan(p\theta_{i})}{1.55}\frac{\sin\left(n\theta_{0}\right)}{0.85}\right)\left(\frac{m_{A,I}^{2}/H^{2}}{0.035}\frac{N_{k}}{50}\right). \label{fNL_rotating}
\end{align}
Similar to the analysis for the oscillating curvaton, there appears
to be a partial cancellation of the NG with the negative contribution
in Eq.~(\ref{fNL_Rotcurv_loc}). This cancellation becomes more pronounced
as the ratio $n/p$ increases while $\theta_{i}\sim\theta_{0}$.
\end{enumerate}
From the above discussion, we note that the self-interaction of the
axial field fluctuations, arising from an explicit $U(1)$ symmetry-breaking potential
term as given in Eq.~(\ref{Vtheta}) can make significant contribution
to the NG. We demonstrate this specifically for axial fields with
a mass $m_{A}^2\sim O(0.02)H^2$ and an initial displacement $\theta_{0}$
such that the deviation from the quadratic form 
is $\lesssim5-10\%$.

In the curvaton scenario, the net positive contribution from the self-interactions
results in a partial cancellation of the local bispectrum signal. This
cancellation can reduce the magnitude of the $f_{{\rm NL}}^{{\rm loc}}$
in certain regions of the parametric space to $\sim O(0.1)$. Similar
to the generation of local NG from curvaton fluctuations during oscillation,
one can also determine the magnitude of the third-order nonlinear
fluctuations by measuring the four-point correlation signal or the
\textit{trispectrum} ($T$). In terms of the quadratic and cubic nonlinear
parameters $f_{{\rm NL}}$ and $g_{{\rm NL}}$, the trispectrum in
the regular tetrahedron limit ($k_{1,2,3,4}=k_{12}=k_{14}=k$) limit
is given by the expression\footnote{$\zeta=\zeta_{g}+3f_{{\rm NL}}^{{\rm loc}}\zeta_{g}^{2}/5+9g_{{\rm NL}}^{{\rm loc}}\zeta_{g}^{3}/25$
where $\zeta_{g}$ is the Gaussian stochastic fluctuation} 
\begin{equation}
T\equiv\left\langle \zeta\zeta\zeta\zeta\right\rangle '\approx\frac{216}{25}\left(2f_{{\rm NL}}^{2}+g_{{\rm NL}}\right)P_{\zeta}^{3}(k).
\end{equation}
For curvature fluctuations generated by a curvaton oscillating in
a quadratic potential, $g_{{\rm NL}}\approx25/12$, in the limit where
the curvaton energy density dominates the universe at the time of
its decay \cite{Sasaki:2006kq}. Conversely, the curvature trispectrum
signal resulting from the quartic self-interaction of the axial field
fluctuations is given as 
\begin{align}
\left\langle \zeta\zeta\zeta\zeta\right\rangle ' & \approx C^{4}\frac{n^{2}}{6S_I^{2}}\times\frac{H^{4}m_{A}^{2}N_{k}}{k^{9}} ,
\end{align}
where at the linear order $\zeta=C\delta A$. For an oscillating curvaton, the quartic self-interaction of the axial field fluctuations yields \( \Delta g_{{\rm NL}} \approx 25/72 \). Unlike the bispectrum, there is no cancellation of the trispectrum signal, as the two contributions combine constructively. Additionally, the contribution from the self-interaction is relatively smaller in comparison.
A similar conclusion holds if the curvature fluctuations are generated
from a rotating curvaton. Thus, if the axial field acts as a curvaton,
the $U(1)$ symmetry-breaking interaction of the underlying complex scalar
spectator field can give rise to suppressed local-type NG but a comparatively
unsuppressed trispectrum. Models featuring suppressed $f_{{\rm NL}} < O(1)$
but much larger $g_{{\rm NL}}$ have been discussed previously. Some
of these models include self-interacting curvatons, where most calculations
have focused on the self-interaction of the \textit{background} curvaton field,
assuming that the contributions from field fluctuations are negligible \cite{Enqvist:2005pg,Enqvist:2008gk,Huang:2008zj,Enqvist:2009zf,Enqvist:2009eq,Enqvist:2009ww,Enqvist:2010dt,Fonseca:2011aa}. For an overview of other models generating large $g_{\rm NL}$, see a review in Ref.~\cite{Takahashi:2014bxa}.

\subsubsection{Effects of background radial oscillations}\label{sec:classical-effects-radial}

In the previous analysis, we assumed that the radial field remained
stationary at its minimum during inflation. We will now examine the
NG generated from the classical oscillations of a heavy background radial field. For quantum-mechanical effects such as from the excitation of the heavy mode, see Refs.~\cite{Chen:2022vzh,Arkani-Hamed:2015bza,Meerburg:2016zdz,Chen:2023txq}. The radial field is assumed to oscillate
around $S_I$ within the quadratic potential with a curvature defined by mass-squared
quantity $m_{S}^{2}$ as given in Eq.~(\ref{mSsq}). The oscillation of the massive background radial field can be induced by an external coupling that provides a kick along the radial direction. Such a coupling could involve an inflaton, with the kick being triggered by sharp features in the inflaton trajectory, as discussed in certain studies  \cite{Chen:2010xka,Chluba:2015bqa,Chen:2011zf,Slosar:2019gvt,Chen:2022vzh}. It is also possible that the radial
field is initially displaced away from $S_I$ at the beginning
of inflation, and then oscillates at the start of inflation \cite{Chung:2021lfg,Chung:2023xcv}. Similarly,
the radial field may be initially trapped at a false minimum before
being released at a time $t_{s}$ during inflation. Thus, we write
the background radial field as $S(t)=S_I+\Delta S(t)$,
and consider the following EoM
\begin{equation}
\ddot{\Delta S}(t)+3H\dot{\Delta S}(t)+\partial_{S}V(S)=0~, \label{eq:Delta_S_EoM}
\end{equation}
where the overdot represents derivative with respect to proper time
$t$. For small amplitude displacements around $S_I$, we approximate the radial potential as quadratic and
obtain $\partial_{S}V(S)\approx m_{S}^{2}\Delta S(t)$. The solution
to the above EoM is given as
\begin{equation}
\Delta S(t)=S_I\beta_{S}\,e^{-\frac{3}{2}H(t-t_{s})} \cos(\mu_{S}H(t-t_{s})+\varphi) \Theta(t-t_{s}), \label{S_osc}
\end{equation}
where we have assumed that the radial field begins oscillation at
time $t_{s}$ with an initial amplitude defined by the parameter $\beta_{S}\ll1$
so that we can perturbatively analyze the effect of the oscillations.
In the above expression, $\varphi$ is an arbitrary phase and $\mu_{S}=\sqrt{m_{S}^{2}/H^{2}-9/4}$.
We shall consider the case where $\mu_{S}>0$ and real. The classical oscillation
of the background radial field induces a scale-dependent primordial standard clock signal on the
scale-invariant power spectrum of the axial field \cite{Chen:2008wn,Chen:2011zf}. These signals arise
from the resonance between the classically oscillating massive field and
the quantum mode functions \cite{Chen:2008wn}. In terms of proper time $t$,
the massless quantum mode function exhibits a time-dependent frequency.
Resonance amplification occurs when this frequency matches the frequency
of the oscillating radial field, resulting in a $k$-dependence of
the clock signal, $\propto k^{i\mu_S}$. In Appendix~\ref{App:clock}, we compare the clock signal in the power spectrum as obtained through the in-in formalism and the EoM approach. 
Based on the results therein, the amplitude of the oscillatory clock signal on the axial power spectrum, for $\mu_S\gg1$ and $2k>k_r$, is given as
\begin{equation}
\frac{\Delta P^{\rm clock}_{A}}{P_{A}}\approx\beta_{S}\sqrt{2\pi\mu_{S}}\left(\frac{2k}{k_{r}}\right)^{-\frac{3}{2}}\sin\left(\mu_{S}\ln\left(\frac{2k}{k_{r}}\right)+\mu_{S}+\pi/4\right),\label{clock-PA}
\end{equation}
where $-k_{r}\tau_{s}=\mu_{S}$. If the axial field functions as a curvaton, existing constraints on the curvature power spectrum, as detailed in Refs.\cite{Braglia:2021rej,Hamann:2021eyw}, suggest that $|\Delta P_{\zeta}/P_{\zeta}| \lesssim O(0.04)$ for $\mu_S \sim O(10)$.\footnote{In Ref.~\cite{Hamann:2021eyw}, the authors quote best-fit parameter values of $|\Delta P_{\zeta}/P_{\zeta}| \approx 0.07$, $\mu_S\sim100$ at $k_r \sim 0.07/\rm{Mpc}$ by analyzing the Planck CMB data.} This imposes the condition $\beta_{S}\sqrt{\mu_{S}}\lesssim O(0.02)$. The Euclid experiment \cite{EUCLID:2011zbd} is anticipated to make significant progress in probing perturbations in the curvature power spectrum, particularly in the presence of primordial features. It is projected to achieve sensitivity levels of \( \left| \Delta P_{\zeta} / P_{\zeta} \right| \approx 0.010 \pm 0.001 \) \cite{Euclid:2023shr}. Conversely, if the axial field acts as CDM, the current isocurvature constraints are weak,  thus the clock signal can be as large as the scale-invariant spectrum, $P_A$.

The oscillation of the background radial field also introduces corrections
to the cubic interaction term in Eq.~(\ref{A3}), resulting in a
high frequency, time-dependent coefficient for the cubic self-interactions.
Up to first order in $\Delta S$, we write the leading correction
to the cubic-interaction Lagrangian term in Eq.~(\ref{A3}) as 
\begin{equation}
\Delta\mathcal{L}_{3}^{(\delta\theta)^{3}}\approx\frac{n^{3}}{3} g^{(4)} \left(\frac{S_I}{\Lambda\sqrt{2}}\right)^{n}\sin(n\theta_{0})\left(\frac{n\Delta S}{S_I}\right)\,(\delta\theta)^{3}.\label{A3-osc}
\end{equation}
To treat the terms in $\left(1+\Delta S/S_I\right)^{n}$
expansion perturbatively, we require that $n\beta_{S}\ll1$. Similar
to how the clock signal is estimated in the power spectrum, we can
use the in-in formalism to determine the correction to the bispectrum
due to the oscillating background radial field. However, it is important
to note that while the signal from the interaction in Eq.~(\ref{A3-osc})
primarily arises from the subhorizon interactions of the field fluctuations
with the oscillating radial field, the bispectrum in
Eq.~(\ref{BA3}) is predominantly influenced by the large contributions
($\propto N_{k}$) from the superhorizon classical evolution of the
fields. Consequently, the contribution
to the bispectrum from the subhorizon interactions can remain much smaller compared to enhancement from the superhorizon evolution.
An approximate expression for the bispectrum signal from
the oscillating radial field in the equilateral limit ($k_{i}\rightarrow k$) is
\begin{equation}
\lim_{\mu_{S}\gg1,k_{\rm T}\gtrsim k_{r}}\Delta B^{{\rm equiv}}(k)\approx c_3n\beta_{S}\frac{3\sqrt{\pi}H^{2}\left(\frac{k_{\rm T}}{k_{r}}\right)^{-\frac{3}{2}}\sin\left(\mu_{S}\left(\log\left(\frac{k_{\rm T}}{k_{r}}\right)\right)+\mu_{S}-\frac{\pi}{4}\right)}{2k^{6}\sqrt{2\mu_{S}}},\label{eq:cubic-bispectrum-oscillating}
\end{equation}
where $k_{\rm T}=3k$ in the equilateral triangle limit and $\tau_{s}$ is the conformal
time corresponding to the start of the radial oscillations at $t_{s}$.
Compared to the leading bispectrum signal in Eq.~(\ref{BA3}), the
oscillating signal is suppressed 
\begin{align}
\lim_{\mu_{S}\gg1,k_{\rm T}\gtrsim k_{r}}\frac{\Delta B^{{\rm equiv}}(k)}{B^{{\rm equiv}}(k)} & \approx\frac{n\beta_{S}\sqrt{\pi}\left(\frac{k_{\rm T}}{k_{r}}\right)^{-\frac{3}{2}}\sin\left(\mu_{S}\left(\log\left(\frac{k_{\rm T}}{k_{r}}\right)\right)+\mu_{S}-\frac{\pi}{4}\right)}{N_{k}\sqrt{2\mu_{S}}}\nonumber \\
 & \approx O(0.025)\times \left(\frac{n}{10}\frac{50}{N_{k}}\frac{10}{\mu_S}\right)\frac{\Delta P^{\rm clock}_{A}}{P_{A}},
\end{align}where in the final expression, $\beta_S \sqrt{2\pi\mu_S} (k_{\rm T}/k_r)^{-3/2}$ was replaced with the relative amplitude of the clock signal in the power spectrum, as given in  Eq.~(\ref{clock-PA}).
For $n\beta_{S}\sim0.1$, the magnitude of the
\textit{normalized} bispectrum from the subhorizon interaction with the oscillating
radial background is $\propto O(1)/\sqrt{\mu_{S}}$ at $k\sim k_{r}$
compared to an $O(N_{k})$ magnitude from the superhorizon evolution.
Note that, unlike the $\sqrt{\mu_{S}}$ resonant amplification of
the clock signal observed in the power spectrum, the bispectrum in Eq.~(\ref{eq:cubic-bispectrum-oscillating}) exhibits
a suppression proportional to $1/\sqrt{\mu_{S}}$. Thus,
while the power spectrum reveals a clock signal, a similar oscillating
signal is either hidden or substantially suppressed in the bispectrum.
Mathematically, the enhancement in the power spectrum and suppression
in the bispectrum can be understood by analyzing the integral in the
in-in formalism, which is used to evaluate the corrections due to
the oscillating radial background. The general form of the integral
can be written as 
\begin{equation}
I\propto\int_{-\infty}^{0}d\tau e^{-ink\tau}\tau^{p}\times\tau^{\frac{3}{2}}\cos(\mu_{S}\ln(\tau/\tau_{s})),\label{eq:general-integral}
\end{equation}
where $n$ counts the number of massless fields in the interaction
Hamiltonian and $\tau^{p}$ is the effective time-scaling of the interaction
derived from the product of powers of scale factor $a(\tau)$ and
the quantum mode function of the interacting fields. The above oscillating
integral can be solved using the method of stationary phase approximation
which states that the primary contributions to the integral come from
points where the phase function is stationary. For a mode $k$, the
stationary point is $\tau_{k}=-\mu_{S}/(nk)$. Hence, quantum modes with
$k>\mu_{S}/(-n\tau_{s})\equiv k_{r}/n$, have $\tau_{k}>\tau_{s}$ implying
that the largest contribution to the integral for these modes occurs
during their subhorizon evolution. Solving the integral in the subhorizon
region yields 
\begin{equation}
I\equiv I_p\propto\tau_{k}^{p+\frac{3}{2}}e^{i\left(\mu_{S}+\mu_{S}\ln\left(\frac{nk}{k_{r}}\right)\right)}\times\sqrt{\mu_{S}},\label{I-soln}
\end{equation}where the subscript $p$ is introduced to emphasize its dependence on this parameter.
The factor $e^{i\left(\mu_{S}+\mu_{S}\ln\left(\frac{nk}{k_{r}}\right)\right)}$
indicates an oscillatory signal. For massless fields, the oscillatory
signal, $I$, scales as $k^{-3/2}$. Hence, the amplitude of the oscillating
signal can be expressed as $I_p \propto (nk/k_{r})^{-3/2}\mu_{S}^{p+1/2}$. For the power spectrum, where $p=0$,
this results in an enhancement proportional to $\sqrt{\mu_{S}}$.
In contrast, for the bispectrum arising from $U(1)$-breaking cubic self-interaction,
$p=-1$ in Eq.~(\ref{eq:general-integral}). Hence, the signal is suppressed by $1/\sqrt{\mu_S}$. 

Based on this general
result, we deduce that a trispectrum signal, where $n=4$ and $p=1$, arising from similar $U(1)$-breaking axial self-interactions, would exhibit a resonant
enhancement similar to that observed in the power spectrum. Likewise,
all higher $n$-point correlation functions ($n\geq4$) of the axial self-interactions
would receive significant enhancements $\propto\mu_{S}^{n-7/2}$. Thus, higher powers of $\mu_S$ and a large radial mass can counteract the $1/N_k$ suppression, thus generating sizable oscillatory signal in higher-order correlation functions.

In the absence of any other massless field(s) coupled
to the axial field, the remaining sources of bispectrum are highly
suppressed compared to the result in Eq.~(\ref{BA3}). In this work, we do not consider contributions to the bispectrum from cosmological-collider interactions
between massive radial and massless axial field quantum modes. These are expected to be smaller than the oscillatory signal derived above.

Finally, it has been noted in Refs.~\cite{Chen:2009zp,Meerburg:2016zdz} that the self-interaction of an intermediate heavy radial field, could lead to unsuppressed signal in the bispectrum of the light axial field. However, in this work, the self-interaction of the heavy radial field is through higher-dimensional terms which are tightly constrained. Assuming $m_S\sim O(H)$, the coefficient for $(\delta S)^3$ term is approximately given as $(2m_S-1)H^2/S_I$, see Eq.~(\ref{Ham-cubic}).  This suppresses the signal by a factor of $H/S_I$ compared to Refs.~\cite{Chen:2009zp,Meerburg:2016zdz} where the coefficient of cubic self- interaction term is order $H$.

In the following section, we consider scenarios where the radial field
can also be a light scalar. This will lead to an enhanced bispectrum signal from $S-\theta$ interactions.

\subsection{Couplings with the light radial partner \label{sec:light-radial-field}}
In this section, we analyze the bispectrum generated by interactions between the axial and radial scalar fields, with particular emphasis on the $U(1)$ symmetry-breaking terms, under the assumption that both fields are nearly massless. A Hubble-induced mass term for the radial field can be vanishing through either a $D$-flat direction as utilized in the $D$-term inflation models~\cite{Binetruy:1996xj,Halyo:1996pp} or a Heisenberg symmetry in the K\"{a}hler potential as in the no-scale inflation models~\cite{Gaillard:1995az,Campbell:1998yi}. We further assume that the radial potential is nearly quadratic ($V_S \sim m_S^2 S^2$ where $m_S\ll H$) such that the growth of superhorizon perturbations, $\delta S/S$, is expected to be negligible after inflation.
We consider that eventually the radial
field decays or thermalizes with the Standard Model sector. Depending on the decay rate and interaction strengths with the thermal bath that we elaborate on in Appendix~\ref{app:Evolution-of-radial}, this radial field can dominate the energy density after inflation and serve as a curvaton; see Refs.~\cite{Enqvist:2002rf,Enqvist:2003mr,Kasuya:2003va,Hamaguchi:2003dc} in the context of supersymmetry. However, we do not pursue this situation in this work and instead focus on the situations where the radial field thermalizes before dominating. 

In terms of the linear field fluctuations $\delta S$ and $\delta A=S_I\delta\theta$,
we write the quadratic and cubic order Lagrangian as 
\begin{align}
a^{-3}\mathcal{L}_{2} & = -\frac{1}{2}g^{\mu\nu}\left(\partial_{\mu}\delta S\partial_{\nu}\delta S+\partial_{\mu}\delta A\partial_{\nu}\delta A\right)-\frac{1}{2}\left(m_{S}^{2}+3\partial_t{\theta_0}^2\right)\left(\delta S\right)^{2}-\frac{1}{2}m_{A}^{2}\left(\delta A\right)^{2}\label{L2}\\
a^{-3}\delta\mathcal{L}_{2} & = 2\partial_{t}\theta_{0}\left(\delta S\partial_{t}\delta A\right)-m_{A,I}^{2}\sin\left(n\theta_{0}\right)\left(\delta S\delta A\right),\label{dL2}
\end{align}
and 
\begin{align}
a^{-3}\delta\mathcal{L}_{3} = & -\frac{1}{S_{0}}\delta Sg^{\mu\nu}\partial_{\mu}\delta A\partial_{\nu}\delta A+\frac{1}{S_{0}}\partial_{t}\theta_{0}\left(\delta S\right)^{2}\partial_{t}\delta A  \label{dL3} \\
 & -\frac{1}{2S_{0}}\left(m_{A,I}^{2}(n-1)\sin\left(n\theta_{0}\right)\right)\left(\delta S\right)^{2}\delta A -\frac{1}{2S_{0}}\left(m_{A}^{2}n\right)\left(\delta A\right)^{2}\delta S \nonumber \\
 & -\frac{1}{6S_{0}}\left((2m-1)m_{S}^{2}-m_{A}^{2}(1-1/n)(n-2)\right)\left(\delta S\right)^{3}
 -\frac{1}{6S_{0}}\left(-m_{A,I}^{2}n\sin\left(n\theta_{0}\right)\right)\left(\delta A\right)^{3}, \nonumber
\end{align}
where we consider
that the radial field is situated at $S_{0}=S_I\gg H$ during
inflation. In the above set of expressions, $\mathcal{L}_{2}$ is the free field Lagrangian while $\delta\mathcal{L}_{2,3}$ are the interactions treated perturbatively. The complete free-field and interaction Hamiltonian densities can be derived starting from the Lagrangian densities and we give the result 
in Appendix \ref{App:Hamiltonian}.

For light radial and axial fields, the leading-order power spectra are scale-invariant and described by Eq.~(\ref{eq:pwrspec-scaleinvariant}). Quadratic interactions between the radial and axial fluctuations, as outlined in Eq.~(\ref{dL2}), introduce corrections to the power spectra of both fields. These corrections place limitations on the strength of the bispectrum signal arising from the cubic interaction terms. In Appendix \ref{App:pwrbispec}, we evaluate the power spectrum corrections from the quadratic Hamiltonian density terms given in Eq.~(\ref{Ham-quad}). Incorporating these contributions, we obtain the leading correction
to the axial power spectrum as 
\begin{equation}
\frac{\Delta P_{A}(k)}{P_{A}(k)}\approx\frac{2\sin^{2}(n\theta_{0})}{9}\left(\frac{m_{A,I}^{2}N_{k}}{H^{2}}\right)^{2}
\end{equation}
which should be $\ll 1$ to treat these corrections perturbativity.
Using the fiducial values $n\theta_0\sim1$, $m_{A,I}^{2}\approx 0.025H^2$ and $N_k\approx50$,
we obtain $\Delta P_{A}/P_{A}\sim0.25\ll1$.

In addition to the cubic self-interaction of the axial field as explored
in Sec.~\ref{subsec:selfintr}, the the $U(1)$ symmetry-breaking interactions between
the massless radial and axial fields will generate additional
contributions to the axial bispectrum. The details of the bispectrum calculations using the in-in formalism are also provided in Appendix \ref{App:pwrbispec}. As detailed therein, new bispectrum contributions arise from a combination
of cubic and quadratic interaction terms. Below, we present the expression
for the total bispectrum by combining the contributions from Eqs.~(\ref{Type1})-(\ref{Type4}).
\begin{align}
\left\langle \delta A_{k_{1}}\delta A_{k_{2}}\delta A_{k_{3}}\right\rangle ' \approx \, &  \frac{H^{4}}{S_I}\sin\left(n\theta_{0}\right)\left(\frac{7n}{72}b^{2}\cos\left(n\theta_{0}\right)-\frac{5}{36}b^{2}\cos\left(n\theta_{0}\right)+\frac{1}{12}b-\frac{1}{6n}b\right) \nonumber \\
& \times \left(\frac{k_{1}^{3}+k_{2}^{3}+k_{3}^{3}}{k_{1}^{3}k_{2}^{3}k_{3}^{3}}\right), \label{massless-radial-bispectrum}
\end{align}
where 
\begin{equation}
b=\frac{m_{A,I}^{2}N_{k}}{H^{2}}.
\end{equation}
The bispectrum signal exhibits a local shape due to the classical
evolution of the fields outside the horizon. For the fiducial values
of $m_{A,I}^{2}/H^{2}\sim0.025$ and $N_{k}\sim50$, we obtain $b\sim O(1)$.
Thus, when $b\sim1$ and $n\gg1$, the bispectrum given in Eq.~(\ref{massless-radial-bispectrum})
is dominated by the first term, which originates  from the $U(1)$
symmetry-breaking cubic and quadratic interaction Hamiltonian densities,
$\mathcal{H}_{3}^{I}(x,t)\propto\delta S\delta A\delta A$ and $\mathcal{H}_{2}^{I}(x,t)\propto\delta S\delta A$,
respectively. Comparing the bispectrum signal in Eqs.~(\ref{BA3})
and (\ref{massless-radial-bispectrum}), we find that their magnitudes
are similar. Consequently, the non-linearity parameter $f_{{\rm NL}}$
as evaluated in Sec.~\ref{subsec:bispectrum} for various scenarios (CDM, oscillating and rotating curvaton), will receive a similar contribution
from the coupled radial-axial interactions. Using the fiducial values
given above and taking $n\theta_{0}\sim1$, we provide the $f_{{\rm NL}}$
values for the CDM and curvaton scenarios by summing the contributions
from axial self-interaction ($\Delta f_{{\rm NL},\left(\delta A\right)^{3}}$) 
and radial-axial interactions ($\Delta f_{{\rm NL},\left(\delta S-\delta A\right)}$)
as obtained from the bispectra in Eqs.~(\ref{BA3}) and Eq.~(\ref{massless-radial-bispectrum})
respectively
\begin{enumerate}
\item CDM 
\begin{equation}
\Delta f_{{\rm NL}}^{\mathcal{S}} \approx\Delta f_{{\rm NL},\left(\delta A\right)^{3}}^{\mathcal{S}}+\Delta f_{{\rm NL},\left(\delta S-\delta A\right)}^{\mathcal{S}}
\approx\Delta f_{{\rm NL},\left(\delta A\right)^{3}}^{\mathcal{S}}\left(1+\frac{7\cos\left(n\theta_{0}\right)b}{6}\right)
\approx\frac{1}{2\omega_{A}}.
\end{equation}
\item Oscillating curvaton 
\begin{equation}
\Delta f_{{\rm NL}}^{{\rm loc}} \approx\frac{5}{4r}.
\end{equation}
\item Rotating curvaton 
\begin{equation}
\Delta f_{{\rm NL}}^{{\rm loc}} \approx\frac{5}{2}\times\frac{3}{2}\left(\frac{n/p}{1}\right)\left(\frac{\tan\left(\frac{p}{n} n\theta_{i}\right)}{1.55}\right).
\end{equation}
\end{enumerate} 
Hence, if the radial field is treated as a light scalar, similar to
the axial field, significant local-shaped bispectrum signals are
generated from the explicit $U(1)$-breaking interactions between the radial and
axial field fluctuations. Within certain parametric region, these
interactions can yield a positive $\Delta f_{{\rm NL}}^{{\rm loc}}\gtrsim O(1)$,
even though they are suppressed by the axial slow-roll parameter $\eta_{{\rm AA}}\equiv m_{A}^{2}/H^{2}$.
As discussed in Sec.~\ref{subsec:bispectrum}, in the CDM scenario, the contributions
to the NG from the interactions during inflation and from non-linear
mapping of $\delta A$ to $\zeta$ add up constructively, resulting
in $f_{{\rm NL}}^{\mathcal{S}}\approx1/\omega_{A}$.

The curvaton scenario is particularly intriguing, as the NG from the
two contributions have opposite signs. The NG generated by the explicit $U(1)$ symmetry-breaking interactions can partially cancel out with the NG resulting from the non-linear mapping of axial field fluctuations to curvature perturbations. For the benchmark values used in this
work, the cancellation can be quite pronounced in both oscillating and rotating curvaton scenarios, potentially reducing the effective NG to a small
magnitude, $|f_{{\rm NL}}^{{\rm loc}}|\lesssim O(0.1)$. In the rotating curvaton scenario, adjusting the ratio $n/p$ to values slightly greater than 1 introduces additional flexibility in fine-tuning the total $f_{{\rm NL}}^{{\rm loc}}$ parameter.

\section{Explicit $U(1)$-breaking couplings with a light scalar $\phi$}\label{Couplings with a light scalar}
We now examine the bispectrum signals that emerge from interactions
between the axial field $A$ and another light scalar $\phi$. Consistent
with assumptions made earlier in this work, we consider that the
$U(1)$ symmetry is spontaneously broken during inflation, with the radial field settling
at a large field value, $S_I\gg S_{\min}$. For
this analysis, the radial field is assumed to be massive. As in previous discussions, we postulate that the explicit $U(1)$ symmetry-breaking terms are introduced by higher-dimensional, non-renormalizable operators.
\subsection{Couplings with the kinetic and potential terms of $\phi$
\label{subsec:modelA}}
We include the following $U(1)$ symmetry-breaking interaction terms between the light scalar $\phi$ and the complex scalar $P$ 
\begin{equation}
a^{-3}\mathcal{L}\supset-c_{K}\frac{(\partial\phi)^{2}}{3}\left(\frac{P}{\Lambda}\right)^{n}-c_{V}\frac{V_{\phi}}{3}\left(\frac{P}{\Lambda}\right)^{n}+{\rm h.c.~,}\label{U(1)-axial-inflaton}
\end{equation}
where $V_{\phi}$ is the potential of $\phi$ and $(\partial\phi)^{2}=g^{\mu\nu}\partial_{\mu}\phi\partial_{\nu}\phi$. These Lagrangian terms can naturally arise within supersymmetric frameworks, including scenarios with supergravity effects. For a detailed example, see Ref.~\cite{Co:2023mhe}. Additionally, we assume that
$\phi$ and $P$ possess canonical kinetic terms. In terms of the radial
and axial fields, we rewrite the above expression as
\begin{equation}
a^{-3}\mathcal{L}\supset-\left(\frac{V_{\phi}}{3}c_{V}+\frac{(\partial\phi)^{2}}{3}c_{K}\right)\left(\frac{S}{\sqrt{2}\Lambda}\right)^{n}2\cos\left(n\theta\right),\label{eq:U(1)-axial-inf-1}
\end{equation}where the term in parentheses can be identified as the dimensionful quantity 
$g^{(4)}$, in comparison with the potential in Eq.~(\ref{Vtheta}).
Since $P$ is a spectator field, we impose the following conditions
on the derived coefficients $q_{V,K}$:
\begin{equation}
|q_{V}|=\left|\frac{2}{3}c_{V}x_{S}^{n}\right|\ll1,\label{eq:inflation-cond-1}
\end{equation}
\begin{equation}
|q_{K}|=\left|\frac{4}{3}c_{K}x_{S}^{n}\right|\ll1,\label{eq:inflation-cond-2}
\end{equation}
to ensure that these higher-dimensional terms from the $P-\phi$
interactions in Eq.~(\ref{U(1)-axial-inflaton}) induce only a small
correction to the potential and kinetic term of the $\phi$ field.
In the above expressions, $x_{S}=S_I/(\sqrt{2}\Lambda)<1$. The
above conditions imply $c_{V},c_{K} < O(r_a)x_{S}^{-n}$
where $r_a\sim0.1$. The Lagrangian in Eq.~(\ref{eq:U(1)-axial-inf-1}) generates a potential for the axial field, from which the mass-squared term during inflation is derived as
\begin{equation}
m_{A,I}^2=-\left(q_{V}V_{\phi}-q_{K}\frac{(\partial_{t}\phi)^{2}}{2}\right)\frac{n^{2}}{S_I^{2}},
\end{equation}
indicating that the expression within parentheses must be negative to ensure a positive mass scalar field. As we will demonstrate later, the largest NG signals are obtained under the conditions $q_K>0$ and $q_V<0$. Additionally, to keep the axial field light,
we impose the following constraint
\begin{equation}
\left|q_{V}V_{\phi}-q_{K}\frac{(\partial_{t}\phi)^{2}}{2}\right|\frac{n^{2}}{S_I^{2}}\ll H^{2}.\label{axial-mass-undisturbed}
\end{equation}
Unless the coefficients $q_{V}$ and $q_{K}$ are precisely tuned
such that the two contributions on the left-hand side of Eq.~(\ref{axial-mass-undisturbed})
cancel each other, it is expected that the magnitudes of each individual
contribution must be bounded by $\approx O(0.01)H^2$.

We now analyze two distinct scenarios, considering \( \phi \) as either a light curvaton or an inflaton. In the case where \( \phi \) acts as a curvaton, the axial field is treated as CDM. Conversely, when \( \phi \) serves as the inflaton, the axial field can play the role of either CDM or a curvaton.

First, we consider the scenario where $\phi$ acts as
a light curvaton governed by a quadratic potential $V_\phi = m_\phi^2 \phi^2/2$ with mass $m_{\phi}\ll H$ and is frozen at a large vev, $\phi_0\gg H$ during inflation. We treat the axial
field as CDM, with the ratio $S_I/\phi$ constrained by the ratio
of the isocurvature power to the curvature power on CMB scales
\begin{equation}
\alpha_{{\rm iso}}=9\left(\frac{\omega_{A}}{r}\right)^{2}\left(\frac{\phi_{0}}{S_I\theta_{0}}\right)^{2},
\end{equation}
where the superhorizon curvature perturbations generated by the curvaton are defined as $\zeta=2r\delta\phi/\left(3\phi_{0}\right)$, and $r$
represents the ratio of the curvaton's energy density to the total energy density of the universe at the time of its decay. The power spectra
of $\delta\phi,\delta A$ fluctuations are assumed to be scale-invariant. From
observations on CMB scales \cite{Planck:2018jri}, $\alpha_{{\rm iso}}=P_{\mathcal{S}}/P_{\zeta}\lesssim0.04$.
In terms of the curvaton and axial field fluctuations, the quadratic
interaction Hamiltonian density between the two fields is given as
\begin{equation}
a^{-3}\delta\mathcal{H}^{(2)}\supset b_{2}\delta A\delta\phi-b_{3}\delta A\partial_{t}\delta\phi\label{eq:quadratic-inflaton-axial-L2}
\end{equation}
with the coefficients 
\begin{align}
\frac{b_{2}}{H^{2}} & =-\frac{nV_{,\phi}}{S_IH^{2}}q_{V}\sin\left(n\theta\right)\approx\left(\frac{-q_V}{1}\right)\frac{nm_{\phi}^{2}\phi}{S_IH^{2}}\sin\left(n\theta\right),\label{b2-eq}\\
\frac{b_{3}}{H} & =-\frac{n\partial_{t}\phi}{S_IH}q_{K}\sin\left(n\theta\right)\approx\left(\frac{q_K}{3}\right)\frac{nm_{\phi}^{2}\phi}{S_IH^2}\sin\left(n\theta\right),\label{b3-eq}
\end{align}
where in the last expression we used the slow-roll of the
curvaton field to replace $\partial_{t}\phi \approx-m_{\phi}^{2}\phi/(3H)$.
When the quadratic order terms in Eq.~(\ref{eq:quadratic-inflaton-axial-L2})
are treated perturbatively, they introduce
corrections to both the axial ($A$) and curvaton ($\phi$) power spectra as
\begin{equation}
\frac{\Delta P_{A}}{P_{A}}\approx\frac{8b_{3}^{2}}{H^{2}}\left(\frac{1}{12}\right)+\frac{8b_{2}^{2}}{H^{4}}\left(\frac{N_{k}^{2}}{36}\right),\qquad \frac{\Delta P_{\phi}}{P_{\phi}}\approx\left(\frac{b_{3}N_{k}}{H}\right)^{2}.
\end{equation}
To treat these interactions as perturbations over the Hamiltonian density $\mathcal{H}_{0}$, we require 
\begin{align}
\left(\frac{\sqrt{2}b_{2}N_{k}}{3H^{2}}\right)^{2}\equiv O(0.1) & \ll1,\label{b2-constraint}\\
\left(\frac{b_{3}N_{k}}{H}\right)^{2}\equiv O(0.1) & \ll1, \label{b3-constraint}
\end{align}
where $N_{k}\approx50-60$. From the expression given
in Eq.~(\ref{b2-eq}), we obtain
\begin{align}
\frac{\sqrt{2}b_{2}N_{k}}{3H^{2}} & \approx-O(0.4)\times\frac{q_{V}r/\omega_{A}}{10}\sqrt{\frac{\alpha_{{\rm iso}}}{0.04}}\frac{m_{\phi}^{2}/H^{2}}{0.02}\frac{N_{k}}{50}
\end{align}
and a similar expression for $b_{3}N_{k}/H$. 
Combining the constraints in Eqs.~(\ref{eq:inflation-cond-1}), (\ref{eq:inflation-cond-2}), (\ref{axial-mass-undisturbed}), (\ref{b2-constraint}), and (\ref{b3-constraint}), we impose the following upper bound on the magnitudes of $q_{V,K}$
\begin{align}
    |q_V| & \lesssim \min \left\{ r_a,\frac{2.55(\omega_A/r)}{\sqrt{\alpha_{\rm iso}}\left(m_{\phi}/H\right)^2 N_k},\frac{0.36(\omega_A/r)^2}{\alpha_{\rm iso}\left(m_{\phi}/H\right)^2}\right\}, \label{qV-bound} \\
    |q_K| & \lesssim \min \left\{ r_a,\frac{3.6(\omega_A/r)}{\sqrt{\alpha_{\rm iso}}\left(m_{\phi}/H\right)^2 N_k},\frac{3.24(\omega_A/r)^2}{\alpha_{\rm iso}\left(m_{\phi}/H\right)^4} \right\}, \label{qK-bound}
\end{align}where we have taken $n\theta_0\sim1$. We will assume $r_a=0.1$ for the rest of our discussion.

From the interactions in Eq.~(\ref{eq:quadratic-inflaton-axial-L2}), the correlated adiabatic-isocurvature power spectrum $P_{\zeta \mathcal{S}}$ can be estimated as
\begin{equation}
    P_{\zeta \mathcal{S}} \propto P_{\phi A} \approx \frac{H^2}{2k^3}\left(\frac{-2b_2 N_k}{3H^2} + \frac{-b_3 N_k}{H}\right).
\end{equation}Using the bounds given above for the parameters $b_{2,3}$, the correlation fraction $|\cos \Delta| =| P_{\zeta \mathcal{S}}/\sqrt{P_\zeta P_\mathcal{S}}|$ (\cite{Gordon:2000hv}) is approximately $\lesssim 0.4$. Since $b_{2,3}>0$ for $q_K>0$ and $q_V<0$, the adiabatic and isocurvature perturbations are expected to be anti-correlated with $\cos \Delta < 0$.

\subsubsection{Bispectrum}
Finally, we estimate the bispectrum signal by considering the following cubic
order Hamiltonian density terms 
\begin{align}
a^{-3}\delta\mathcal{H}_{3} & \approx h_{1}\delta A\left(-\left(\partial_{t}\phi\right)^{2}+\frac{\left(\partial_{i}\phi\right)^{2}}{a^{2}}\right)+h_{2}(\delta A)^{2}\partial_{t}\delta\phi+h_{3}\delta A(\delta\phi)^{2}+h_{4}(\delta A)^{2}\delta\phi+h_{5}(\delta A)^3, \label{cubic-Inf-axial}
\end{align}
where the coefficients $h_{1-5}$ are 
\begin{align}
h_{1} & = \frac{-1}{2S_I}q_{K}n\sin\left(n\theta\right), \qquad
h_{2}  = \frac{\partial_t{\phi}}{2S_I^{2}}q_{K}n^{2}\cos\left(n\theta\right),\nonumber \\
h_{3} & =\frac{-V_{,\phi\phi}}{2S_I}q_{V}n\sin\left(n\theta\right), \qquad
h_{4}  =\frac{-V_{,\phi}}{2S_I^{2}}q_{V}n^{2}\cos\left(n\theta\right),\nonumber \\
& \hspace{-1cm} \hspace{2.5cm} h_{5}  = \left(q_{V}V_{\phi}-q_{K}\frac{(\partial_t{\phi})^{2}}{2}\right)\frac{n^{3}\sin(n\theta)}{6S_I^{3}}. \label{h1to5}
\end{align}

\textbf{Mixed bispectra}: The interactions with coefficients $h_{1-4}$ generate mixed adiabatic-isocurvature NG signals, where the curvaton and axial fields contribute to the curvature and isocurvature fluctuations respectively. From the first four interaction terms in Eq.~(\ref{cubic-Inf-axial}), we obtain the following three-point correlation functions between the axial and $\phi$ fields 
\begin{align}
\left\langle \delta\phi_{k_{1}}\delta\phi_{k_{2}}\delta A_{k_{3}}\right\rangle ' & \approx \left(-\frac{h_{1}}{2}-\frac{N_k h_3}{H^2}\right)\frac{H^{4}}{2k_{1}^{3}k_{2}^{3}k_{3}^{3}}\sum_{i=1}^{3}k_{i}^{3},\label{h1h3-mixed-bispec}\\
\left\langle \delta A_{k_{1}}\delta A_{k_{2}}\delta\phi_{k_{3}}\right\rangle ' & \approx \left(\frac{N_k h_{2}}{H}-\frac{N_k h_4}{H^2}\right)\frac{H^{4}}{2k_{1}^{3}k_{2}^{3}k_{3}^{3}}\sum_{i=1}^{3} k_{i}^{3}.\label{h2h4-mixed-bispec}
\end{align}
Notably, the bispectrum in Eq.~(\ref{h1h3-mixed-bispec}) does not feature the superhorizon enhancement proportional to $N_k$ for the $h_1$ coefficient, which is present in the other terms. 

Based on the mixed bispectra signals derived above, the non-linearity parameters $f_{{\rm NL}}^{\zeta\zeta \mathcal{S}}$ and $f_{{\rm NL}}^{\mathcal{S}\mathcal{S}\zeta}$ are found to be suppressed by powers of $\alpha_{{\rm iso}}$. For instance, the correlation function $\left\langle \delta\phi_{k_{1}}\delta\phi_{k_{2}}\delta A_{k_{3}}\right\rangle _{\tau}'$ which contributes to the $\zeta\zeta \mathcal{S}$ bispectrum signal, yields $f_{{\rm NL}}^{\zeta\zeta \mathcal{S}}\lesssim O(0.01)$ within the parametric region where $r,\omega_{A}\sim1$.
In the equilateral triangle limit, we find
\begin{align}
\Delta f_{{\rm NL}}^{\zeta\zeta \mathcal{S}} & = \lim_{k_{i}\rightarrow k}\frac{\left\langle \zeta_{k_{1}}\zeta_{k_{2}} \mathcal{S}_{k_{3}}\right\rangle _{\tau}'}{P_{\zeta}^{2}} \nonumber \\
& = \frac{3 \alpha_{\rm iso}} {4r \left(\frac{\omega_A}{r} \right)} \left( q_K + 2 q_V N_k \left(\frac{m_\phi}{H} \right)^2  \right) n \theta_0 \sin(n \theta_0) \label{eq:zzs} \\
& \approx \frac{0.3}{r}\left(\frac{0.01}{\omega_A/r}\right)\left(\frac{q_{K}}{0.1}+\frac{q_{V}}{0.1}\left(\frac{m_{\phi}^{2}}{0.01H^{2}}\frac{N_{k}}{50}\right)\right)\left(\frac{\alpha_{{\rm iso}}}{0.04}\right)\left(\frac{n\theta_{0}\sin\left(n\theta_{0}\right)}{1}\right) , \nonumber
\end{align}
where in the above expression we have used the normalization from Refs.~\cite{Langlois:2011hn,Langlois:2012tm} to define the mixed adiabatic-isocurvature $f_{\rm NL}$ parameters. Considering $|q_{\{K,V\}}|\sim0.1$ and $r\sim1$, a magnitude of $O(1)$ can be obtained for $\omega_{A}<0.01$ for the fiducial values used in the above expression.
Likewise, we evaluate the non-linearity parameter $f_{{\rm NL}}^{\mathcal{S}\mathcal{S}\zeta}$
associated with the bispectrum signal $\left\langle \delta A_{k_{1}}\delta A_{k_{2}}\delta\phi_{k_{3}}\right\rangle _{\tau}'$. In this case, the signal is suppressed by $\alpha_{{\rm iso}}^{2}$, yielding 
\begin{align}
\Delta f_{{\rm NL}}^{\mathcal{S}\mathcal{S}\zeta} & =\lim_{k_{i}\rightarrow k}\frac{\left\langle \mathcal{S}_{k_{1}}\mathcal{S}_{k_{2}} \zeta_{k_{3}}\right\rangle _{\tau}'}{P_{\zeta}^{2}} \nonumber \\
& = \frac{N_k \alpha_{\rm iso}^2} {6r \left(\frac{\omega_A}{r} \right)^2} \left(\frac{m_\phi}{H} \right)^2 \left( q_K - 3 q_V  \right)  n^2 \theta_0^2 \cos(n \theta_0) \label{eq:sszeta} \\
& \approx -\frac{0.1}{r}\left(\frac{0.01}{\omega_A/r}\right)^{2}\left(\frac{q_{K}}{0.1}-3\frac{q_{V}}{0.1}\right)\left(\frac{\alpha_{{\rm iso}}}{0.04}\right)^{2}\left(\frac{m_{\phi}^{2}}{0.01H^{2}}\frac{N_{k}}{50}\right)\left(\frac{n^{2}\theta_{0}^{2}\cos\left(n\theta_{0}\right)}{0.5}\right). \nonumber
\end{align}

\textbf{Pure bispectra}: 
Since Eq.~(\ref{cubic-Inf-axial}) lacks a cubic term in $\delta\phi$, a pure curvaton  bispectra is generated by the quadratic and cubic interactions between the axial and curvaton fields during inflation.
Specifically, Eqs.~(\ref{eq:quadratic-inflaton-axial-L2})
and (\ref{cubic-Inf-axial}) contain two quadratic and four cubic order terms, whose combinations collectively contribute to the total bispectrum signal. Hence, the
pure bispectrum of the curvaton field $\left\langle \delta\phi_{k_{1}}\delta\phi_{k_{2}}\delta\phi_{k_{3}}\right\rangle '$
can be expressed as a sum of the following leading-order contributions:
\begin{equation}
\left\langle \delta\phi_{k_{1}}\delta\phi_{k_{2}}\delta\phi_{k_{3}}\right\rangle '\approx O(h_{1}b_{2})+O(h_{1}b_{3})+O(h_{3}b_{2})+O(h_{3}b_{3}) ,
\end{equation}
where $O(h_{i}b_{j})$ represents the contribution from the cubic
and quadratic interaction vertices proportional to $h_{i}$ and $b_{j}$
in Eqs.~(\ref{eq:quadratic-inflaton-axial-L2}) and (\ref{cubic-Inf-axial}),
respectively. The bispectrum contributions from the interaction of two light scalar fields are evaluated in Appendix \ref{App:pwrbispec}
for various combinations of cubic and quadratic interactions. Using
the expression therein, we obtain 
\begin{align}
\left\langle \delta\phi_{k_{1}}\delta\phi_{k_{2}}\delta\phi_{k_{3}}\right\rangle ' & \approx\left(h_{1}b_{2}\frac{N_{k}}{H^{2}}-h_{1}b_{3}\frac{-3N_{k}}{H}+h_{3}b_{2}\frac{7N_{k}^{2}}{3H^{4}}-h_{3}b_{3}\frac{-5N_{k}^{2}}{H^{3}}\right)\frac{H^{4} \sum_{i=1}^{3} k_{i}^{3}}{12k_{1}^{3}k_{2}^{3}k_{3}^{3}} , 
\end{align}
which yields the contribution to the NG parameter $f_{{\rm NL}}^{\zeta}\equiv f_{{\rm NL}}^{\zeta\zeta\zeta}$ as
\begin{align}
\Delta f_{{\rm NL}}^{\zeta\zeta\zeta}&=\frac{\left\langle \zeta_{k_{1}}\zeta_{k_{2}}\zeta_{k_{3}}\right\rangle _{\tau}'}{3P_{\zeta}^{2}} \nonumber \\
& = \frac{N_k \alpha_{{\rm iso}}}{108r\left(\frac{\omega_A}{r}\right)^2} \left(\frac{m_\phi}{H}\right)^2  \left(-3 q_K^2+\left(3-5\left(\frac{m_\phi}{H}\right)^2 N_k\right) q_K q_V+7\left(\frac{m_\phi}{H}\right)^2 N_k q_V^2\right) \label{zzz} \\
&\approx-\frac{0.05}{r}\left(\frac{0.01}{\omega_A/r}\right)^{2}\left(\frac{q_{K}}{0.1}\right)^{2}\left(\frac{m_{\phi}^{2}/H^{2}}{0.01}\frac{N_{k}}{50}\right)\left(\frac{\alpha_{{\rm iso}}}{0.04}\right)\left(\frac{n\theta_0\sin\left(n\theta_0\right)}{1}\right)^{2} , \nonumber
\end{align}
where, in the last line, terms proportional to \(q_V\) have been neglected due to their subdominant contribution.

The dominant contribution to the three-point correlation function of the
axial field fluctuation given by the interaction term $\propto h_5(\delta A)^3$ in Eq.~(\ref{cubic-Inf-axial}) is given as
\begin{align}
\left\langle \delta A_{k_{1}}\delta A_{k_{2}}\delta A_{k_{3}}\right\rangle '_{\tau} & \approx h_5H^2\frac{(-N_k)\sum_{i=1}^{3}k_{i}^{3}}{2\left(k_{1}k_{2}k_{3}\right)^{3}}.
\end{align}The corresponding contribution to the isocurvature NG parameter $f_{{\rm NL}}^{\mathcal{S}\mathcal{S}\mathcal{S}}$ is
\begin{align}
\Delta f_{{\rm NL}}^{\mathcal{S}\mathcal{S}\mathcal{S}}&=\frac{\left\langle  \mathcal{S}_{k_{1}}\mathcal{S}_{k_{2}}\mathcal{S}_{k_{3}}\right\rangle _{\tau}'}{3P_{\mathcal{\zeta}}^{2}} \nonumber \\
& = \frac{a_{\rm iso}^3 N_k }{972 r \left(\frac{\omega_A}{r} \right)^3} \left(\frac{m_\phi}{H}\right)^2 \left(q_K \left(\frac{m_\phi}{H}\right)^2 - 9 q_V \right) n^3\theta^3 \sin(n \theta) \label{sss} \\
& \approx \frac{0.03}{r}\left(\frac{0.01}{\omega_{A}/r}\right)^{3}\left(10^{-3}\frac{q_{K}}{0.1}\frac{m_{\phi}^{2}}{0.01H^{2}} - \frac{q_{V}}{0.1}\right)\left(\frac{\alpha_{{\rm iso}}}{0.04}\right)^3\left(\frac{m_{\phi}^{2}}{0.01H^{2}}\frac{N_{k}}{50}\right)\left(\frac{n^{3}\theta_0^{3}\sin(n\theta_0)}{1}\right). \nonumber
\end{align}
We remark that the NG parameter $f_{\rm NL}^{\mathcal{S}\mathcal{S}\mathcal{S}}$ is related to $f_{\rm NL}^{\mathcal{S}}$ (defined through Eq.~(\ref{fNLi})) by $f_{\rm NL}^{\mathcal{S}\mathcal{S}\mathcal{S}}=\alpha_{\rm iso}^2f_{\rm NL}^{\mathcal{S}}$. This difference arises due to the choice of the normalizing power spectrum. Similar to the evaluation of the curvature NG, one could add higher-order contributions from combinations of quadratic and cubic vertices to the isocurvature NG estimation. However, these only contribute a small fraction compared to the above estimate.
\begin{figure}[th]
\begin{centering}
\includegraphics[width=0.49\columnwidth]{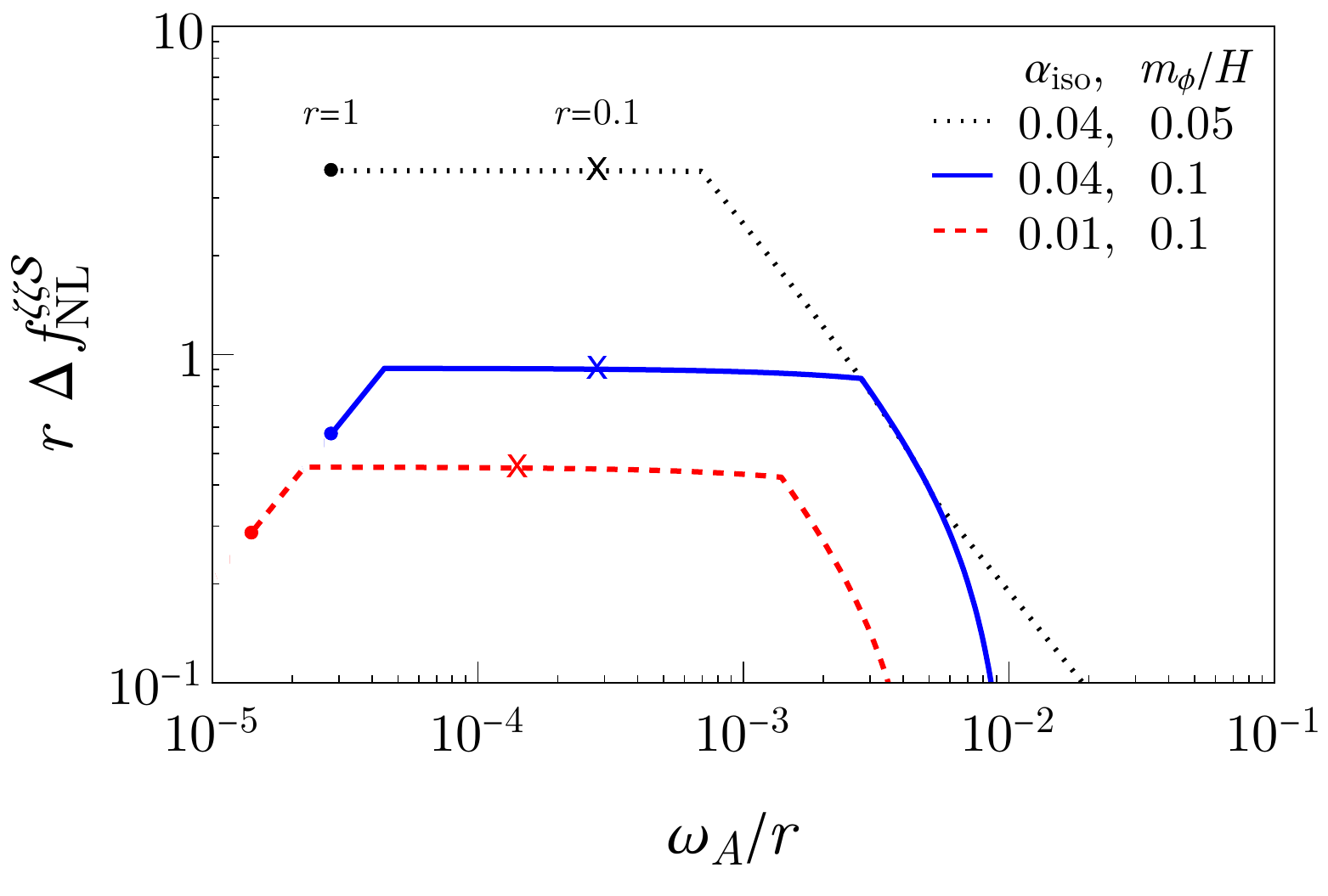}\hspace{0.2cm}{\includegraphics[width=0.49\columnwidth]{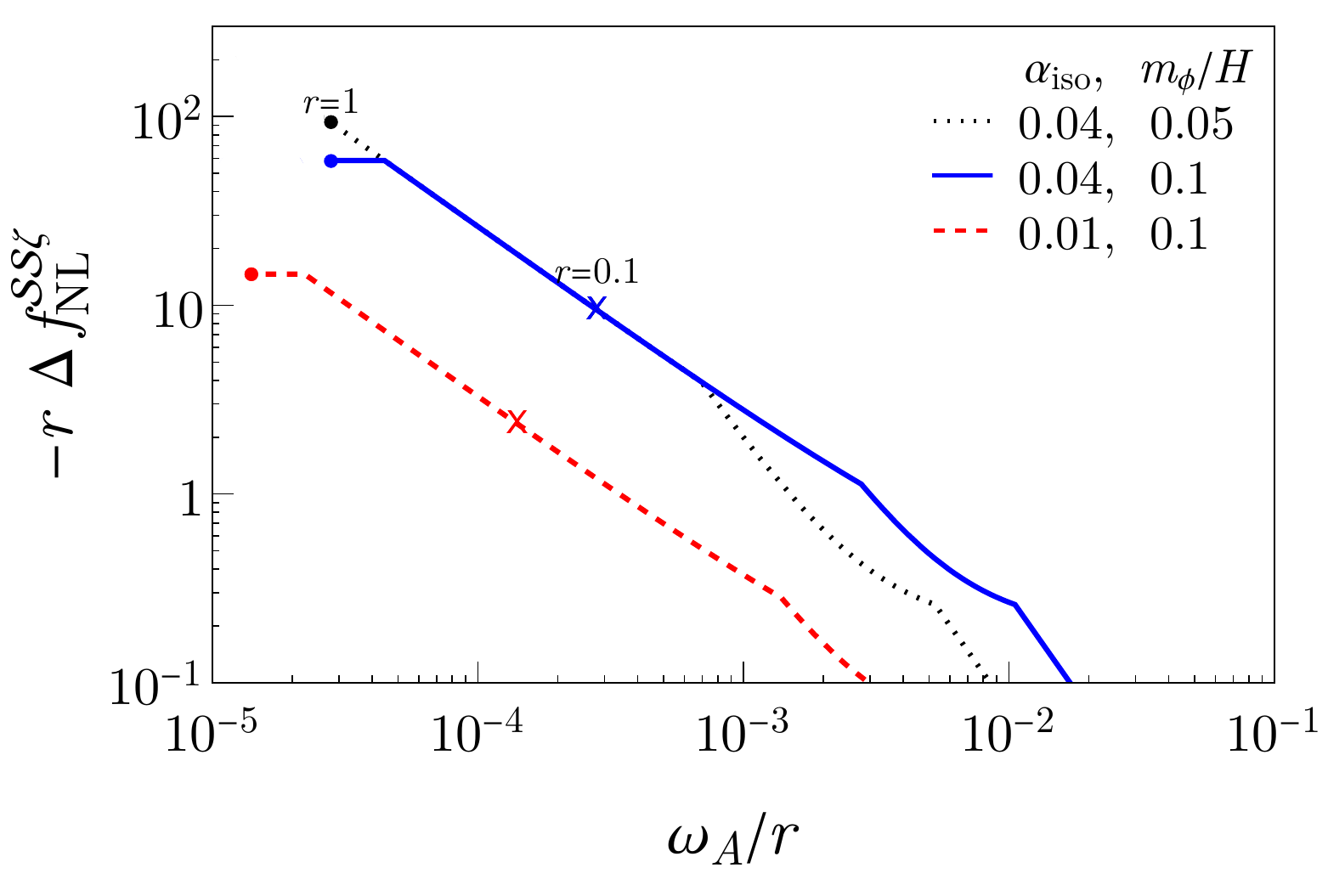}}\hspace{0.2cm}{\includegraphics[width=0.49\columnwidth]{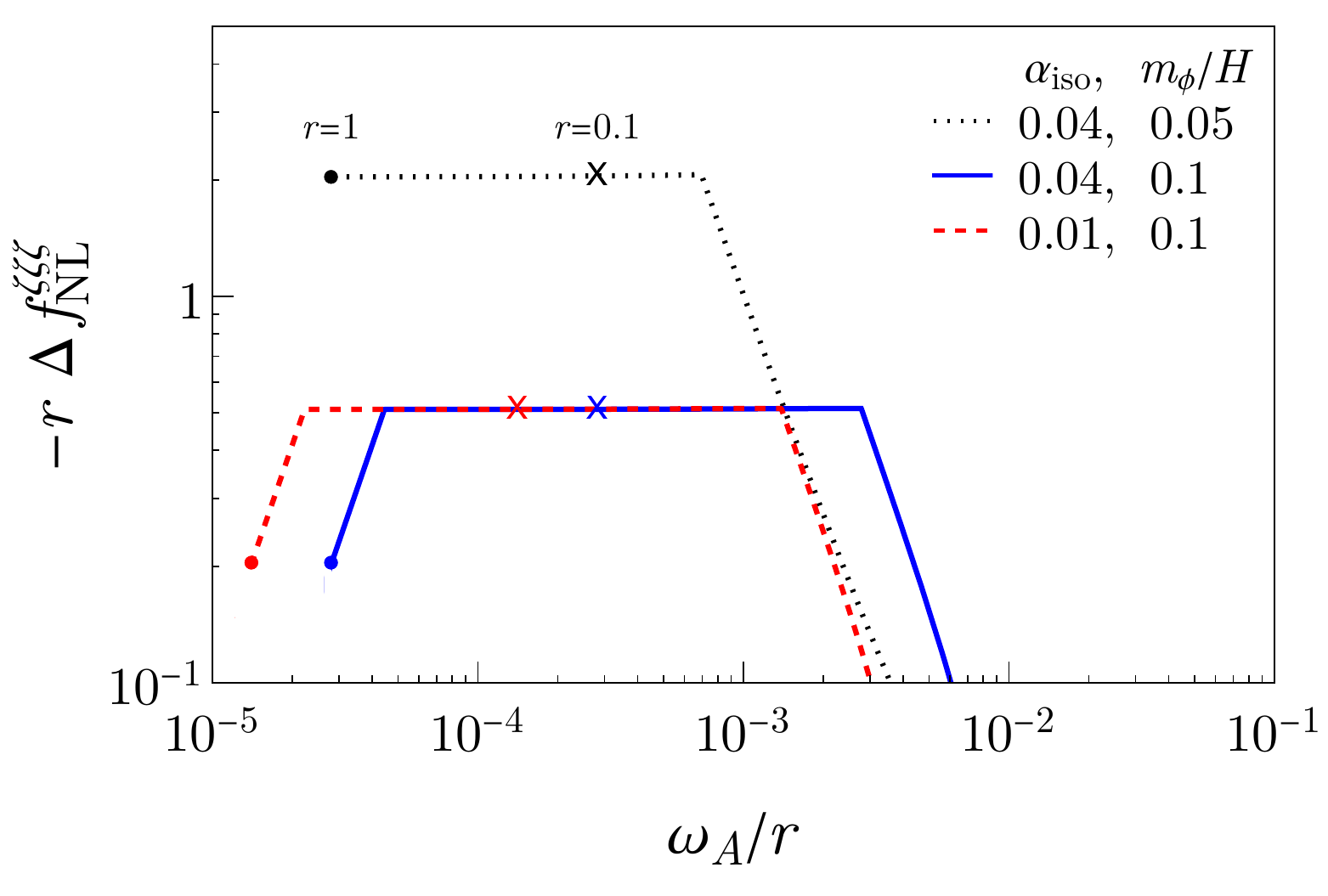}}
{\includegraphics[width=0.49\columnwidth]{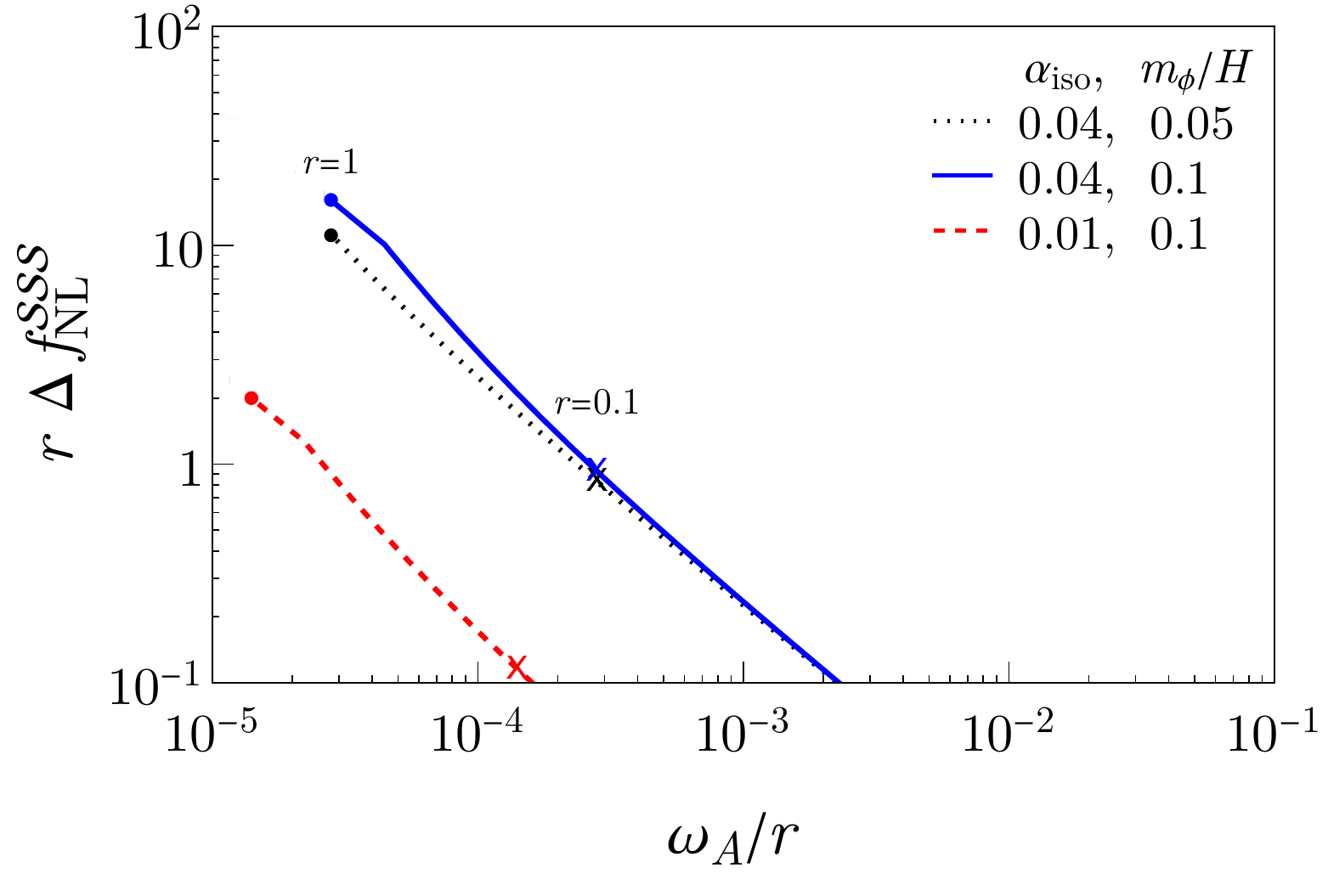}}
\par\end{centering}
\caption{\label{fig:bispec-A-phi-1}  From top left to bottom right, figures of the upper bound of NG parameters, $r\Delta f_{{\rm NL}}^{\zeta\zeta \mathcal{S}}$, $-r\Delta f_{{\rm NL}}^{ \mathcal{S}\mathcal{S}\zeta}$, $-r\Delta f_{{\rm NL}}^{\zeta\zeta\zeta}$ and $r\Delta f_{{\rm NL}}^{\mathcal{S}\mathcal{S}\mathcal{S}}$ (normalized with $r$, the fraction of curvaton's energy density to the
total energy at the time of the decay) as functions of $\omega_A/r$ with $\omega_A$ as the axial-field dark matter fraction, for various choices of curvaton mass $m_{\phi}$ and isocurvature fraction $\alpha_{\rm iso}$ while $N_k$ is fixed at $50$, $n\sim10$ and $n\theta_0 \sim 1$. These bounds are obtained from Eqs.~(\ref{eq:zzs}), (\ref{eq:sszeta}), (\ref{zzz}) and (\ref{sss}) respectively by saturating the magnitude of $q_{V,K}$ as defined in Eqs.~(\ref{qV-bound}) and (\ref{qK-bound}). We take $q_K>0$ and $q_V<0$ for positivity of the axial mass. The curves have been truncated as marked by dot(cross) for $r=1(0.1)$ to ensure that $S_I/H \geq 10$. The peak magnitude of the pure and mixed curvature bispectra are primarily dominated by the $q_K$-dependent terms and exhibit an inverse dependence on curvaton mass-squared. The curve for isocurvature NG parameter $f_{\rm NL}^{\mathcal{S}\mathcal{S}\mathcal{S}}$ displays a typical $1/\omega_A$ dependence. The kinks in the curves are due to the piecewise functions given in Eqs.~(\ref{qV-bound}) and (\ref{qK-bound}).}
\end{figure}

Based on the preceding analysis, we conclude that the adiabatic-isocurvature bispectra from the axial-curvaton interactions are $\ll 1$ for $\{r,\omega_{A}\}\sim1$. In Fig.~\ref{fig:bispec-A-phi-1}, we show upper bounds on the magnitude of the various adiabatic-isocurvature mixed and pure NG parameters derived in Eqs.~(\ref{eq:zzs}), (\ref{eq:sszeta}), (\ref{zzz}) and (\ref{sss}) while saturating the constraint on the magnitude of $q_{V,K}$ in Eqs.~(\ref{qV-bound}) and (\ref{qK-bound}). The magnitudes are normalized with parameter $r$. For the fiducial values of $\alpha_{\rm iso}\sim0.04$ and $N_k\sim50$, the NG is predominantly driven by the contribution from $q_K$.  
For significantly smaller dark matter fractions corresponding to $\omega_{A}\lesssim0.01$, these bispectra signals can be enhanced to
$O(1)$. Notably, $f_{{\rm NL}}^{\mathcal{S}\mathcal{S}\zeta}$ and $f_{{\rm NL}}^{\mathcal{S}\mathcal{S}\mathcal{S}}$ exhibit the typical $1/\omega_A$ enhancement as observed in Eq.~(\ref{CDM_fNL_1}), and thus become much larger in magnitude for sufficiently small $\omega_A$. As shown in the figure, for \(\omega_A < 0.01\) (assuming \(r \sim 1\)), the curve for \(f_{\rm NL}^{\mathcal{S}\mathcal{S}\mathcal{S}}\) can be well approximated by the expression \(0.5(\alpha_{\rm iso}/0.04)^2/(2\omega_A)\). Comparing this result with the expression in Eq.~(\ref{CDM_fNL_1}), we infer that the induced mass-squared of the axial field is approximately \(O(0.03H^2)\). For a fixed \( \alpha_{\rm iso} \), the parameter \( \omega_A \) is subject to a lower bound due to the constraint \( S_I/H \geq 10 \). Consequently, in Figure~\ref{fig:bispec-A-phi-1}, the curves are truncated at the minimum \( \omega_A \) values, indicated by dots for \( r = 1 \) and crosses for \( r = 0.1 \), to ensure compliance with this condition. 

The bound on the parameters $q_{V,K}$ as given in Eqs.~(\ref{qV-bound}-\ref{qK-bound}) transition across three distinct piecewise regions. Consequently, the plots in Fig.~\ref{fig:bispec-A-phi-1} display curves with three characteristic slopes, reflecting the change in the dominant terms governing the upper bounds. Notably, with the exception of the pure isocurvature NG, all pure and mixed curvature NG signals scale inversely with the square of the curvaton mass when the bound on $q_K$ transitions to the second or third piecewise regions.\footnote{The black-dotted curve for $-r\Delta f_{\rm NL}^{\mathcal{S}\mathcal{S}\zeta}$ must be extended beyond the truncated point to observe this behavior in Fig.~\ref{fig:bispec-A-phi-1}.} This behavior arises from the parametric dependence of $q_{K}$ on the curvaton mass within the three piecewise regions, combined with the functional dependence of the NG parameters on $q_K$ and other variables such as $\omega_A/r$. 
As illustrated through these figures, for a fiducial curvaton mass $m_{\phi}\sim0.05H$ and $r\sim1$, it is feasible to generate sizable curvature NG signals with magnitudes corresponding to $f_{\rm NL}^{\zeta\zeta \mathcal{S}}\sim 5$, $f_{\rm NL}^{ \mathcal{S}\mathcal{S}\zeta}\sim 100$, $f_{\rm NL}^{\zeta\zeta\zeta}\sim 2$ under the condition that the axial field contributes negligibly to the CDM density. Thus, the explicit $U(1)$ symmetry-breaking interaction considered in Eq.~(\ref{U(1)-axial-inflaton}) can generate sizable adiabatic-isocurvature mixed and pure NG signals.

The current sensitivity from CMB data analysis, as reported by \cite{Planck:2019kim}, and the projected future sensitivity of experiments such as LiteBIRD~\cite{Hazumi:2012gjy,LiteBIRD:2020tzb,LiteBIRD:2023iiy} and CMB-S4~\cite{Abazajian:2019eic,CMB-S4:2016ple}, as discussed in Ref.~\cite{Montandon:2020kuk}, indicate that $\sigma(f_{{\rm NL}}^{\zeta,\zeta \mathcal{S}})\sim5$,  $\sigma(f_{{\rm NL}}^{\zeta,\mathcal{S}\mathcal{S}})\sim100$ and $\sigma(f_{{\rm NL}}^{\mathcal{S}\mathcal{S}\mathcal{S}})\sim100$ are achievable. These sensitivity levels suggest that for $\omega_A\lesssim 0.01$, the enhanced  bispectrum signals may become detectable in future experiments. We note that to estimate the amplitude of the mixed adiabatic-isocurvature NG parameters, the bispectrum templates used in Refs.~\cite{Planck:2019kim} and \cite{Montandon:2020kuk} followed the definitions in Refs.~\cite{Langlois:2011hn,Langlois:2012tm}.
However, as noted in Ref.~\cite{Geller:2024upd}, the shapes of the mixed bispectrum signals described in Eqs.~(\ref{h1h3-mixed-bispec}) and (\ref{h2h4-mixed-bispec}) differ from the local bispectrum templates defined in Refs.~\cite{Langlois:2011hn,Langlois:2012tm}. Consequently, a focused search using the templates specifically tailored to the signals in Eqs.~(\ref{h1h3-mixed-bispec}) and (\ref{h2h4-mixed-bispec}) could significantly enhance the sensitivity of these analyses, improving the potential for detecting mixed bispectrum signals.
A key distinction between the inflaton and curvaton scenarios is that the inflaton field has a significantly large energy density ($\rho_{\rm inf}\sim H^2M^2_{\rm Pl}$) compared to a spectator curvaton field. Accordingly, when we assume the $\phi$ field as an inflaton, then couplings similar to those in Eq.~(\ref{eq:U(1)-axial-inf-1}) require that the coefficients $q_{V,K}$ be kept sufficiently small to ensure that the axial field remains light during inflation. Following the reasoning outlined earlier in this section, the bounds on \( |q_K| \) and \( |q_V| \) are now expressed as
\begin{eqnarray}
|q_V| &\lesssim& \min\left\{r_a, ,\frac{0.27\omega_A}{\sqrt{\alpha_{\rm iso}}N_k},\frac{0.027\omega_A^2(\partial_t\phi)^2}{\alpha_{\rm iso} H^2M^2_{\rm Pl}}\right\}, \\
|q_K| &\lesssim& \min\left\{r_a,\frac{0.8\omega_A}{\sqrt{\alpha_{\rm iso}}N_k}, \frac{0.16\omega_A^2}{\alpha_{\rm iso} }\right\}.
\end{eqnarray} Compared to a similar set of expressions derived in Eqs.~(\ref{qV-bound}) and (\ref{qK-bound}), we observe that the factors of $m_\phi /H$ are absent in the denominators. Consequently, the enhancement associated with $m_\phi/H \ll 1$ does not occur. As a result, the magnitudes of the pure and mixed curvature bispectrum signals originating from axial-inflaton couplings are relatively suppressed, reaching an approximate upper bound of $O(0.1)$ for $\mathcal{S}\mathcal{S}\zeta$ and $O(0.01)$ for the remaining two, even for small values of $\omega_A$. For pure isocurvature NG, which is approximately independent of $m_\phi$ for $\omega_A\ll 1$,  we obtain 
 a maximum $\Delta f_{\rm NL}^{\mathcal{S}\mathcal{S}\mathcal{S}} \sim O(10)$.

In models (such as in Ref.~\cite{Chen:2023txq}) where the axial and inflaton fields are not coupled directly due to the lack of $U(1)$-breaking interactions, the coupling can be mediated indirectly through a heavy radial field. Unlike our scenario, in these cases, the interaction coefficients are not constrained by the lightness of axial mass. For instance, in Ref.~\cite{Chen:2023txq}, \(\omega_A\) serves as a free parameter that can significantly enhance the signal when \(\omega_A \ll 1\). However, in our analysis, for \(\omega_A \lesssim 0.01\), the bound on $q_K$ transitions to \( |q_K| \lesssim 4\omega_A^2 \), as inferred from the fiducial values. Thus for \(\omega_A \ll 0.01\), the resulting \( f_{\rm NL} \) values become independent of \(\omega_A\) and there are no additional parameters available to enhance \( f_{\rm NL} \).

If the axial field is a curvaton, an axial mass-squared of approximately $0.03 H^2$ can be achieved for  $S_I \lesssim 0.1 ~\partial_t \phi/H$, while satisfying all other constraints. Under these conditions, the axial bispectrum arising from the self-interaction term proportional to $h_5$ in Eq.~(\ref{cubic-Inf-axial}) can generate local NG similar in magnitude to those obtained in Eqs.~(\ref{fNL_NRcurv}) and (\ref{fNL_rotating}) for oscillating and rotating-type curvaton models.

\subsubsection{Effects of background radial oscillations}\label{subsec:oscille-radial-field}
Similar to the analysis in Sec.~\ref{sec:classical-effects-radial}, we now examine
the scenario where the background radial field undergoes small
amplitude oscillations around its local minimum, $S_I$. These oscillations are assumed to follow the solution described in Eq.~(\ref{S_osc}). If the radial oscillation is excited by interactions with the $\phi$ field, the amplitude $\beta_S$ in Eq.~(\ref{S_osc}) is determined by the interaction/coupling strength between the two fields. Since experimental searches for primordial clock signals typically assume a free-amplitude, we will treat $\beta_S$ as a free-parameter in this analysis. 

The oscillating
background will give rise to correlated clock signals within the curvaton/inflaton and axial power spectra. The dominant contribution to the clock signal in the axial power spectrum
is given in Eq.~(\ref{clock-PA}).  If the curvature fluctuations are given by the $\phi$ field and the dominant $\phi$ interactions are given by the $U(1)$-breaking Lagrangian term in Eq.~(\ref{U(1)-axial-inflaton}), the correlated clock signal in curvature power spectrum is given as
\begin{equation}
    \frac{\Delta P^{\rm clock}_{\zeta}}{P_{\zeta}}=\frac{\Delta P^{\rm clock}_{\phi}}{P_{\phi}}\approx0.03\times\frac{q_{K}}{0.05} \frac{n\beta_{S}}{0.2}\sqrt{\frac{\mu_{S}}{10}}\left(\frac{2k}{k_{r}}\right)^{-\frac{3}{2}}\sin\left(\mu_{S}\ln\left(\frac{2k}{k_{r}}\right)+\mu_{S}+\pi/4\right), \label{clock-phi}
\end{equation}where the contribution from the term $\propto q_V$ is comparatively subdominant.\footnote{Conversely, if $\phi$ is an inflaton field and the axial field functions as a spectator curvaton, the constraints on the amplitude of the clock signal in curvature power spectrum from \cite{Braglia:2021rej,Hamann:2021eyw} imply that $\beta_S\sqrt{\mu_S}\lesssim O(0.02)$.}
For the fiducial parameter values considered above, the amplitude of the clock signal in the CDM (axial) isocurvature power spectrum is estimated to be approximately $30\%$ of the scale-invariant power spectrum. In comparison, we observe that the clock signal in the $\phi$-field's power spectrum is limited by the factor $n q_K / 2 $, with the parameter $q_K$ subject to the constraint in Eq.~(\ref{qK-bound}). Consequently, the clock signal is typically expected to contribute less significantly to the curvature power spectrum than to the isocurvature power spectrum. Clock signals in the correlated mixed power spectrum \( P_{\zeta \mathcal{S}} \) can also arise due to the interactions specified in Eq.~(\ref{eq:quadratic-inflaton-axial-L2}). By comparing with the integral solution \( I_p \) in Eq.~(\ref{I-soln}), we find that the parameter \( p \) takes the values \( p = -2 \) and \( p = -1 \) for the interactions proportional to \( b_2 \) and \( b_3 \), respectively, in Eq.~(\ref{eq:quadratic-inflaton-axial-L2}). Consequently, for massive radial fields, we find that clock signals in \( P_{\zeta \mathcal{S}} \) are not expected, as their amplitudes are suppressed by factors of \( \mu_S^{p + 1/2} \).

To assess the amplitude of the oscillatory bispectrum signal, we note that the coefficients $h_i$ are proportional to $S^n$. Thus, we
modify the cubic interaction vertices in Eq.~(\ref{cubic-Inf-axial})
to $h'_{i}$ where $h'_{i}=h_{i}+\Delta h_{i}$, and at the linear order
in oscillations, $\Delta h_{i}/h_{i}=n\Delta S/S_I$. From the discussion in Sec.~\ref{sec:classical-effects-radial}, we know that the majority of the signal from the oscillating background is produced while the mode functions remain subhorizon. Given that the radial field's oscillations decay as $\exp(-3H(t-t_{s})/2)$, their influence persists for only a few e-folds while the mode functions are subhorizon.

In scenarios where the bispectrum is predominantly generated by prolonged
superhorizon interactions of the mode functions such
as in Eq.~(\ref{BA3})—where the bispectrum is proportional to
powers of $N_{k}$, the subhorizon effects of an oscillating background
can be negligible unless amplified by powers of the frequency $\mu_{S}$.
This was noted in Sec.~\ref{sec:classical-effects-radial} where we showed that the bispectrum
in Eq.~(\ref{eq:cubic-bispectrum-oscillating}) is subdominant compared
to that in Eq.~(\ref{BA3}). Among the various bispectrum signals
evaluated in the preceding section, the mixed $\phi\phi A$ three-point
correlation proportional to $h_1$ in Eq.~(\ref{h1h3-mixed-bispec}) is the only signal that
is independent of $N_{k}$ since it is not enhanced by the superhorizon
evolution of the mode functions. From the discussion following Eq.~(\ref{eq:general-integral}), we predict that the contribution
from the $\Delta h_{1}$ term, $a^{-3}\mathcal{H}_{3}\supset\Delta h_{1}\delta A\left(\partial\delta\phi\right)^{2}\approx h_{1}\frac{n\Delta S(t)}{S_I}\delta A\left(\partial\delta\phi\right)^{2}$,
will induce an oscillatory bispectrum signal with an enhancement $\propto\mu_{S}^{3/2}$.
In quantitative terms, in the equilateral triangle limit, the signal in Eq.~(\ref{h1h3-mixed-bispec})
receives a correction which has a simple analytical expression in
the limit $\mu_{S}\gg1,\,3k\gtrsim k_{r}= -\mu_S/(\tau_s)$
\begin{align}
\lim_{k_{i}\rightarrow k}\left\langle \delta\phi_{k_{1}}\delta\phi_{k_{2}}\delta A_{k_{3}}\right\rangle ' & \approx-h_{1}\,\frac{n\beta_{S}}{4k^{6}}\sqrt{\frac{\pi\mu_{S}^{3}}{2}}H^{4}\left(\frac{3k}{k_{r}}\right)^{-\frac{3}{2}}\cos\left(\mu_{S}\ln\left(\frac{3k}{k_{r}}\right)+\mu_{S}+\frac{\pi}{4}\right).
\end{align}
When compared with the strength of the total three-point correlation signal from Eqs.~(\ref{h1h3-mixed-bispec}), treating $\phi$ as a curvaton, the relative amplitude of the oscillatory signal is given by
\begin{align}
\lim_{k_{i}\rightarrow k}\frac{\Delta B^{\zeta\zeta \mathcal{S}}}{B^{\zeta\zeta \mathcal{S}}}&\approx \frac{n\beta_{S}\sqrt{2\pi\mu^3_S}}{6}\left(\frac{3k}{k_{r}}\right)^{-\frac{3}{2}} \nonumber \\
&\approx 0.54\left(\frac{0.1}{q_{K}}\right)\left(\frac{\mu_{S}}{10}\right)\frac{\Delta P_{\zeta}^{{\rm clock}}}{0.03P_{\zeta}},\label{osc-ZZS}
\end{align}where in the final expression the relative strength of the primordial clock signal in the curvature power spectrum is introduced as a constraint. The oscillating clock signal in the $\zeta\zeta \mathcal{S}$ bispectrum signal can be significantly enhanced for a highly massive background radial field, $\mu_S > 10$ and $q_K<0.1$. This enhancement is driven by the resonant amplification as discussed in Sec.~\ref{sec:classical-effects-radial}.

A similar analysis of the mixed \( \mathcal{S}\mathcal{S}\zeta \) bispectrum signal shows that the relative amplitude of the oscillatory component arising from the \(\Delta h_{2,4}\) interaction vertices scales as \(\sqrt{\mu_S}/N_k\). This suppression by \(N_k\) significantly diminishes the signal, reducing it to \(O(0.003)\) compared to the $O(0.5)$ relative magnitude given in Eq.~(\ref{osc-ZZS}). Similarly, the relative amplitudes of the oscillatory bispectrum signals in the curvature and isocurvature bispectra are also suppressed by a factor proportional to the inverse of $N_k$. Thus, for the model analyzed in this section, while correlated clock signals are present in both the curvature and isocurvature power spectra, a comparable clock signal is likely to be measurable only in the \(\zeta\zeta \mathcal{S}\) bispectrum. Additionally, the oscillatory nature of these bispectrum
signals makes them easily distinguishable from a local-shape bispectrum
of similar magnitude. Therefore, constraints derived from local-shape
analyses may not be particularly relevant for these unique signals.

\subsection{Kinetic mixing with $\phi$}\label{Sec:KineticMixing}
Lastly, we will study the NG signals generated from a $U(1)$ symmetry-breaking kinetic mixing between the axial field $A$ and another light scalar $\phi$. To generate
such a Lagrangian term, we consider the following contribution to
the K\"{a}hler potential $K$ within supergravity
\begin{equation}
K\supset\left(I+I^{*}\right)^{2}+|P|^{2}+ic_{M}\left(I+I^{*}\right)\frac{P^{n}}{\Lambda^{n-1}}+{\rm h.c.} ~,
\end{equation}
where $c_{M}$ is real and we decompose the complex scalar $I$ as
$I=(\chi+i\phi)/\sqrt{2}$. The K\"{a}hler potential enjoys a shift
symmetry in the field $\phi$. This approach is similar to that used in
Refs.~\cite{Kawasaki:2000yn,Kallosh:2010ug} to avoid the $\eta$-problem in inflation and thus ensures that $\phi$ remains light. Meanwhile,  the heavier
partner $\chi$ is driven to a field value close to zero. 

In terms of $\epsilon=c_{M}nx_{S}^{n-1}$ where $x_{S}=S_I/(\sqrt{2}\Lambda)$, the last term in the K\"{a}hler potential generates the following kinetic term
\begin{equation}
\mathcal{L}_{{\rm kin}}  \supset -a^{3} \epsilon \, g^{\mu\nu}\partial_{\mu}\phi\left(\sin\left(n\theta\right)\partial_{\nu}A-\cos\left(n\theta\right)\partial_{\nu}S\right) ,
\end{equation}
where the $\sin  (n\theta)$ and $\cos (n\theta)$ terms explicitly break the shift symmetry. For the following analysis, we assume that $\phi$ functions as an inflaton field, with the results for the curvaton scenario discussed later.

If the radial field is heavy
and $\partial_{t}S=0$, the EoM for the background axial field is
\begin{equation}
\partial_{t}^{2}A+3H\partial_{t}A+\partial_{A}V_{{\rm \cancel{U(1)}}}\approx-\epsilon\sin(n\theta_{0})3H\partial_{t}\phi \, , \label{eq:axial-EoM-mixing}
\end{equation}
where we have assumed slow-roll for the inflaton field\footnote{An equivalent expression
for the EoM of the inflaton is given as
\begin{equation}
\partial_{t}^{2}\phi+3H\partial_{t}\phi+\partial_{\phi}V_{{\rm \phi}}\approx-\epsilon\sin(n\theta_{0})3H\partial_{t}A.
\end{equation}
To ensure that the axial field does not influence the inflaton dynamics, we require $\epsilon^{2}\ll1$.} and the term on the right-hand side generates a finite mass-squared term, $3\epsilon n\partial_{t}\phi/(HS_I)\,H^2$, for the axial field. We note that $\epsilon\,\partial_{t}\phi$ must be $>0$ for a downward
roll of the axial field in the region where $|n\theta|<\pi/2$.

For simplicity of our analysis, we will consider that the axial slow-roll
is dominated by the $\partial_{t}\phi/S_I$ term rather than
an axial mass induced by an explicit $U(1)$-breaking potential. As the
inflationary era ends and the inflaton begins oscillation, $\partial_{t}\phi$ rapidly
redshifts, allowing the axial field to settle into the $U(1)$ symmetry-breaking potential. To ensure that the axial field also undergoes
a slow-roll motion we require that
\begin{equation}
n\epsilon\frac{\partial_{t}\phi}{S_IH} \lesssim  0.01, \label{eq:KM-constraint-1}
\end{equation}where as established previously we take $\epsilon \partial_t \phi>0$.

In terms of the inflaton and axial field fluctuations and up to linear order in $\epsilon$, the quadratic and cubic order interaction Hamiltonian density terms  arising from the kinetic mixing are given as
\begin{align}
a^{-3}\mathcal{H}_{{\rm int}}^{(2)} \approx \ & \epsilon\sin(n\theta)g^{\mu\nu}\partial_{\mu}\delta\phi\partial_{\nu}\delta A  \label{eq:cM-Ham-quadratic} \\
& + \epsilon n\cos(n\theta)\partial_{t}\theta\delta A\partial_{t}\delta\phi+\epsilon n\cos(n\theta)\frac{\partial_{t}\phi}{S}\delta A\partial_{t}\delta A+\epsilon n^{2}\sin(n\theta)\frac{\partial_{t}\phi}{S}\partial_{t}\theta\frac{(\delta A)^{2}}{2},\nonumber
\end{align}and
\begin{align}
a^{-3}\mathcal{H}_{{\rm int}}^{(3)} \approx \, &  -\frac{\epsilon n}{S_I}\cos(n\theta)g^{\mu\nu}\partial_{\mu}\delta\phi\partial_{\nu}\delta A\delta A+\frac{\epsilon n^{2}}{2S_I}\sin(n\theta)\partial_{t}\theta(\delta A)^{2}\partial_{t}\delta\phi\nonumber \\
 & +\frac{\epsilon n^{2}}{2S_I}\sin(n\theta)\frac{\partial_{t}\phi}{S_I}(\delta A)^{2}\partial_{t}\delta A-\frac{\epsilon n^{3}}{6S_I}\cos(n\theta)\frac{\partial_{t}\phi}{S_I}\partial_{t}\theta(\delta A)^{3}.\label{eq:cM-Ham-Cubic}
\end{align}
The first term in the Eq.~(\ref{eq:cM-Ham-quadratic}) couples the
inflaton and the axial field fluctuations at the quadratic order
through a $U(1)$-breaking kinetic mixing. For a weak mixing, we require
\begin{equation}
|\epsilon\sin\left(n\theta_{0}\right)|\ll1, \label{eq:cM-constraint}
\end{equation}
which allows us to utilize the Bunch-Davies vacuum solution for the
mode functions of the fields and to treat the mixing terms perturbatively. For $\sin(n\theta)\sim O(1)$, we expect $|\epsilon|\lesssim 0.05$.
The last term in Eq.~(\ref{eq:cM-Ham-quadratic}) generates a finite
mass for the axial field fluctuations. For light axial field fluctuations,
we set the constraint
\begin{equation}
\left| \epsilon n^{2}\sin(n\theta_{0})\frac{\partial_{t}\phi}{S_I}\partial_{t}\theta_{0}\right|\ll H^{2},
\end{equation}
which is readily satisfied by the conditions in Eq.~(\ref{eq:KM-constraint-1}).

The remaining two terms
in Eq.~(\ref{eq:cM-Ham-quadratic}) introduce corrections to the
power spectra of the inflaton and axial fields. Similar to the analysis in the preceding subsection, we determine that maintaining perturbative control requires the following constraints
\begin{align}
\frac{\Delta P_{\phi}}{P_{\phi}} & \approx\left(\frac{\epsilon n\cos(n\theta_{0})\partial_{t}\theta_{0}N_{k}}{H}\right)^{2}\ll1,\\
\frac{\Delta P_{A}}{P_{A}} & \approx\left(\frac{\epsilon n\cos(n\theta_{0})\partial_{t}\phi N_{k}}{S_I H}\right)^{2}\ll1.\label{eq:PA-constraint-KM}
\end{align}
From Eq.~(\ref{eq:axial-EoM-mixing}),  $\partial_{t}\theta_{0} \approx -\epsilon\, \partial_t \phi/S_I$, so the constraint in Eq.~(\ref{eq:PA-constraint-KM}) is more restrictive. 
\subsubsection{Bispectrum}
We now turn to the evaluation of the three-point correlation functions.
Since $\partial_{t}\theta\ll H$, the dominant axial self-interaction
bispectrum signal is derived from the $\partial_{t}\delta A\delta A^{2}$
interaction term in Eq.~(\ref{eq:cM-Ham-Cubic}) and is given by
\begin{equation}
\left\langle \delta A_{k_{1}}\delta A_{k_{2}}\delta A_{k_{3}}\right\rangle '\approx\left(\frac{\epsilon n^{2}}{2S_I}\sin(n\theta_0)\frac{\partial_{t}\phi}{S_IH}\right)\frac{N_{k}H^{4}}{2k_{1}^{3}k_{2}^{3}k_{3}^{3}} \sum_{i=1}^{3} k_{i}^{3}.\label{AAA-KM}
\end{equation}
Since we take $\phi$ as the inflaton, the axial field can serve as a curvaton. For an oscillating curvaton, the three-point correlation in the above expression yields curvature bispectrum with the corresponding local $f_{{\rm NL}}$ parameter given as \begin{align}
\Delta f_{{\rm NL}}^{{\rm loc}} & \approx\frac{5}{6}C_{\zeta}^{-1}2\left(\frac{\epsilon n^{2}}{2S_I}\sin(n\theta_{0})\frac{\partial_{t}\phi}{S_I H}N_{k}\right)\\
 & \approx \frac{5}{4r}\times0.5\left(\frac{n\theta_{0}\sin(n\theta_{0})}{1}\right)\left(\frac{\epsilon n \partial_{t}\phi}{0.01S_I H}\right)\left(\frac{N_{k}}{50}\right),
\end{align}
where $C_{\zeta}=2r/(3\theta_{0}S_I)$. In the above expression,
we took fiducial values $N_{k}=50$, $n\theta_{0}\sim1$ and $\epsilon n \partial_{t}\phi/(S_IH)\sim0.01$.
These fiducial values satisfy the constraints listed in Eqs.~(\ref{eq:KM-constraint-1}) and (\ref{eq:PA-constraint-KM}). To satisfy the weak-mixing criteria in Eq.~(\ref{eq:cM-constraint}), we require $|\epsilon|\lesssim 0.05$, implying that $n |\partial_{t}\phi|/(S_IH)\lesssim 1$. Since $\epsilon\partial_{t}\phi>0$
for the downward roll of the axial field, the NG parameter $\Delta f_{{\rm NL}}^{{\rm loc}}>0$. Similarly, in the rotating curvaton
scenario,
\begin{align}
\Delta f_{{\rm NL}}^{{\rm loc}} & \approx\frac{5}{2}\times 0.8\left(\frac{n/p}{1}\frac{\tan\left(\frac{p}{n}n\theta_{0}\right)}{1.55}\right)\left(\frac{\sin(n\theta_{0})}{1}\right)\left(\frac{\epsilon n \partial_{t}\phi}{0.01S_IH}\right)\left(\frac{N_{k}}{50}\right),
\end{align}
where we take $\theta_{i}\approx\theta_{0}$. 

If the axial field rather functions as CDM, the axial field bispectrum in Eq.~(\ref{AAA-KM}) gives the following isocurvature NG parameter
\begin{align}
\Delta f^{\mathcal{S}\mathcal{S}\mathcal{S}}_{\rm NL} 
 & =\frac{\alpha_{\rm iso}^2}{2\omega_{A}}\times 0.5\left(\frac{n\theta_{0}\sin\left(n\theta_{0}\right)}{1}\right)\left(\frac{\epsilon n\partial_{t}\phi}{0.01S_IH}\right)\left(\frac{N_{k}}{50}\right).\label{KM-SSS}
\end{align} In this scenario, one obtains the relation $\partial_{t}\phi/(S_IH) = \theta_{0}\sqrt{\alpha_{{\rm iso}}}/(2\omega_{A})$, which leads to
\begin{equation}
    \frac{\epsilon n\partial_{t}\phi}{S_IH}	\approx O(0.01)\times\frac{\epsilon}{0.01}\frac{0.1}{\omega_{A}}\sqrt{\frac{\alpha_{{\rm iso}}}{0.04}}\frac{n\theta_{0}}{1}.
\end{equation}
Thus, $\omega_A\ll0.1$ requires a correspondingly smaller value of $\epsilon$ or $\alpha_{\rm iso}$ while keeping $n\theta\sim1$.

In all of the above scenarios, significant NG is generated in the axial field fluctuations due to the \(U(1)\)-breaking kinetic mixing. The cubic interaction Hamiltonian density in Eq.~(\ref{Ham-cubic}) includes terms that are linear in \(\delta \phi\). Consequently, generating a three-point correlation for \(\phi\) requires at least three interaction vertices. As a result, no substantial NG for \(\delta \phi\) is expected from the kinetic mixing interaction.

The terms in the first line of Eq.~(\ref{eq:cM-Ham-Cubic}) contribute to the $\phi AA$ correlation function presented below, in the same order as they appear in Eq.~(\ref{eq:cM-Ham-Cubic})
\begin{align}
\left\langle \delta\phi_{k_{1}}\delta A_{k_{2}}\delta A_{k_{3}}\right\rangle ' &\approx\left(\frac{\epsilon n}{S_I}\cos(n\theta_0)+\frac{\epsilon n^{2}}{S_I H}\sin(n\theta_0)\partial_{t}\theta N_{k}\right)\frac{H^{4}}{4k_{1}^{3}k_{2}^{3}k_{3}^{3}}\sum_{i=1}^{3}k_{i}^{3}.
\end{align}
Consequently, the only mixed adiabatic-isocurvature bispectrum NG signal at leading order in $\epsilon$ is $\mathcal{S}\mathcal{S}\zeta$ and the corresponding $f_{\rm NL}^{\mathcal{S}\mathcal{S}\zeta}$ parameter is evaluated as 
\begin{align}
\lim_{k_{i}\rightarrow k}\frac{\left\langle \mathcal{S}_{k_{1}}\mathcal{S}_{k_{2}}\zeta_{k_{3}}\right\rangle _{\tau}'}{P_{\zeta}^{2}} & \approx-3\alpha_{{\rm iso}}\left(\sin(n\theta_{0})\frac{\partial_{t}(n\theta)}{H}N_{k} + \cos(n\theta_{0})\right)\frac{\epsilon n\partial_{t}\phi}{S_{I}H}\\
 & \approx 10^{-3}\left(\frac{\alpha_{{\rm iso}}}{0.04}\right)\left(\frac{\sin^{2}(n\theta_{0})}{2}\frac{\epsilon n\partial_{t}\phi}{0.01S_{I}H}\frac{N_{k}}{50} - \cos(n\theta_{0}) \right)\left(\frac{\epsilon n\partial_{t}\phi}{0.01S_{I}H}\right),\label{KM-zetaSS}
\end{align}
where we used Eq.~(\ref{eq:axial-EoM-mixing})
to simplify $\partial_{t}(n\theta)$. Even if one were to neglect the cancellation between the two contributions, the resulting NG amplitude is too small to be detectable.

Finally, if the $\phi$ field functions as a curvaton, the majority of the previous discussion and constraints considering axial field as a CDM remain applicable. In this case, the ratio $\epsilon n\partial_{t}\phi/(S_IH)$ is given as
\begin{equation}
    \frac{\epsilon n\partial_{t}\phi}{S_IH}\approx-\frac{\epsilon n m_{\phi}^{2}\phi}{3S_IH^{2}}\approx O(0.01)\frac{-\epsilon}{0.01}\frac{m_{\phi}^{2}}{0.01H^{2}}\frac{0.0002}{\omega_{A}/r}\sqrt{\frac{\alpha_{{\rm iso}}}{0.04}}\frac{n\theta_0}{1}.
\end{equation}Thus, to saturate the bound on $\epsilon n\partial_{t}\phi/(S_IH)$ at $O(0.01)$ while satisfying the constraint in Eq.~(\ref{eq:KM-constraint-1}), we require $\omega_{A}/r \sim O(10^{-4})$. For $r\sim1$, this implies that the axial field must contribute only a subdominant fraction to the CDM. Hence, for $\omega_{A}/r \lesssim O(10^{-4})$ such that $\epsilon n\partial_{t}\phi/(S_IH) \approx 0.01$, the magnitudes of the NG parameters 
$f_{\rm NL}^{\mathcal{S}\mathcal{S}\mathcal{S}}$ and $f_{\rm NL}^{\mathcal{S}\mathcal{S}\zeta}$ are comparable to those derived in Eqs.~(\ref{KM-SSS}) and (\ref{KM-zetaSS}).  

\subsubsection{Effects of background radial oscillations}
We conclude this section with a discussion on the effects of an oscillating background radial field for the kinetic mixing scenario. 
We remark that significant radial oscillations cannot be generated solely through the kinetic mixing with the inflaton field while simultaneously satisfying the constraints on a light axial scalar field.\footnote{An order-of-magnitude estimate suggests that $\beta_{S}\sim O(10^{-5})$ if the oscillations are excited through the kinetic mixing term due to a sharp feature in the inflaton trajectory.} 
If kinetic mixing is the primary interaction term, the radial field oscillations must be initiated by a different mechanism, potentially through coupling with another field. In this case, a correlated clock signal in the inflaton field is unlikely to exist or would be heavily suppressed. Conversely, if the radial oscillations are induced by the inflaton, the Lagrangian must include interaction terms beyond kinetic mixing. Under this scenario, a correlated clock signal in the inflaton power spectrum becomes possible.

Since the parameter \(\epsilon \propto S^{n-1}\), the interaction vertices receive a time-dependent correction during radial oscillations given by \(\Delta \epsilon/\epsilon \approx  (n-1)\Delta S(t)/S_I\). Furthermore, the presence of an oscillating background radial field implies \(\partial_t S \neq 0\), necessitating the inclusion of contributions from the kinetic mixing term \(\propto \partial \phi \partial S\). These effects collectively contribute to the observed oscillatory behavior in the system.
In terms of axial and $\phi$ field fluctuations, the additional contributions to the interaction Hamiltonian density at quadratic and cubic order from \( \partial \phi \partial S\) mixing are 
\begin{equation}
a^{-3}\mathcal{H}^{(2)}\supset-\epsilon n\sin\left(n\theta\right)\frac{\partial_{t}S}{S_I}\partial_{t}\delta\phi\delta A-\frac{1}{2}\epsilon n^{2}\cos\left(n\theta\right)\frac{\partial_{t}S}{S_I}\frac{\partial_{t}\phi}{S_I}\left(\delta A\right)^{2}\label{dtS-quad-Ham}
\end{equation}
and 
\begin{equation}
a^{-3}\mathcal{H}^{(3)}\supset-\frac{1}{2S_I}\epsilon n^{2}\cos\left(n\theta\right)\frac{\partial_{t}S}{S_I}\partial_{t}\delta\phi\left(\delta A\right)^{2}+\frac{1}{6S_I}\epsilon n^{3}\sin\left(n\theta\right)\frac{\partial_{t}S}{S_I}\frac{\partial_{t}\phi}{S_I}\left(\delta A\right)^{3}, \label{dtS-cubic-Ham}
\end{equation} where in the above expressions
\begin{align}
    \partial_t S \approx-\beta_{S}S_Ie^{-\frac{3}{2}H(t-t_{s})}H\mu_{S}\cos\left(\mu_{S}H(t-t_{s})+\varphi'\right)\Theta(t-t_{s})
\end{align} 
for $\mu_S\gg 3/2$ and phase $\varphi' = \varphi - \tan^{-1}(2\mu_S/3)$. 

The combined $\delta\phi$-$\delta A$ interaction terms in the quadratic Hamiltonian density, as given in Eqs.~(\ref{dtS-quad-Ham}) and (\ref{eq:cM-Ham-quadratic}), generate clock signals in both the axial field and the \( \phi \)-field. These clock signals arise at higher order in the perturbation theory requiring two vertices and are consequently suppressed by the square of the interaction vertices, \( (\epsilon n \beta_S)^2 \). However, the clock signal in the axial field is predominantly generated by the term in Eq.~(\ref{clock-PA}). As a result, a significant clock signal can be observed in the isocurvature power spectrum, while a signal in the curvature power spectrum may remain too small to be detected. The last term in Eq.~(\ref{dtS-quad-Ham}) generates a finite oscillating mass term for the axial field. For $\mu_{S}\gg1$, and $\beta_{S}\lesssim O(0.1)$, the effect of this oscillating mass term is suppressed by the requirement of a light axial field. 

The mixed inflaton-axial three-point correlation in Eq.~(\ref{KM-zetaSS}) receives contributions from the oscillating background radial field, through its effect on the $\phi AA$ interaction terms in Eqs.~(\ref{eq:cM-Ham-Cubic}) and (\ref{dtS-cubic-Ham}). 
Accounting for all contributions, the relative strength of the oscillatory signal in the $\mathcal{S}\mathcal{S}\zeta$ bispectrum signal is analogous to the expression in Eq.~(\ref{osc-ZZS}) derived in Sec.~\ref{subsec:oscille-radial-field} for the $\zeta \zeta \mathcal{S}$ bispectrum, as the underlying interaction terms in both cases are similar. 
Consequently, a substantial oscillatory bispectrum signal in \(\mathcal{S}\mathcal{S}\zeta\) is anticipated if \(\mu_S\) is large and the clock signal in the isocurvature power spectrum is pronounced. However, given that \(f_{\rm NL}^{\mathcal{S}\mathcal{S}\zeta}\) is of the order of 0.001, the prospects for detection may not be promising. 

Lastly, when considering the three-point correlation of the axial field, we note that the magnitude of the NG parameter $\Delta f^{\rm loc}_{\rm NL}$ derived in this section from $U(1)$-breaking kinetic mixing is comparable to that obtained in Sec.~\ref{sec:1} from a $U(1)$-breaking potential. Importantly, the interaction term responsible for the dominant bispectrum signal in this section is proportional to $(\delta A)^2\partial_t \delta A $, in contrast to the $(\delta A)^3$ term discussed in Sec.~\ref{sec:1}. Beyond the similarity in magnitude, both bispectrum shapes are also approximately local.

As discussed in Sec.~\ref{sec:classical-effects-radial}, the oscillating bispectrum signal from the \((\delta A)^3\) interaction term is suppressed by a factor of \(1/\sqrt{\mu_S}\). In contrast, the \((\delta A)^2\partial_t \delta A \) interaction term in Eq.~(\ref{eq:cM-Ham-Cubic})  yields a signal similar to the  integral solution \( I_p \) in Eq.~(\ref{I-soln}) with  \( p = 0 \). Hence, the amplitude of the oscillatory bispectrum signal scales as $\propto(k/k_{r})^{-3/2}\mu_{S}^{1/2}$, indicating an overall amplification by a factor of $\sqrt{\mu_{S}}$. The relative strength of the oscillatory bispectrum compared to the result in Eq.~(\ref{AAA-KM}) is given as
\begin{align}
    \lim_{k_i \rightarrow k}\frac{\Delta B^{AAA}}{B^{AAA}}&\approx O(0.5)\frac{n}{N_{k}}\beta_S\sqrt{2\pi\mu_{s}}\left(\frac{3k}{k_{r}}\right)^{-\frac{3}{2}} \nonumber \\
    \lim_{k_i \rightarrow k}\frac{\Delta B^{AAA}}{B^{AAA}}&\approx O(0.1)\frac{n}{10}\frac{50}{N_{k}}\frac{\Delta P_{A}^{{\rm clock}}}{P_{A}}.
\end{align}
If the axial field functions as a curvaton, the clock signal in the power spectrum is constrained to be \( \lesssim 0.04 \), rendering the oscillatory bispectrum signal, as estimated in the above expression, negligible. However, if the axial field serves as CDM, the clock signal in the isocurvature spectrum can be much larger rendering the oscillatory isocurvature bispectrum signal sizable for large \( n \). Therefore, in scenarios involving massive oscillating radial field, the detection of an oscillatory component in the isocurvature bispectrum with an amplitude around \( 10\% \) of the background may serve as one of the distinguishing features between the \( (\delta A)^2\partial_t \delta A  \) and \( (\delta A)^{3} \) interaction terms.

\section{Conclusion}
\label{sec:conclusion}

In this paper, we explored the NG generated by explicit \(U(1)\)-breaking interactions involving the light axial component $A$ of a complex scalar field. We considered scenarios where the axial field acts as a curvaton or as CDM after inflation, demonstrating that sizable NG signals can arise in both cases. The most stringent constraints on the amplitude of NG arise from the requirement that the axial field remain light.

We began with an analysis of the \(U(1)\)-breaking cosine potential, which introduces a self-interaction term \(\propto (\delta A)^3\) (Sec.~\ref{subsec:selfintr}). In the curvaton paradigm, we evaluated NG contributions from curvaton fluctuations developing during inflation and identified parameter regions where axial self-interactions can generate sizable NG. Interestingly, in this parameter space, a positive NG generated during inflation from nonlinear $U(1)$-breaking interactions can partially cancel with the conventional negative NG generated through the curvaton mechanism after inflation. This can suppress the overall local curvature NG signal to \(f_{{\rm NL}} \lesssim O(0.1)\). However, while the NG is reduced, the trispectrum remains largely unaffected, as the contribution from nonlinear interactions is subdominant. For a heavy radial partner (\(m_S \gg H\)), we analyzed scenarios where the background radial field oscillates during inflation. As is well known, these classical oscillations generate resonantly amplified oscillatory (clock) signals in the power spectrum, with the amplitude scaling as \(\sqrt{m_S}\). For the self-interaction term \(\propto (\delta A)^3\) (Sec.~\ref{sec:classical-effects-radial}), a complementary oscillatory bispectrum signal is suppressed by \(1/\sqrt{m_S}\), while the trispectrum retains a \(\sqrt{m_S}\) enhancement. Due to their distinctive shape, these oscillatory signals can be differentiated from standard local NG templates. If the radial field is rather taken to be a light scalar (Sec.~\ref{sec:light-radial-field}), we find that the \(U(1)\)-breaking interactions between the radial and axial fields can amplify NG signals and enhance bispectra. This amplification highlights a pronounced suppression of NG in the curvaton scenario.

In Sec.~\ref{Couplings with a light scalar}, we explored two models where a light scalar field \(\phi\) (either an inflaton or curvaton) interacts with the axial field. In the first model (Sec.~\ref{subsec:modelA}), we analyzed the interactions of the complex scalar field with the potential and kinetic  energy terms of \(\phi\). When \(\phi\) acts as a curvaton, sizable curvature ($\zeta$) bispectrum signals can be generated, provided that the axial field contributes subdominantly to the CDM density. Similarly, the total isocurvature ($\mathcal{S}$) NG is enhanced by approximately a factor of two. These NG signals can be large enough to be within the sensitivity thresholds of upcoming LSS and CMB experiments. If the radial field oscillates during inflation, the model predicts large oscillatory signals in the \(\zeta\zeta \mathcal{S}\) bispectrum, enhanced by \(\propto m_S^{3/2}\), alongside correlated clock signals in the adiabatic and isocurvature power spectra. In contrast, the clock signal in $P_{\zeta \mathcal{S}}$ and an oscillatory signal in the \(\mathcal{S}\mathcal{S}\zeta\) bispectrum remain suppressed in this model. To the contrary, if \( \phi \) serves as the inflaton, the axial field can act either as CDM or as a curvaton. When the axial-inflaton coupling generates an axial mass-squared of approximately $O(0.01)H^2$, it can yield \( O(1) \) local NG when the axial field is a curvaton and \( O(10) \) isocurvature NG when it is CDM.
In the second model (Sec.~\ref{Sec:KineticMixing}), we examined NG arising from \(U(1)\)-breaking kinetic mixing interactions between the axial field and the inflaton \(\phi\). Similar results are obtained when \(\phi\) acts as a curvaton. Here, the dominant axial cubic self-interaction term involves a derivative coupling (\(\mathcal{L} \supset  (\delta A)^2 \partial_t \delta A \)), as opposed to the \((\delta A)^3\) interaction term analyzed in Sec.~\ref{sec:1}. In the weak-mixing regime, NG in the axial field is comparable to that obtained in Sec.~\ref{sec:1} for similar parameter values. A key distinction is that an oscillating radial field generates an oscillatory axial bispectrum signal with a \(\sqrt{m_S}\) enhancement. In contrast, for the $\delta A^3$ interaction, the oscillatory bispectrum signal is suppressed by $1/\sqrt{m_S}$. This difference in the behavior of the bispectrum provides a clear distinguishing feature between the two interactions, particularly when the axial field is CDM. Also, compared to the model in Sec.~\ref{subsec:modelA}, negligible NG is generated in the perturbations of the $\phi$ field.

Overall, our analysis highlights how \(U(1)\)-breaking interactions, coupled with oscillatory dynamics, can generate bispectrum signals that offer distinct signatures for observational probes in future cosmological surveys.
Finally, while the isocurvature power spectrum remains undetected experimentally, it represents a theoretically compelling scenario. Detecting such signals in the future will require significant advancements in the sensitivity and precision of cosmological surveys. In particular, next-generation experiments must achieve tighter constraints on isocurvature contributions and possess sufficient resolution to distinguish oscillatory bispectrum features from background noise. If the isocurvature power spectrum is dominated by oscillatory features, as proposed in Refs.~\cite{Chung:2021lfg,Chung:2023xcv} or driven by a large clock signal as recently explored in Ref.~\cite{Chen:2023txq} and this work, a substantial oscillatory bispectrum signal could also emerge. This highlights the critical need for dedicated searches in both existing and future datasets for oscillatory features in the power spectrum and bispectrum. Indeed, this would be an interesting follow-up work to Ref.~\cite{Chung:2023syw} to forecast sensitivities for oscillatory signals as we generically expect the oscillations to break degeneracies in the fit that exist in vanilla isocurvature signals.

\section*{Acknowledgments}
This work was supported by the Department of Energy under Grant No.~DE-SC0025611 at Indiana University. We thank Yohei Ema and Soubhik Kumar for helpful suggestions and insightful comments on the manuscript.

\appendix

\section{Evolution of the radial mode\label{app:Evolution-of-radial}}
We need to pay particular attention to the evolution of the radial
mode due to the moduli problem it may create. The energy density of
the radial mode easily overcloses the universe unless it is thermalized.
Furthermore, if the energy density of the radial mode dominates before thermalization,
thermalization must occur before BBN to avoid the disruption to nucleosynthesis
from the associated entropy production. Even for the case of axion
rotations, the dynamics of the field $P$ begins with both angular
and radial components, where the energy density associated with the
radial mode is comparable to or greater than the rotational energy.
For a (nearly) quadratic potential for the radial mode, the energy
density scales as matter whether it is an axion rotation with $S>S_{\min}$
or an radial mode oscillation. For axion rotations, the energy density
scales as kination when $S$ approaches $S_{\min}$. As a result, if
thermalization occurs before $S$ reaches $S_{\min}$, the elliptical
rotation transforms into a circular rotation, and there may be a kination-dominated
era following a matter-dominated era.

In the simplest scenario, we assume a Yukawa interaction between the
radial mode and fermions $\psi$ and $\bar{\psi}$, which interact
with the thermal bath via 
\begin{equation}
\mathcal{L}\supset y_{\psi}S\psi\bar{\psi}.
\end{equation}
The simplest case involves a fermion charged under Standard Model gauge interactions, though a dark sector fermion is also viable. The thermalization rate is expressed as~\cite{Mukaida:2012qn}
\begin{equation}
\Gamma_{S\psi\bar{\psi}}=by_{\psi}^{2}T,
\end{equation}
where $b$ is a constant of order $\mathcal{O}(0.1)$ when the fermion’s
interaction with the thermal bath is $\mathcal{O}(1)$. Early in the
universe, the fermion mass is large due to the significant value of
the radial field, $m_{\psi}=y_{\psi}S$, but the fermions must be
present in the thermal bath to thermalize the radial mode at the temperature
$T_{{\rm th}}$. This condition, $y_{\psi}S_{{\rm th}}\le T_{{\rm th}}$,
imposes an upper limit on both the Yukawa coupling and the thermalization
rate 
\begin{equation}
\Gamma_{S\psi\bar{\psi}}\le\frac{bT_{{\rm th}}^{3}}{S_{{\rm th}}^{2}}.
\end{equation}
A similar constraint applies for the interactions between the radial
mode and a scalar. In the case of gauge boson couplings, which arise
from integrating out charged fermions or scalars, the thermalization
rate is approximately $10^{-5}T^{3}/S^{2}$~\cite{Mukaida:2012qn}.
To adjust for this scenario, one can substitute $b=10^{-5}$ into
the relevant equations to determine the parameter space constraints.

We now discuss the relevant temperatures. The dominant source of the
energy density may transition from radiation to matter at 
\begin{align}
T_{{\rm RM}}(Y) & =\frac{4}{3}m_sY\simeq10^{7}\GeV\left(\frac{m_s}{100\TeV}\right)\left(\frac{Y}{100}\right),\label{eq:TRM}
\end{align}
where the yield $Y\equiv n/s$ is given by that of the radial mode
$Y_{S}=\frac{1}{2}m_sS/s(T)$ or by that of the rotation $Y_{\theta}=\dot{\theta}S^{2}/s(T)$
with $s(T)$ the entropy density at $T$. Specifically, there are
two chances for the complex field $P$ to dominate the energy density.
If thermalization occurs after $T<T_{{\rm RM}}(Y_{S})$, the radial
mode oscillations lead to a matter-dominated era. Upon thermalization,
a large amount of entropy is then injected into the thermal bath,
which reheats the universe and dilutes the initial $Y_{\theta,i}$
into a small yield $Y_{\theta}$. At this time, if the temperature
is still larger than $T_{{\rm RM}}(Y_{\theta})$ and if the rotation
radius $S$ is still greater than $S_{\min}$, the rotation energy density
starts to dominate at $T=T_{{\rm RM}}(Y_{\theta})$. The evolution
transitions from matter to kination scaling when $S$ reaches the
minimum at $S_{\min}$ at the temperature 
\begin{align}
    \label{eq:TMK}
    T_{\rm MK} & = 
    \left(\frac{45}{2\pi^2 g_*} \frac{m_s S_{\min}^2}{Y_{\theta}}\right)^{ \scalebox{1.01}{$\frac{1}{3}$} } \simeq  3 \times 10^6 \GeV \left(\frac{m_s}{100 \TeV} \frac{100}{Y_\theta} \frac{g_{*,{\rm SM}}}{g_*} \right)^{ \scalebox{1.01}{$\frac{1}{3}$} } 
    \left(\frac{S_{\min}}{10^9 \GeV}\right)^{ \scalebox{1.01}{$\frac{2}{3}$} }.
\end{align}
If the energy density of the rotation is dominating at $T_{{\rm MK}}$,
the universe enters a kination-dominated era. Eventually, the kination
scaling allows the rotation energy density to redshift away and the
universe returns to radiation domination at the temperature 
\begin{align}
    \label{eq:TKR}
    T_{\rm KR} & = \frac{3 \sqrt{15}}{2 \sqrt{g_*}\pi } \frac{S_{\min}}{Y_{\theta}} 
    \simeq   2 \times 10^6 \GeV \left( \frac{S_{\min}}{10^9 \GeV} \right) \left(\frac{100}{Y_\theta} \right) \left(\frac{g_{*,{\rm SM}}}{g_*}\right)^{ \scalebox{1.01}{$\frac{1}{2}$} }.
\end{align}
On the other hand, if $T_{{\rm RM}}(Y_{\theta})<T_{{\rm MK}}$, the energy density redshifts as kination before dominating, and thus the epochs of domination by the rotation
will be absent.

We now comment on the possibilities where the radial or angular mode
serves as the curvaton, which creates the curvature perturbation by
reheating the universe. For the radial mode to be a curvaton, the
thermalization of the complex field should occur after $T=T_{{\rm RM}}(Y_{S})$
and there should be no subsequent production of entropy by the angular
mode or other fields. On the other hand, if the radial mode is thermalized
before dominating or if thermalization occurs before $T=T_{{\rm RM}}(Y_{\theta})$,
the rotation becomes circular and may dominate the energy density
if $T_{{\rm RM}}(Y_{\theta})>T_{{\rm KR}}$. If the rotation is (partially)
washed out to reheat the universe during the matter/kination-dominated
era, then the angular mode serves as a curvaton~\cite{Co:2022qpr}.
In any of these cases, the remaining axion rotation, if not completely
washed out, can provide the necessary PQ charge asymmetry for kinetic
misalignment so that the angular mode accounts for the dark matter
abundance.

\section{Light scalar fields}
\label{App:light-scalar} In this Appendix, we evaluate and analyze
the superhorizon two- and three-point correlation functions for a
light scalar field during inflation. By \textit{light scalars}, we refer
to the fields with mass-squared $m^{2}\lesssim O(0.02)H^{2}$ during inflation.
We demonstrate that for these light scalar fields, the correlation
functions can be studied using massless mode functions.

In de-Sitter spacetime, the positive frequency mode function solution
for the massive fields ($m\leq3/2H$) is given as 
\begin{equation}
v_{k}(\tau)=-ie^{i(\nu+1/2)\frac{\pi}{2}}\frac{\sqrt{\pi}H}{2}(-\tau)^{3/2}H_{\nu}^{(1)}(-k\tau), \label{massive-mode-fn}
\end{equation}
where $\nu=\sqrt{9/4-m^{2}/H^{2}}$ \cite{Baumann:2018muz}. For light scalar fields with
$m^{2}\ll H^{2}$, we can approximate the late-time superhorizon power
spectrum as 
\begin{equation}
P_{\nu}(k,\tau_{e})=\lim_{\tau\rightarrow\tau_{e}}v_{k}^{*}v_{k}\approx\frac{H^{2}}{2k^{3}}\times\gamma^{2}, \label{eq:massive_pwr_spec}
\end{equation}
where 
\begin{equation}
\gamma=\left(\frac{2^{1-x}\left(-k\tau_{e}\right)^{x}\Gamma\left(\frac{3}{2}-x\right)}{\sqrt{\pi}}\right)\label{dilution}
\end{equation}
is the dilution factor corresponding to each massive mode function
and $\tau_{e}$ is the conformal time at the end of the quasi de-Sitter
phase. In the above expression, $x=3/2-\nu$ such that $0<x\ll1$
for light scalar fields. Hence, in comparison to the power spectrum
of a massless scalar field, which is given by $P_{\nu=3/2}(k,\tau_{e})=H^{2}/(2k^{3})$,
the power spectrum for a massive field dilutes over time, as indicated
by the $\left(-k\tau_{e}\right)^{x}$ factor in Eq.~(\ref{dilution}) where
$x>0$ for a massive field. For instance, consider $m^{2}/H^{2}=0.02$
which yields $\gamma^{2}\approx0.5$ and $0.44$ for $N_{k}=50$ and
$60$ respectively.

\begin{figure}
\centering \includegraphics[width=0.5\linewidth]{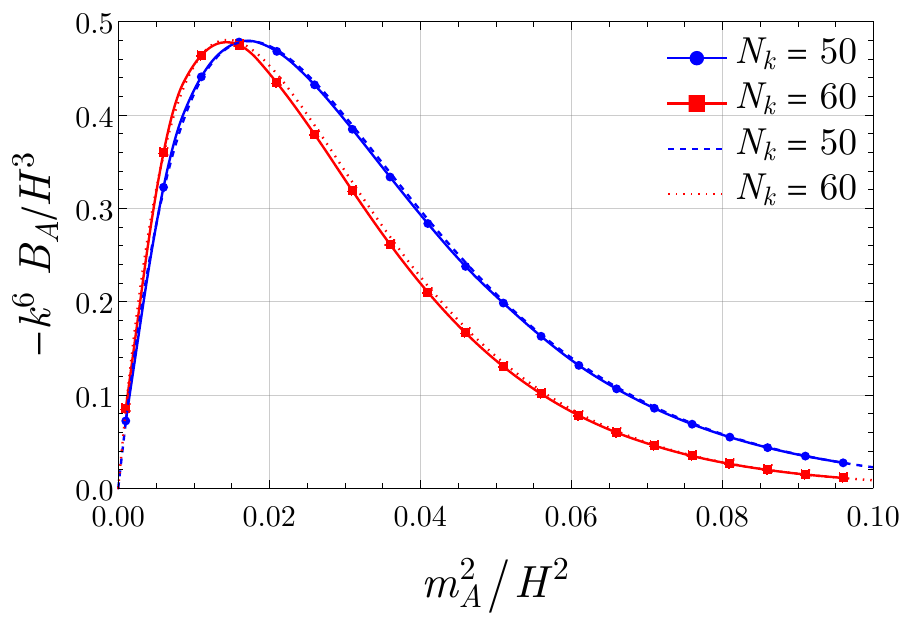}
\caption{The figure shows $k^{6}H^{-3}\langle \delta A \delta A\delta A\rangle'$ as a function of
$m_{A}^{2}/H^{2}$ for a light scalar field $A$ with the coefficient
$c_3$ set to $H\times m_{A}^{2}/H^{2}$. The solid blue and red
curves represent numerical results for $N_{k}=50$ and $60$, respectively.
The blue dashed and red dotted curves correspond to the analytical
approximations as given in Eq.~(\ref{eq:bispec-light-scalar}), by setting
 $\epsilon=0.4$ for
empirical agreement.}
\label{fig:bispec-light-scalar} 
\end{figure}

Now consider the bispectrum of the fluctuations of the light scalar
field resulting from a cubic self-interaction term as described in
Eq.~(\ref{A3}). As discussed in Sec.~\ref{sec:1}, this interaction
term can arise from an explicit $U(1)$-breaking potential given in Eq.~(\ref{Vtheta}).
Since the interaction term is proportional to the mass-squared of
the light scalar, it vanishes in the $m\rightarrow0$ limiting
scenario. For a light scalar field fluctuation $\delta A$ with non-zero mass $m_{A}$, the
bispectrum due to the cubic self-interaction term (with a coefficient
$c_3$ as given in Eq.~(\ref{eq:c3-coeff})), in the equilateral triangle limit can be approximated
as 
\begin{equation}
k^{6}H^{-3}\langle \delta A \delta A\delta A\rangle'\approx\frac{c_3}{H}\frac{3}{2}\ln(-k\tau_{e})\times\gamma^{3+\epsilon}, \label{eq:bispec-light-scalar}
\end{equation}
where $-k\tau_{e}\approx\exp(-N_{k})$ with $N_{k}$ as the number
of e-folds from the horizon exit of mode $k$ at time $t_{k}$ until
the end of inflationary quasi de-Sitter phase at $t_{e}$. For CMB
modes $N_{k}\sim50-60$. Note the bispectrum signal dilutes by a factor
of $\gamma^{3+\epsilon}$. The additional factor of $\gamma^{\epsilon}$
represents the approximate cumulative effect of integrating over the
cube of the massive mode functions when evaluating the bispectrum
in Eq.~(\ref{A3-corr}) using the in-in formalism. The coefficient
$c_3$ in Eq.~(\ref{eq:bispec-light-scalar}) is proportional
to the mass-squared quantity $m_{A}^{2}/H^{2}$ of the light scalar
$A$. To highlight the functional dependence of the bispectrum on
the mass of the light scalar, we set the coefficient $c_3=H\times m_{A}^{2}/H^{2}$,
and show $k^{6}H^{-3}\langle \delta A \delta A\delta A\rangle'$ as a function of $m_{A}^{2}/H^{2}$
in Fig.~\ref{fig:bispec-light-scalar} for two distinct values of
$N_{k}$. Numerical results are shown with solid curves, while our analytical
approximations from Eq.~(\ref{eq:bispec-light-scalar}) are depicted with dash-dot curves. 

From Fig.~\ref{fig:bispec-light-scalar}, we observe that the normalized
bispectrum quantity $k^{6}H^{-3}\langle \delta A \delta A\delta A\rangle'$ for light scalar fields peaks within the mass range $0.01<m_{A}^{2}/H^{2}<0.03$, and falls sharply outside of this mass-range. Accordingly, we will concentrate our study on the bispectrum arising from the self-interaction of light scalar fields with masses falling within this range. 

If the axial field perturbations can be approximated as contributing
to the curvature perturbations at the linear order by $\zeta\approx C_\zeta\delta A$,
for a constant $C_\zeta$, then the non-linearity parameter $f_{{\rm NL}}^{{\zeta}}$
can be estimated as 
\begin{eqnarray}
f_{{\rm NL}}^{{\zeta}} & = & C_\zeta^{-1}\frac{\langle \delta A \delta A\delta A\rangle'}{P_{A}(k_{1})P_{A}(k_{2})+\mbox{2 perms.}} \approx  \frac{\gamma^{\epsilon}}{\gamma C_\zeta}\frac{c_3N_{k}}{2H},\label{fA}
\end{eqnarray}
where in the last expression we have utilized the fact that the bispectrum
shape $\langle \delta A \delta A\delta A\rangle'$ is local. Given that the amplitude of
the curvature perturbation is well-constrained, the dilution of the
mode function $\delta A$ by the factor $\gamma$ can be absorbed
into the constant $C_\zeta$, resulting in $\gamma C_\zeta=$constant. Since $\epsilon=0.4$
and $\gamma\sim0.7$, the remaining dilution factor $\gamma^{0.4}\sim O(1)$
and can be neglected. Therefore, to a good approximation, the bispectrum and the associated
non-linearity parameter for the light scalar fields can be studied using
the massless mode functions.

\section{Power spectrum correction from oscillating background radial field\label{App:clock}} 
Consider the quadratic Lagrangian for a massless $\delta\theta$
field 
\begin{equation}
\mathcal{L}_{2}=a^{3}\frac{(S_I+\Delta S(t))^{2}}{2}\left[(\partial_t{\delta\theta})^{2}-a^{-2}\left(\partial_{i}\delta\theta\right)^{2}\right].\label{eq:dtheta_Lag}
\end{equation}Here, $\Delta S$ is the oscillating component of the background radial
field whose solution is given in Eq.~(\ref{S_osc}). When the radial field is stationary ($\Delta S=0$), the axial
field is canonically normalized and massless. In the scenarios
where $\Delta S\neq0$, the oscillation of the background radial field
induces corrections to the axial mode function, resulting in the clock
signal on the power spectrum.
\subsection{Using in-in formalism}
For small-amplitude oscillations $|\beta_{S}|\ll1$, we can evaluate
the correction using the in-in formalism. The clock signal in this
case arises from corrections to the kinetic term of the axial field.
For further details, see Ref.~\cite{Chen:2014cwa}. Here, we briefly reproduce their analysis. By defining the massless axial field
as $\delta A=S_I\delta\theta$, the leading order correction
comes from the $2\Delta S(t)/S_I$ term in Eq.~(\ref{eq:dtheta_Lag}).
The relevant interaction Hamiltonian is 
\begin{eqnarray*}
H_{K}^{(2)}(t) & = & \int d^{3}\vec{x}\mathcal{H}_{K}^{(2)}
   =  -\frac{\Delta S}{S_I}a(\tau)\int_{p}\left(\partial_{\tau}\delta A_{p}\partial_{\tau}\delta A_{-p}-\int_{p}p^{2}\delta A_{p}\delta A_{-p}\right)_{\tau}, 
\end{eqnarray*}
which allows us to compute the clock signal
on the axial power spectrum as
\begin{eqnarray}
\Delta P_{A} & = & -2{\rm Im}\left[\int_{0}^{t}dt'\left\langle 0\right|H(t')W(t)\left|0\right\rangle \right]\nonumber \\
 & = & 2\frac{1}{S_I}\int_{p}{\rm Im}\left[\int_{-\infty}^{\tau}d\tau'a(\tau')^{2}\Delta S(\tau')\left\langle 0\right|\left(\partial_{\tau}\delta A_{p}\partial_{\tau}\delta A_{-p}-p^{2}\delta A_{p}\delta A_{-p}\right)_{\tau'}\delta A_{k_{1}}(\tau)\delta A_{k_{2}}(\tau)\left|0\right\rangle \right]\nonumber \\
 & = & 4\frac{1}{S_I}\left(2\pi\right)^{3}\delta_{\vec{k_{1}}+\vec{k_{2}}}^{3}u_{k_{1}}^{*}(\tau)u_{k_{2}}^{*}(\tau){\rm Im}\left[\int_{-\infty}^{\tau}d\tau'a^{2}(\tau')\Delta S(\tau')\left(\left(\partial_{\tau}u_{k_{1}}\right)^{2}-k_{1}^{2}u_{k_{1}}^{2}\right)_{\tau'}\right].\label{inin-eqn}
\end{eqnarray}
where $u_{k}$ is the massless mode function defined in Eq.~(\ref{massless_modefn}).
Taking $\Delta S$ from Eq.~(\ref{S_osc}), we obtain a simple expression for the correction
for $2k\gtrsim k_{r}$ in the limit $\mu_{S}\gg1$ as 
\begin{equation}
\frac{\Delta P_{A}}{P_{A}}\approx\beta_{S}\sqrt{2\pi\mu_{S}}\left(\frac{2k}{k_{r}}\right)^{-3/2}\sin\left(\mu_{S}\ln\left(\frac{2k}{k_{r}}\right)+\mu_{S}+\pi/4\right),\label{clock-inin}
\end{equation}
where $k_{r}=-\mu_{S}/\tau_{s}$. Note that the amplitude of the
\textit{oscillating} clock signal increases with $\mu_{S}$. This is due
to the \textit{resonance} effect. For more discussion on the resonant enhancement of the oscillatory signal, we refer the readers to Ref.~\cite{Chen:2011zf}.

\subsection{Using equations of motion}
Rather than treating $\Delta S(t)$ as a perturbative correction,
we can reformulate our theory by defining the axial field as $\delta A=S\delta\theta\equiv(S_I+\Delta S(t))\delta\theta$. In the limit $\dot{\theta}\rightarrow0$, the radial and axial field
fluctuations remain kinetically decoupled. The radial field's oscillations
will only appear as a correction to the mass of the axial field. Thus, we write the axial field fluctuation's EoM as 
\begin{eqnarray}
\partial^2_t{\delta A}+3H\partial_t{\delta A}+\left(\frac{k^{2}}{a^{2}}-\left(\frac{\partial^2_t{S}+3H\partial_t{S}}{S}\right)\right)\delta A & =0,\nonumber \\
\partial^2_t{\delta A}+3H\partial_t{\delta A}+\left(\frac{k^{2}}{a^{2}}+\frac{m_{S}^{2}\Delta S/S_I}{(1+\Delta S/S_I)}\right)\delta A & =0, \label{Clock-eom}
\end{eqnarray}
where in the last expression above we used the background radial field's EoM in Eq.~(\ref{eq:Delta_S_EoM}) to introduce radial mass-squared quantity $m^2_{S}$. Due to a finite induced mass of the axial field, the massless mode function defined in Eq.~(\ref{massless_modefn})
may not be a suitable solution, and we must solve the above EoM to obtain
the correct time-dependence of the mode function for the axial field.
For instance, if we consider $m_{S}\gtrsim O(10)H$ and $\beta_{S}\sim O(0.1)$,
the oscillatory mass-squared term $\propto m^2_{S}\Delta S/S_I$ appears dominant even when the modes are subhorizon, i.e. when $k/a\gtrsim1$.
However, as shown in the Appendix B of Ref.~\cite{Chung:2024ctx}, if the ratio
of the amplitude to frequency-squared of the oscillatory mass-squared
term is $\ll k^{2}/a^{2}$, the solution to the Eq.~(\ref{Clock-eom})
is a superposition of states. For the oscillatory mass-squared term $\propto m^2_{S}\Delta S/S_I$, this ratio is $\approx \beta_S$ which is $\ll 1$ since the frequency of oscillation is $\propto m_S$. In this scenario, the dominant state
is similar to a massless mode function with a frequency $k$ in conformal
time coordinates, while the sub-dominant states have frequencies comparable
to that of the oscillatory mass term. Therefore, to ensure that the
axial field remains light during subhorizon evolution until $k/a\sim\beta_{S}$,
it is sufficient to impose that $\beta_{S}\ll1$. 
\begin{figure}[t]
\centering \includegraphics[width=0.6\linewidth]{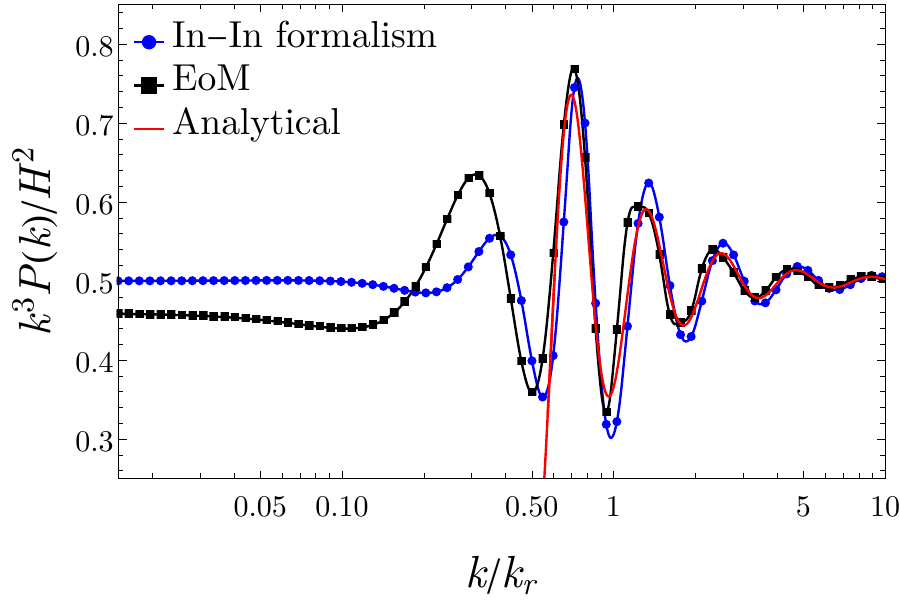}
\caption{Clock signal on the \textquotedblleft normalized" axial power spectrum
arising from an oscillating background radial field. The blue (circular
marker) and the black (box marker) curves are obtained by numerically
solving Eqs.~(\ref{inin-eqn}) and (\ref{Clock-eom}) respectively.
The solid red curve is the analytical result taken from Eq.~(\ref{clock-inin})
and is restricted to scales $k/k_{r} \gtrsim 1/2$. In this figure, the amplitude of the normalized scale-invariant power spectrum (without any oscillations)
is $1/2$. We fix $m_{S}^{2}=100H^{2}$ and $\beta_{S}=0.1$. The
suppression of power on long wavelengths from solving the EoM is explained
by the expression in Eq.~(\ref{pwr_suppr}).}
\label{fig:clock-signal} 
\end{figure}

In Fig.~\ref{fig:clock-signal}, we illustrate the power spectra obtained
by numerically solving the integral in Eq.~(\ref{inin-eqn}) from the in-in formalism, and solving the EoM in Eq.~(\ref{Clock-eom}) with the appropriate Bunch-Davies (BD)
vacuum boundary conditions. For this plot, we set $m_{S}^{2}=100H^{2}$, $\beta_S=0.1$
and $t_{s}/t_{0}=1$ where $t_{0}$ denotes the start of inflation.
Note that modes that are well outside the horizon when the oscillation
begins do not display any $k$-dependent feature; instead, the amplitude of the power spectrum is uniformly suppressed on
these scales. This suppression can be attributed to the integrated
effect of a finite non-zero positive mass arising from the oscillating
mass-squared term in Eq.~(\ref{Clock-eom}). By retaining only the
leading order term in $\Delta S$, the power suppression can be approximated
as follows 
\begin{equation}
\frac{\Delta P_{A}}{P_{A}}(k\ll k_{r})\approx1-\exp\left(\frac{-2m_{S}^{2}}{3}\int_{t_{s}}^{\infty}Hdt'\Delta S(t')/S_I\right)\approx1-e^{-\beta_{S}}\sim\beta_{S}\label{pwr_suppr}
\end{equation}
for $\beta_{S}\ll1$. As seen in Fig.~\ref{fig:clock-signal}, this
suppression is absent in our leading-order perturbative calculation
using the in-in formalism. We remark that while the amplitude of the
oscillating clock signal for $k\sim k_{r}$ is proportional to $\beta_{S}\sqrt{\mu_{S}}$,
the suppression on long wavelengths $k\ll k_{r}$ is only dependent
upon $\beta_{S}$.

\section{Hamiltonian density}\label{App:Hamiltonian} 
In this Appendix, we give the free-field and interaction Hamiltonian densities in the interaction picture for
the radial and axial field fluctuations derived from the Lagrangian density given in Eqs.~(\ref{L2})-(\ref{dL3}). The Hamiltonian density in terms of interaction
picture fields (we drop the subscript $I$ for brevity) is given as
\begin{align}
a^{-3}\mathcal{H}_{0}  = \ & \frac{1}{2}\left(\partial_{t}\delta S\right)^{2}+\frac{1}{2a^2}\left(\partial_{i}\delta S\right)^{2}+\frac{1}{2}\left(\partial_{t}\delta A\right)^{2}+\frac{1}{2a^2}\left(\partial_{i}\delta A\right)^{2} \label{Ham-free}  \\
& +\frac{1}{2}m_{A}^{2}(\delta A)^{2}+\frac{1}{2}\left(m_{S}^{2}+3\left(\partial_{t}\theta_0\right)^{2}\right)(\delta S)^{2} \nonumber \\
a^{-3}\delta\mathcal{H}_{2}  = \ & m_{A,I}^{2}\sin\left(n\theta_0\right)\delta A\delta S-2\partial_{t}\theta_0\delta S\partial_{t}\delta A \label{Ham-quad}\\
a^{-3}\delta\mathcal{H}_{3} = \ & -\frac{1}{S}\left(\partial_{t}\delta A\right)^{2}\delta S+\frac{1}{S_{0}}a^{-2}\left(\partial_{i}\delta A\right)^{2}\delta S+\frac{3}{S_{0}}\partial_{t}\theta_0\left(\partial_{t}\delta A\right)(\delta S)^{2} \label{Ham-cubic} \\
 & +\frac{1}{2S_{0}}\left(m_{A,I}^{2}(n-1)\sin\left(n\theta_0\right)\right)\left(\delta S^{2}\right)\delta A+\frac{1}{2S_{0}}\left(m_{A}^{2}n\right)\left(\delta A\right)^{2}\delta S \nonumber \\
 & +\frac{1}{6S_{0}}\left((2m-1)m_{S}^{2}-m_{A}^{2}(n-1)(n-2)-12\left(\partial_{t}\theta_0\right)^{2}\right)\left(\delta S\right)^{3} \nonumber \\
 & -\frac{1}{6S_{0}}\left(m_{A,I}^{2}n\sin\left(n\theta_0\right)\right)\left(\delta A\right)^{3}\nonumber. 
\end{align}

\section{Detailed calculations for the power spectrum and bispectrum}\label{App:pwrbispec} 
In this Appendix, we evaluate the two- and
three-point correlation functions of the field $A$ from interactions resulting from
the quadratic and cubic Hamiltonian densities involving two generic light scalar fields $S$ and $A$, as shown in Eqs.~(\ref{Ham-quad}) and (\ref{Ham-cubic}).
Considering nearly massless axial and radial fields, we have $m_{S}^{2}, m_{A}^{2}\sim O(0.01)H^{2}$.
Consequently, we will describe these fields through the massless mode
function provided in Eq.~(\ref{massless_modefn}).

\subsection{Power spectrum}
In the in-in formalism, we evaluate the two-point function $\left\langle \delta A\delta A\right\rangle $
as 
\begin{equation}
\left\langle \delta A(t)\delta A(t)\right\rangle =\left\langle 0\right|\left[\bar{T}\exp\left(i\int_{t_{0}}^{t}dt'H^{I}(t')\right)\right]\delta A_{I}^{2}(t)\left[T\exp\left(-i\int_{t_{0}}^{t}dt''H^{I}(t'')\right)\right]\left|0\right\rangle.
\end{equation}
Writing terms up to the second order in perturbation, we obtain the leading
correction to the power spectrum $\Delta P=\left\langle \delta A(\vec{p},t)\delta A(\vec{q},t)\right\rangle '-\left\langle 0\right|\delta A(\vec{p},t)\delta A(\vec{q},t)\left|0\right\rangle' $
as 
\begin{equation}
\Delta P=i^{2}\int_{t_{0}}^{t}dt_{1}\int_{t_{0}}^{t_{1}}dt_{2}\left\langle 0\right|\left[H^{I}(t_{2}),\left[H^{I}(t_{1}),\delta A(\vec{p},t)\delta A(\vec{q},t)\right]\right]\left|0\right\rangle '.
\end{equation}
The above expression can be simplified to 
\begin{align}
\Delta P & =-2{\rm Re}\left[\int_{-\infty}^{\tau}d\tau_{1}a(\tau_{1})\int_{-\infty}^{\tau_{1}}d\tau_{2}a(\tau_{2})\left\langle 0\right|H^{I}(\tau_{2})H^{I}(\tau_{1})W(\tau)-H^{I}(\tau_{2})W(\tau)H^{I}(\tau_{1})\left|0\right\rangle '\right],\label{in-in-mixedformula}
\end{align}
where $W(\tau)=\delta A(\vec{p},\tau)\delta A(\vec{q},\tau)$. The
interaction terms relevant for power spectrum correction at
second order are obtained from Eq.~(\ref{Ham-quad}) 
\begin{equation}
\mathcal{H}_{2}^{I}=a^{3}\left[c_{1}(\delta S\partial_{t}\delta A)+c_{2}\left(\delta S\delta A\right)\right],\label{mixedpower}
\end{equation}
where we define, with $\partial_{t}\theta$ taken from Eq.~(\ref{angular velocity}),
\begin{equation}
c_{1}\approx2\left(\frac{m_{A,I}^{2}}{H^{2}}\right)\frac{H\sin\left(n\theta_0\right)}{3n},~\mbox{and}\quad c_{2}\approx\left(\frac{m_{A,I}^{2}}{H^{2}}\right)H^{2}\sin\left(n\theta_0\right).\label{c1c2}
\end{equation}
Thus, we obtain the net quadratic interaction Hamiltonian as
\begin{align}
H_{2}^{I}(t) & =\int d^{3}\vec{x}\mathcal{H}_{2}^{I}=a^{3}(t)\left(c_{1}\int_{k}\delta S_{k}\partial_{t}\delta A_{-k}+c_{2}\int_{k}\delta S_{k}\delta A_{-k}\right)\\
 & =c_{1}a^{2}(\tau)\int_{k}\delta S_{k}\partial_{\tau}\delta A_{-k}+c_{2}a^{3}(\tau)\int_{k}\delta S_{k}\delta A_{-k}\\
 & =H^{1}(\tau)+H^{2}(\tau),\label{HI(1,2)}
\end{align}
where all fields are in the interaction picture. Due to the mixing of $H^{1}$ and $H^{2}$
pairings, we obtain three
distinct contributions. Using the massless mode functions as the solution to the
free field EoM for both fields, we obtain the result for the corrections
to the power spectrum of the axial field from $H^{(1,2)}$ interaction
Hamiltonian densities in Eq.~(\ref{HI(1,2)}) respectively as \begin{align}
\Delta P_{AA}^{(1)}(k) & \approx k^{-3}\left(\frac{c_{1}N_{k}}{\sqrt{2}}\right)^{2}\\
\Delta P_{AA}^{(1,2)}(k) & \approx k^{-3}\left(\frac{c_1c_2N_{k}^2}{2H}\right)\\
\Delta P_{AA}^{(2)}(k) & \approx k^{-3}\left(\frac{c_{2}N_{k}}{3H}\right)^{2}. 
\end{align}

Similarly, the correlated power spectrum for axial-radial fluctuations from the interaction Hamiltonian in Eq.~(\ref{HI(1,2)}) is given as
\begin{eqnarray}
    \Delta P_{AS}^{(1)}(k) & \approx k^{-3}c_1\frac{H N_k}{2}, \qquad
    \Delta P_{AS}^{(2)}(k)  \approx k^{-3}c_2\frac{-N_k}{3}.\label{PAS}
\end{eqnarray}
From the expressions given in Eq.~(\ref{c1c2}) for
the coefficients $c_{1,2}$, we note that the ratio $c_2/c_1 = 3nH/2$. Thus, when $n\gg 1$, the $U(1)$-breaking interaction, $\propto c_2 \delta S \delta A$,
gives a dominant contribution to the power spectrum corrections.

\subsection{Bispectrum \label{App: bispectrum-contributions-H2H3}}

\begin{figure}
\centering \begin{tikzpicture}  
\draw[thick] (0,0) -- (5,0);  
\node at (5.2,0) {$\tau$};
\node at (0.7,-1) {$\delta A$};
\node at (2.1,-1) {$\delta A$};
\node at (2.9,-2) {$\delta S$};
\node at (4.5,-1) {$\delta A$};
\node at (2,-2.4) {$H_{3}^{I}$};       
\draw[thick] (0.5,0) -- (2,-2);     
\draw[thick] (1.5,0) -- (2,-2);  
\draw[thick] (2,-2) -- (3.5,-2.5);          
\draw[thick] (4.5,0) -- (3.5,-2.5);         
\node at (3.5,-2.8) {$H_{2}^{I}$};       
\end{tikzpicture} \caption{Feynman diagram representation of the bispectrum signal in Eq.~(\ref{W2}) from the cubic
and quadratic interaction Hamiltonians involving radial and axial field fluctuations.}
\label{fig:bispectrum-mixedRA} 
\end{figure}
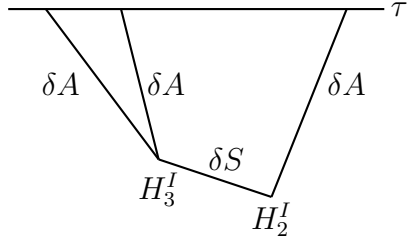

We now analyze the bispectrum signal for axial field fluctuations
resulting from various cubic interaction terms involving both radial
and axial fields. Fig.~\ref{fig:bispectrum-mixedRA} provides a diagrammatic
representation of these interactions. The contributions to
the three-point correlation function of the operator $W(t)=\delta A_{k_{1}}\delta A_{k_{2}}\delta A_{k_{3}}$ are computed using a similar
expression as on the right-hand side of Eq.~(\ref{in-in-mixedformula}).
In this case, the interaction Hamiltonian must include one cubic
term and one quadratic term. Hence, we obtain two contributions as
\begin{align}
\left\langle W(\tau)\right\rangle _{1} & =-2{\rm Re}\left[\int_{-\infty}^{0}d\tau_{1}a(\tau_{1})\int_{-\infty}^{\tau_{1}}d\tau_{2}a(\tau_{2})\left\langle 0\right|H_{3}(\tau_{2})H_{2}(\tau_{1})W(\tau)-H_{3}(\tau_{2})W(\tau)H_{2}(\tau_{1})\left|0\right\rangle \right] , \\
\left\langle W(\tau)\right\rangle _{2} & =-2{\rm Re}\left[\int_{-\infty}^{0}d\tau_{1}a(\tau_{1})\int_{-\infty}^{\tau_{1}}d\tau_{2}a(\tau_{2})\left\langle 0\right|H_{2}(\tau_{2})H_{3}(\tau_{1})W(\tau)-H_{2}(\tau_{2})W(\tau)H_{3}(\tau_{1})\left|0\right\rangle \right].\label{W2}
\end{align}
Below, we present the bispectra for various combinations
of cubic and quadratic interactions. 
\begin{enumerate}
\item $\mathcal{H}_{3}^{I}(x,t)=a^{3}d_{1}\delta S\delta A\delta A$ and
$\mathcal{H}_{2}^{I}(x,t)=a^{3}c_{2}\delta S\delta A$  
\begin{eqnarray}
\left\langle W(\tau)\right\rangle ' & = & \left\langle W(\tau)\right\rangle '_{1}+\left\langle W(\tau)\right\rangle '_{2}\nonumber \\
 & \approx & d_{1}c_{2}\frac{7N_{k}^{2}\left(k_{1}^{3}+k_{2}^{3}+k_{3}^{3}\right)}{36k_{1}^{3}k_{2}^{3}k_{3}^{3}}.\label{Type1}
\end{eqnarray}
In the equilateral triangle limit, $\left\langle W(\tau)\right\rangle '\rightarrow d_{1}c_{2}7N_{k}^{2}/\left(12k^{6}\right)$. 
\item $\mathcal{H}_{3}^{I}(x,t)=a^{3}d_{1}\delta S\delta A\delta A$ and
$\mathcal{H}_{2}^{I}(x,t)=a^{3}c_{1}\delta S\partial_{t}\delta A$
\begin{eqnarray}
\left\langle W(\tau)\right\rangle '\approx d_{1}c_{1}\frac{-5HN_{k}^{2}\left(k_{1}^{3}+k_{2}^{3}+k_{3}^{3}\right)}{12k_{1}^{3}k_{2}^{3}k_{3}^{3}}.\label{Type2}
\end{eqnarray}
In the equilateral triangle limit, $\left\langle W(\tau)\right\rangle '\rightarrow-Hd_{1}c_{1}5N_{k}^{2}/\left(4k^{6}\right)$. 
\item $\mathcal{H}_{3}^{I}(x,t)=a^{3}\frac{1}{S_{0}}\delta S\left(-\left(\partial_{t}\delta A\right)^{2}+\left(a^{-1}\partial_{i}\delta A\right)^{2}\right)$
and $\mathcal{H}_{2}^{I}(x,t)=a^{3}c_{2}\delta S\delta A$ 
\begin{eqnarray}
\left\langle W(\tau)\right\rangle '\approx\frac{c_{2}}{S_{0}}\frac{H^{2}N_{k}\left(k_{1}^{3}+k_{2}^{3}+k_{3}^{3}\right)}{12k_{1}^{3}k_{2}^{3}k_{3}^{3}}.\label{Type3}
\end{eqnarray}
In the equilateral triangle limit, $\left\langle W(\tau)\right\rangle '\rightarrow H^{2}c_{2}N_{k}/\left(4S_{0}k^{6}\right)$. 
\item $\mathcal{H}_{3}^{I}(x,t)=a^{3}\frac{1}{S_{0}}\delta S\left(-\left(\partial_{t}\delta A\right)^{2}+\left(a^{-1}\partial_{i}\delta A\right)^{2}\right)$
and $\mathcal{H}_{2}^{I}(x,t)=a^{3}c_{1}\delta S\partial_{t}\delta A$
\begin{eqnarray}
\left\langle W(\tau)\right\rangle '\approx\frac{c_{1}}{S_{0}}\frac{-H^{3}N_{k}\left(k_{1}^{3}+k_{2}^{3}+k_{3}^{3}\right)}{4k_{1}^{3}k_{2}^{3}k_{3}^{3}}.\label{Type4}
\end{eqnarray}
In the equilateral triangle limit, $\left\langle W(\tau)\right\rangle '\rightarrow-H^{3}c_{1}3N_{k}/\left(4S_{0}k^{6}\right)$. 
\end{enumerate}

\nocite{apsrev41Control}
\bibliographystyle{apsrev4-1}
\bibliography{refs,refs1,refs2}

\begin{thebibliography}{100}%
\makeatletter
\providecommand \@ifxundefined [1]{%
 \@ifx{#1\undefined}
}%
\providecommand \@ifnum [1]{%
 \ifnum #1\expandafter \@firstoftwo
 \else \expandafter \@secondoftwo
 \fi
}%
\providecommand \@ifx [1]{%
 \ifx #1\expandafter \@firstoftwo
 \else \expandafter \@secondoftwo
 \fi
}%
\providecommand \natexlab [1]{#1}%
\providecommand \enquote  [1]{``#1''}%
\providecommand \bibnamefont  [1]{#1}%
\providecommand \bibfnamefont [1]{#1}%
\providecommand \citenamefont [1]{#1}%
\providecommand \href@noop [0]{\@secondoftwo}%
\providecommand \href [0]{\begingroup \@sanitize@url \@href}%
\providecommand \@href[1]{\@@startlink{#1}\@@href}%
\providecommand \@@href[1]{\endgroup#1\@@endlink}%
\providecommand \@sanitize@url [0]{\catcode `\\12\catcode `\$12\catcode `\&12\catcode `\#12\catcode `\^12\catcode `\_12\catcode `\%12\relax}%
\providecommand \@@startlink[1]{}%
\providecommand \@@endlink[0]{}%
\providecommand \url  [0]{\begingroup\@sanitize@url \@url }%
\providecommand \@url [1]{\endgroup\@href {#1}{\urlprefix }}%
\providecommand \urlprefix  [0]{URL }%
\providecommand \Eprint [0]{\href }%
\providecommand \doibase [0]{http://dx.doi.org/}%
\providecommand \selectlanguage [0]{\@gobble}%
\providecommand \bibinfo  [0]{\@secondoftwo}%
\providecommand \bibfield  [0]{\@secondoftwo}%
\providecommand \translation [1]{[#1]}%
\providecommand \BibitemOpen [0]{}%
\providecommand \bibitemStop [0]{}%
\providecommand \bibitemNoStop [0]{.\EOS\space}%
\providecommand \EOS [0]{\spacefactor3000\relax}%
\providecommand \BibitemShut  [1]{\csname bibitem#1\endcsname}%
\let\auto@bib@innerbib\@empty
\bibitem [{\citenamefont {Riotto}(2003)}]{Riotto:2002yw}%
  \BibitemOpen
  \bibfield  {author} {\bibinfo {author} {\bibfnamefont {A.}~\bibnamefont {Riotto}},\ }\bibfield  {title} {\enquote {\bibinfo {title} {{Inflation and the theory of cosmological perturbations}},}\ }\href@noop {} {\bibfield  {journal} {\bibinfo  {journal} {ICTP Lect. Notes Ser.}\ }\textbf {\bibinfo {volume} {14}},\ \bibinfo {pages} {317--413} (\bibinfo {year} {2003})},\ \Eprint {http://arxiv.org/abs/hep-ph/0210162} {arXiv:hep-ph/0210162} \BibitemShut {NoStop}%
\bibitem [{\citenamefont {Linde}(2008)}]{Linde:2007fr}%
  \BibitemOpen
  \bibfield  {author} {\bibinfo {author} {\bibfnamefont {A.~D.}\ \bibnamefont {Linde}},\ }\bibfield  {title} {\enquote {\bibinfo {title} {{Inflationary Cosmology}},}\ }\href {\doibase 10.1007/978-3-540-74353-8_1} {\bibfield  {journal} {\bibinfo  {journal} {Lect. Notes Phys.}\ }\textbf {\bibinfo {volume} {738}},\ \bibinfo {pages} {1--54} (\bibinfo {year} {2008})},\ \Eprint {http://arxiv.org/abs/0705.0164} {arXiv:0705.0164 [hep-th]} \BibitemShut {NoStop}%
\bibitem [{\citenamefont {Bartolo}\ \emph {et~al.}(2004{\natexlab{a}})\citenamefont {Bartolo}, \citenamefont {Komatsu}, \citenamefont {Matarrese},\ and\ \citenamefont {Riotto}}]{Bartolo:2004if}%
  \BibitemOpen
  \bibfield  {author} {\bibinfo {author} {\bibfnamefont {N.}~\bibnamefont {Bartolo}}, \bibinfo {author} {\bibfnamefont {E.}~\bibnamefont {Komatsu}}, \bibinfo {author} {\bibfnamefont {S.}~\bibnamefont {Matarrese}}, \ and\ \bibinfo {author} {\bibfnamefont {A.}~\bibnamefont {Riotto}},\ }\bibfield  {title} {\enquote {\bibinfo {title} {{Non-Gaussianity from inflation: Theory and observations}},}\ }\href {\doibase 10.1016/j.physrep.2004.08.022} {\bibfield  {journal} {\bibinfo  {journal} {Phys. Rept.}\ }\textbf {\bibinfo {volume} {402}},\ \bibinfo {pages} {103--266} (\bibinfo {year} {2004}{\natexlab{a}})},\ \Eprint {http://arxiv.org/abs/astro-ph/0406398} {arXiv:astro-ph/0406398} \BibitemShut {NoStop}%
\bibitem [{\citenamefont {Akrami}\ \emph {et~al.}(2020{\natexlab{a}})\citenamefont {Akrami} \emph {et~al.}}]{Planck:2019kim}%
  \BibitemOpen
  \bibfield  {author} {\bibinfo {author} {\bibfnamefont {Y.}~\bibnamefont {Akrami}} \emph {et~al.} (\bibinfo {collaboration} {Planck}),\ }\bibfield  {title} {\enquote {\bibinfo {title} {{Planck 2018 results. IX. Constraints on primordial non-Gaussianity}},}\ }\href {\doibase 10.1051/0004-6361/201935891} {\bibfield  {journal} {\bibinfo  {journal} {Astron. Astrophys.}\ }\textbf {\bibinfo {volume} {641}},\ \bibinfo {pages} {A9} (\bibinfo {year} {2020}{\natexlab{a}})},\ \Eprint {http://arxiv.org/abs/1905.05697} {arXiv:1905.05697 [astro-ph.CO]} \BibitemShut {NoStop}%
\bibitem [{\citenamefont {Abazajian}\ \emph {et~al.}(2019)\citenamefont {Abazajian} \emph {et~al.}}]{Abazajian:2019eic}%
  \BibitemOpen
  \bibfield  {author} {\bibinfo {author} {\bibfnamefont {K.}~\bibnamefont {Abazajian}} \emph {et~al.},\ }\bibfield  {title} {\enquote {\bibinfo {title} {{CMB-S4 Science Case, Reference Design, and Project Plan}},}\ }\href@noop {} {\  (\bibinfo {year} {2019})},\ \Eprint {http://arxiv.org/abs/1907.04473} {arXiv:1907.04473 [astro-ph.IM]} \BibitemShut {NoStop}%
\bibitem [{\citenamefont {Mueller}\ \emph {et~al.}(2022)\citenamefont {Mueller} \emph {et~al.}}]{eBOSS:2021jbt}%
  \BibitemOpen
  \bibfield  {author} {\bibinfo {author} {\bibfnamefont {E.-M.}\ \bibnamefont {Mueller}} \emph {et~al.} (\bibinfo {collaboration} {eBOSS}),\ }\bibfield  {title} {\enquote {\bibinfo {title} {{Primordial non-Gaussianity from the completed SDSS-IV extended Baryon Oscillation Spectroscopic Survey II: measurements in Fourier space with optimal weights}},}\ }\href {\doibase 10.1093/mnras/stac812} {\bibfield  {journal} {\bibinfo  {journal} {Mon. Not. Roy. Astron. Soc.}\ }\textbf {\bibinfo {volume} {514}},\ \bibinfo {pages} {3396--3409} (\bibinfo {year} {2022})},\ \Eprint {http://arxiv.org/abs/2106.13725} {arXiv:2106.13725 [astro-ph.CO]} \BibitemShut {NoStop}%
\bibitem [{\citenamefont {Cabass}\ \emph {et~al.}(2022)\citenamefont {Cabass}, \citenamefont {Ivanov}, \citenamefont {Philcox}, \citenamefont {Simonovi\'c},\ and\ \citenamefont {Zaldarriaga}}]{Cabass:2022wjy}%
  \BibitemOpen
  \bibfield  {author} {\bibinfo {author} {\bibfnamefont {G.}~\bibnamefont {Cabass}}, \bibinfo {author} {\bibfnamefont {M.~M.}\ \bibnamefont {Ivanov}}, \bibinfo {author} {\bibfnamefont {O.~H.~E.}\ \bibnamefont {Philcox}}, \bibinfo {author} {\bibfnamefont {M.}~\bibnamefont {Simonovi\'c}}, \ and\ \bibinfo {author} {\bibfnamefont {M.}~\bibnamefont {Zaldarriaga}},\ }\bibfield  {title} {\enquote {\bibinfo {title} {{Constraints on Single-Field Inflation from the BOSS Galaxy Survey}},}\ }\href {\doibase 10.1103/PhysRevLett.129.021301} {\bibfield  {journal} {\bibinfo  {journal} {Phys. Rev. Lett.}\ }\textbf {\bibinfo {volume} {129}},\ \bibinfo {pages} {021301} (\bibinfo {year} {2022})},\ \Eprint {http://arxiv.org/abs/2201.07238} {arXiv:2201.07238 [astro-ph.CO]} \BibitemShut {NoStop}%
\bibitem [{\citenamefont {D'Amico}\ \emph {et~al.}(2022)\citenamefont {D'Amico}, \citenamefont {Lewandowski}, \citenamefont {Senatore},\ and\ \citenamefont {Zhang}}]{DAmico:2022gki}%
  \BibitemOpen
  \bibfield  {author} {\bibinfo {author} {\bibfnamefont {G.}~\bibnamefont {D'Amico}}, \bibinfo {author} {\bibfnamefont {M.}~\bibnamefont {Lewandowski}}, \bibinfo {author} {\bibfnamefont {L.}~\bibnamefont {Senatore}}, \ and\ \bibinfo {author} {\bibfnamefont {P.}~\bibnamefont {Zhang}},\ }\bibfield  {title} {\enquote {\bibinfo {title} {{Limits on primordial non-Gaussianities from BOSS galaxy-clustering data}},}\ }\href@noop {} {\  (\bibinfo {year} {2022})},\ \Eprint {http://arxiv.org/abs/2201.11518} {arXiv:2201.11518 [astro-ph.CO]} \BibitemShut {NoStop}%
\bibitem [{\citenamefont {Dor\'e}\ \emph {et~al.}(2014)\citenamefont {Dor\'e} \emph {et~al.}}]{SPHEREx:2014bgr}%
  \BibitemOpen
  \bibfield  {author} {\bibinfo {author} {\bibfnamefont {O.}~\bibnamefont {Dor\'e}} \emph {et~al.} (\bibinfo {collaboration} {SPHEREx}),\ }\bibfield  {title} {\enquote {\bibinfo {title} {{Cosmology with the SPHEREX All-Sky Spectral Survey}},}\ }\href@noop {} {\  (\bibinfo {year} {2014})},\ \Eprint {http://arxiv.org/abs/1412.4872} {arXiv:1412.4872 [astro-ph.CO]} \BibitemShut {NoStop}%
\bibitem [{\citenamefont {Schlegel}\ \emph {et~al.}(2019)\citenamefont {Schlegel} \emph {et~al.}}]{Schlegel:2019eqc}%
  \BibitemOpen
  \bibfield  {author} {\bibinfo {author} {\bibfnamefont {D.~J.}\ \bibnamefont {Schlegel}} \emph {et~al.},\ }\bibfield  {title} {\enquote {\bibinfo {title} {{Astro2020 APC White Paper: The MegaMapper: a z \ensuremath{>} 2 Spectroscopic Instrument for the Study of Inflation and Dark Energy}},}\ }\href@noop {} {\bibfield  {journal} {\bibinfo  {journal} {Bull. Am. Astron. Soc.}\ }\textbf {\bibinfo {volume} {51}},\ \bibinfo {pages} {229} (\bibinfo {year} {2019})},\ \Eprint {http://arxiv.org/abs/1907.11171} {arXiv:1907.11171 [astro-ph.IM]} \BibitemShut {NoStop}%
\bibitem [{\citenamefont {Abell}\ \emph {et~al.}(2009)\citenamefont {Abell} \emph {et~al.}}]{LSSTScience:2009jmu}%
  \BibitemOpen
  \bibfield  {author} {\bibinfo {author} {\bibfnamefont {P.~A.}\ \bibnamefont {Abell}} \emph {et~al.} (\bibinfo {collaboration} {LSST Science, LSST Project}),\ }\bibfield  {title} {\enquote {\bibinfo {title} {{LSST Science Book, Version 2.0}},}\ }\href@noop {} {\  (\bibinfo {year} {2009})},\ \Eprint {http://arxiv.org/abs/0912.0201} {arXiv:0912.0201 [astro-ph.IM]} \BibitemShut {NoStop}%
\bibitem [{\citenamefont {Meerburg}\ \emph {et~al.}(2017)\citenamefont {Meerburg}, \citenamefont {M\"unchmeyer}, \citenamefont {Mu\~noz},\ and\ \citenamefont {Chen}}]{Meerburg:2016zdz}%
  \BibitemOpen
  \bibfield  {author} {\bibinfo {author} {\bibfnamefont {P.~D.}\ \bibnamefont {Meerburg}}, \bibinfo {author} {\bibfnamefont {M.}~\bibnamefont {M\"unchmeyer}}, \bibinfo {author} {\bibfnamefont {J.~B.}\ \bibnamefont {Mu\~noz}}, \ and\ \bibinfo {author} {\bibfnamefont {X.}~\bibnamefont {Chen}},\ }\bibfield  {title} {\enquote {\bibinfo {title} {{Prospects for Cosmological Collider Physics}},}\ }\href {\doibase 10.1088/1475-7516/2017/03/050} {\bibfield  {journal} {\bibinfo  {journal} {JCAP}\ }\textbf {\bibinfo {volume} {03}},\ \bibinfo {pages} {050} (\bibinfo {year} {2017})},\ \Eprint {http://arxiv.org/abs/1610.06559} {arXiv:1610.06559 [astro-ph.CO]} \BibitemShut {NoStop}%
\bibitem [{\citenamefont {Abazajian}\ \emph {et~al.}(2016)\citenamefont {Abazajian} \emph {et~al.}}]{CMB-S4:2016ple}%
  \BibitemOpen
  \bibfield  {author} {\bibinfo {author} {\bibfnamefont {K.~N.}\ \bibnamefont {Abazajian}} \emph {et~al.} (\bibinfo {collaboration} {CMB-S4}),\ }\bibfield  {title} {\enquote {\bibinfo {title} {{CMB-S4 Science Book, First Edition}},}\ }\href@noop {} {\  (\bibinfo {year} {2016})},\ \Eprint {http://arxiv.org/abs/1610.02743} {arXiv:1610.02743 [astro-ph.CO]} \BibitemShut {NoStop}%
\bibitem [{\citenamefont {Ade}\ \emph {et~al.}(2019)\citenamefont {Ade} \emph {et~al.}}]{SimonsObservatory:2018koc}%
  \BibitemOpen
  \bibfield  {author} {\bibinfo {author} {\bibfnamefont {P.}~\bibnamefont {Ade}} \emph {et~al.} (\bibinfo {collaboration} {Simons Observatory}),\ }\bibfield  {title} {\enquote {\bibinfo {title} {{The Simons Observatory: Science goals and forecasts}},}\ }\href {\doibase 10.1088/1475-7516/2019/02/056} {\bibfield  {journal} {\bibinfo  {journal} {JCAP}\ }\textbf {\bibinfo {volume} {02}},\ \bibinfo {pages} {056} (\bibinfo {year} {2019})},\ \Eprint {http://arxiv.org/abs/1808.07445} {arXiv:1808.07445 [astro-ph.CO]} \BibitemShut {NoStop}%
\bibitem [{\citenamefont {Komatsu}\ and\ \citenamefont {Spergel}(2001)}]{Komatsu:2001rj}%
  \BibitemOpen
  \bibfield  {author} {\bibinfo {author} {\bibfnamefont {E.}~\bibnamefont {Komatsu}}\ and\ \bibinfo {author} {\bibfnamefont {D.~N.}\ \bibnamefont {Spergel}},\ }\bibfield  {title} {\enquote {\bibinfo {title} {{Acoustic signatures in the primary microwave background bispectrum}},}\ }\href {\doibase 10.1103/PhysRevD.63.063002} {\bibfield  {journal} {\bibinfo  {journal} {Phys. Rev. D}\ }\textbf {\bibinfo {volume} {63}},\ \bibinfo {pages} {063002} (\bibinfo {year} {2001})},\ \Eprint {http://arxiv.org/abs/astro-ph/0005036} {arXiv:astro-ph/0005036} \BibitemShut {NoStop}%
\bibitem [{\citenamefont {Enqvist}\ and\ \citenamefont {Nurmi}(2005)}]{Enqvist:2005pg}%
  \BibitemOpen
  \bibfield  {author} {\bibinfo {author} {\bibfnamefont {K.}~\bibnamefont {Enqvist}}\ and\ \bibinfo {author} {\bibfnamefont {S.}~\bibnamefont {Nurmi}},\ }\bibfield  {title} {\enquote {\bibinfo {title} {{Non-gaussianity in curvaton models with nearly quadratic potential}},}\ }\href {\doibase 10.1088/1475-7516/2005/10/013} {\bibfield  {journal} {\bibinfo  {journal} {JCAP}\ }\textbf {\bibinfo {volume} {10}},\ \bibinfo {pages} {013} (\bibinfo {year} {2005})},\ \Eprint {http://arxiv.org/abs/astro-ph/0508573} {arXiv:astro-ph/0508573} \BibitemShut {NoStop}%
\bibitem [{\citenamefont {Enqvist}\ and\ \citenamefont {Takahashi}(2008)}]{Enqvist:2008gk}%
  \BibitemOpen
  \bibfield  {author} {\bibinfo {author} {\bibfnamefont {K.}~\bibnamefont {Enqvist}}\ and\ \bibinfo {author} {\bibfnamefont {T.}~\bibnamefont {Takahashi}},\ }\bibfield  {title} {\enquote {\bibinfo {title} {{Signatures of Non-Gaussianity in the Curvaton Model}},}\ }\href {\doibase 10.1088/1475-7516/2008/09/012} {\bibfield  {journal} {\bibinfo  {journal} {JCAP}\ }\textbf {\bibinfo {volume} {09}},\ \bibinfo {pages} {012} (\bibinfo {year} {2008})},\ \Eprint {http://arxiv.org/abs/0807.3069} {arXiv:0807.3069 [astro-ph]} \BibitemShut {NoStop}%
\bibitem [{\citenamefont {Huang}(2008)}]{Huang:2008zj}%
  \BibitemOpen
  \bibfield  {author} {\bibinfo {author} {\bibfnamefont {Q.-G.}\ \bibnamefont {Huang}},\ }\bibfield  {title} {\enquote {\bibinfo {title} {{Curvaton with Polynomial Potential}},}\ }\href {\doibase 10.1088/1475-7516/2008/11/005} {\bibfield  {journal} {\bibinfo  {journal} {JCAP}\ }\textbf {\bibinfo {volume} {11}},\ \bibinfo {pages} {005} (\bibinfo {year} {2008})},\ \Eprint {http://arxiv.org/abs/0808.1793} {arXiv:0808.1793 [hep-th]} \BibitemShut {NoStop}%
\bibitem [{\citenamefont {Enqvist}\ \emph {et~al.}(2009)\citenamefont {Enqvist}, \citenamefont {Nurmi}, \citenamefont {Rigopoulos}, \citenamefont {Taanila},\ and\ \citenamefont {Takahashi}}]{Enqvist:2009zf}%
  \BibitemOpen
  \bibfield  {author} {\bibinfo {author} {\bibfnamefont {K.}~\bibnamefont {Enqvist}}, \bibinfo {author} {\bibfnamefont {S.}~\bibnamefont {Nurmi}}, \bibinfo {author} {\bibfnamefont {G.}~\bibnamefont {Rigopoulos}}, \bibinfo {author} {\bibfnamefont {O.}~\bibnamefont {Taanila}}, \ and\ \bibinfo {author} {\bibfnamefont {T.}~\bibnamefont {Takahashi}},\ }\bibfield  {title} {\enquote {\bibinfo {title} {{The Subdominant Curvaton}},}\ }\href {\doibase 10.1088/1475-7516/2009/11/003} {\bibfield  {journal} {\bibinfo  {journal} {JCAP}\ }\textbf {\bibinfo {volume} {11}},\ \bibinfo {pages} {003} (\bibinfo {year} {2009})},\ \Eprint {http://arxiv.org/abs/0906.3126} {arXiv:0906.3126 [astro-ph.CO]} \BibitemShut {NoStop}%
\bibitem [{\citenamefont {Enqvist}\ and\ \citenamefont {Takahashi}(2009)}]{Enqvist:2009eq}%
  \BibitemOpen
  \bibfield  {author} {\bibinfo {author} {\bibfnamefont {K.}~\bibnamefont {Enqvist}}\ and\ \bibinfo {author} {\bibfnamefont {T.}~\bibnamefont {Takahashi}},\ }\bibfield  {title} {\enquote {\bibinfo {title} {{Effect of Background Evolution on the Curvaton Non-Gaussianity}},}\ }\href {\doibase 10.1088/1475-7516/2009/12/001} {\bibfield  {journal} {\bibinfo  {journal} {JCAP}\ }\textbf {\bibinfo {volume} {12}},\ \bibinfo {pages} {001} (\bibinfo {year} {2009})},\ \Eprint {http://arxiv.org/abs/0909.5362} {arXiv:0909.5362 [astro-ph.CO]} \BibitemShut {NoStop}%
\bibitem [{\citenamefont {Enqvist}\ \emph {et~al.}(2010)\citenamefont {Enqvist}, \citenamefont {Nurmi}, \citenamefont {Taanila},\ and\ \citenamefont {Takahashi}}]{Enqvist:2009ww}%
  \BibitemOpen
  \bibfield  {author} {\bibinfo {author} {\bibfnamefont {K.}~\bibnamefont {Enqvist}}, \bibinfo {author} {\bibfnamefont {S.}~\bibnamefont {Nurmi}}, \bibinfo {author} {\bibfnamefont {O.}~\bibnamefont {Taanila}}, \ and\ \bibinfo {author} {\bibfnamefont {T.}~\bibnamefont {Takahashi}},\ }\bibfield  {title} {\enquote {\bibinfo {title} {{Non-Gaussian Fingerprints of Self-Interacting Curvaton}},}\ }\href {\doibase 10.1088/1475-7516/2010/04/009} {\bibfield  {journal} {\bibinfo  {journal} {JCAP}\ }\textbf {\bibinfo {volume} {04}},\ \bibinfo {pages} {009} (\bibinfo {year} {2010})},\ \Eprint {http://arxiv.org/abs/0912.4657} {arXiv:0912.4657 [astro-ph.CO]} \BibitemShut {NoStop}%
\bibitem [{\citenamefont {Enqvist}(2011)}]{Enqvist:2010dt}%
  \BibitemOpen
  \bibfield  {author} {\bibinfo {author} {\bibfnamefont {K.}~\bibnamefont {Enqvist}},\ }\bibfield  {title} {\enquote {\bibinfo {title} {{The self-interacting curvaton}},}\ }\href {\doibase 10.1143/PTPS.190.62} {\bibfield  {journal} {\bibinfo  {journal} {Prog. Theor. Phys. Suppl.}\ }\textbf {\bibinfo {volume} {190}},\ \bibinfo {pages} {62--74} (\bibinfo {year} {2011})},\ \Eprint {http://arxiv.org/abs/1012.1711} {arXiv:1012.1711 [astro-ph.CO]} \BibitemShut {NoStop}%
\bibitem [{\citenamefont {Fonseca}\ and\ \citenamefont {Wands}(2011)}]{Fonseca:2011aa}%
  \BibitemOpen
  \bibfield  {author} {\bibinfo {author} {\bibfnamefont {J.}~\bibnamefont {Fonseca}}\ and\ \bibinfo {author} {\bibfnamefont {D.}~\bibnamefont {Wands}},\ }\bibfield  {title} {\enquote {\bibinfo {title} {{Non-Gaussianity and Gravitational Waves from Quadratic and Self-interacting Curvaton}},}\ }\href {\doibase 10.1103/PhysRevD.83.064025} {\bibfield  {journal} {\bibinfo  {journal} {Phys. Rev. D}\ }\textbf {\bibinfo {volume} {83}},\ \bibinfo {pages} {064025} (\bibinfo {year} {2011})},\ \Eprint {http://arxiv.org/abs/1101.1254} {arXiv:1101.1254 [astro-ph.CO]} \BibitemShut {NoStop}%
\bibitem [{\citenamefont {Kawasaki}\ \emph {et~al.}(2008)\citenamefont {Kawasaki}, \citenamefont {Nakayama}, \citenamefont {Sekiguchi}, \citenamefont {Suyama},\ and\ \citenamefont {Takahashi}}]{Kawasaki:2008sn}%
  \BibitemOpen
  \bibfield  {author} {\bibinfo {author} {\bibfnamefont {M.}~\bibnamefont {Kawasaki}}, \bibinfo {author} {\bibfnamefont {K.}~\bibnamefont {Nakayama}}, \bibinfo {author} {\bibfnamefont {T.}~\bibnamefont {Sekiguchi}}, \bibinfo {author} {\bibfnamefont {T.}~\bibnamefont {Suyama}}, \ and\ \bibinfo {author} {\bibfnamefont {F.}~\bibnamefont {Takahashi}},\ }\bibfield  {title} {\enquote {\bibinfo {title} {{Non-Gaussianity from isocurvature perturbations}},}\ }\href {\doibase 10.1088/1475-7516/2008/11/019} {\bibfield  {journal} {\bibinfo  {journal} {JCAP}\ }\textbf {\bibinfo {volume} {11}},\ \bibinfo {pages} {019} (\bibinfo {year} {2008})},\ \Eprint {http://arxiv.org/abs/0808.0009} {arXiv:0808.0009 [astro-ph]} \BibitemShut {NoStop}%
\bibitem [{\citenamefont {Hikage}\ \emph {et~al.}(2013)\citenamefont {Hikage}, \citenamefont {Kawasaki}, \citenamefont {Sekiguchi},\ and\ \citenamefont {Takahashi}}]{Hikage:2012be}%
  \BibitemOpen
  \bibfield  {author} {\bibinfo {author} {\bibfnamefont {C.}~\bibnamefont {Hikage}}, \bibinfo {author} {\bibfnamefont {M.}~\bibnamefont {Kawasaki}}, \bibinfo {author} {\bibfnamefont {T.}~\bibnamefont {Sekiguchi}}, \ and\ \bibinfo {author} {\bibfnamefont {T.}~\bibnamefont {Takahashi}},\ }\bibfield  {title} {\enquote {\bibinfo {title} {{CMB constraint on non-Gaussianity in isocurvature perturbations}},}\ }\href {\doibase 10.1088/1475-7516/2013/07/007} {\bibfield  {journal} {\bibinfo  {journal} {JCAP}\ }\textbf {\bibinfo {volume} {07}},\ \bibinfo {pages} {007} (\bibinfo {year} {2013})},\ \Eprint {http://arxiv.org/abs/1211.1095} {arXiv:1211.1095 [astro-ph.CO]} \BibitemShut {NoStop}%
\bibitem [{\citenamefont {Geller}\ \emph {et~al.}(2024)\citenamefont {Geller}, \citenamefont {Kumar},\ and\ \citenamefont {Wang}}]{Geller:2024upd}%
  \BibitemOpen
  \bibfield  {author} {\bibinfo {author} {\bibfnamefont {M.}~\bibnamefont {Geller}}, \bibinfo {author} {\bibfnamefont {S.}~\bibnamefont {Kumar}}, \ and\ \bibinfo {author} {\bibfnamefont {L.-T.}\ \bibnamefont {Wang}},\ }\bibfield  {title} {\enquote {\bibinfo {title} {{Probing Dark Matter Isocurvature with Primordial Non-Gaussianity}},}\ }\href {\doibase https://doi.org/10.1088/1475-7516/2024/12/018} {\bibfield  {journal} {\bibinfo  {journal} {JCAP}\ }\textbf {\bibinfo {volume} {12}},\ \bibinfo {pages} {018} (\bibinfo {year} {2024})},\ \Eprint {http://arxiv.org/abs/2405.09607} {arXiv:2405.09607 [astro-ph.CO]} \BibitemShut {NoStop}%
\bibitem [{\citenamefont {Peccei}\ and\ \citenamefont {Quinn}(1977{\natexlab{a}})}]{Peccei:1977hh}%
  \BibitemOpen
  \bibfield  {author} {\bibinfo {author} {\bibfnamefont {R.~D.}\ \bibnamefont {Peccei}}\ and\ \bibinfo {author} {\bibfnamefont {H.~R.}\ \bibnamefont {Quinn}},\ }\bibfield  {title} {\enquote {\bibinfo {title} {{CP Conservation in the Presence of Instantons}},}\ }\href {\doibase 10.1103/PhysRevLett.38.1440} {\bibfield  {journal} {\bibinfo  {journal} {Phys. Rev. Lett.}\ }\textbf {\bibinfo {volume} {38}},\ \bibinfo {pages} {1440--1443} (\bibinfo {year} {1977}{\natexlab{a}})}\BibitemShut {NoStop}%
\bibitem [{\citenamefont {Peccei}\ and\ \citenamefont {Quinn}(1977{\natexlab{b}})}]{Peccei:1977ur}%
  \BibitemOpen
  \bibfield  {author} {\bibinfo {author} {\bibfnamefont {R.~D.}\ \bibnamefont {Peccei}}\ and\ \bibinfo {author} {\bibfnamefont {H.~R.}\ \bibnamefont {Quinn}},\ }\bibfield  {title} {\enquote {\bibinfo {title} {{Constraints Imposed by CP Conservation in the Presence of Instantons}},}\ }\href {\doibase 10.1103/PhysRevD.16.1791} {\bibfield  {journal} {\bibinfo  {journal} {Phys. Rev. D}\ }\textbf {\bibinfo {volume} {16}},\ \bibinfo {pages} {1791--1797} (\bibinfo {year} {1977}{\natexlab{b}})}\BibitemShut {NoStop}%
\bibitem [{\citenamefont {Weinberg}(1978)}]{Weinberg:1977ma}%
  \BibitemOpen
  \bibfield  {author} {\bibinfo {author} {\bibfnamefont {S.}~\bibnamefont {Weinberg}},\ }\bibfield  {title} {\enquote {\bibinfo {title} {{A New Light Boson?}}}\ }\href {\doibase 10.1103/PhysRevLett.40.223} {\bibfield  {journal} {\bibinfo  {journal} {Phys. Rev. Lett.}\ }\textbf {\bibinfo {volume} {40}},\ \bibinfo {pages} {223--226} (\bibinfo {year} {1978})}\BibitemShut {NoStop}%
\bibitem [{\citenamefont {Wilczek}(1978)}]{Wilczek:1977pj}%
  \BibitemOpen
  \bibfield  {author} {\bibinfo {author} {\bibfnamefont {F.}~\bibnamefont {Wilczek}},\ }\bibfield  {title} {\enquote {\bibinfo {title} {{Problem of Strong $P$ and $T$ Invariance in the Presence of Instantons}},}\ }\href {\doibase 10.1103/PhysRevLett.40.279} {\bibfield  {journal} {\bibinfo  {journal} {Phys. Rev. Lett.}\ }\textbf {\bibinfo {volume} {40}},\ \bibinfo {pages} {279--282} (\bibinfo {year} {1978})}\BibitemShut {NoStop}%
\bibitem [{\citenamefont {Marsh}(2016)}]{Marsh:2015xka}%
  \BibitemOpen
  \bibfield  {author} {\bibinfo {author} {\bibfnamefont {D.~J.~E.}\ \bibnamefont {Marsh}},\ }\bibfield  {title} {\enquote {\bibinfo {title} {{Axion Cosmology}},}\ }\href {\doibase 10.1016/j.physrep.2016.06.005} {\bibfield  {journal} {\bibinfo  {journal} {Phys. Rept.}\ }\textbf {\bibinfo {volume} {643}},\ \bibinfo {pages} {1--79} (\bibinfo {year} {2016})},\ \Eprint {http://arxiv.org/abs/1510.07633} {arXiv:1510.07633 [astro-ph.CO]} \BibitemShut {NoStop}%
\bibitem [{\citenamefont {Di~Luzio}\ \emph {et~al.}(2020)\citenamefont {Di~Luzio}, \citenamefont {Giannotti}, \citenamefont {Nardi},\ and\ \citenamefont {Visinelli}}]{DiLuzio:2020wdo}%
  \BibitemOpen
  \bibfield  {author} {\bibinfo {author} {\bibfnamefont {L.}~\bibnamefont {Di~Luzio}}, \bibinfo {author} {\bibfnamefont {M.}~\bibnamefont {Giannotti}}, \bibinfo {author} {\bibfnamefont {E.}~\bibnamefont {Nardi}}, \ and\ \bibinfo {author} {\bibfnamefont {L.}~\bibnamefont {Visinelli}},\ }\bibfield  {title} {\enquote {\bibinfo {title} {{The landscape of QCD axion models}},}\ }\href {\doibase 10.1016/j.physrep.2020.06.002} {\bibfield  {journal} {\bibinfo  {journal} {Phys. Rept.}\ }\textbf {\bibinfo {volume} {870}},\ \bibinfo {pages} {1--117} (\bibinfo {year} {2020})},\ \Eprint {http://arxiv.org/abs/2003.01100} {arXiv:2003.01100 [hep-ph]} \BibitemShut {NoStop}%
\bibitem [{\citenamefont {Lu}(2022)}]{Lu:2021gso}%
  \BibitemOpen
  \bibfield  {author} {\bibinfo {author} {\bibfnamefont {S.}~\bibnamefont {Lu}},\ }\bibfield  {title} {\enquote {\bibinfo {title} {{Axion isocurvature collider}},}\ }\href {\doibase 10.1007/JHEP04(2022)157} {\bibfield  {journal} {\bibinfo  {journal} {JHEP}\ }\textbf {\bibinfo {volume} {04}},\ \bibinfo {pages} {157} (\bibinfo {year} {2022})},\ \Eprint {http://arxiv.org/abs/2103.05958} {arXiv:2103.05958 [hep-th]} \BibitemShut {NoStop}%
\bibitem [{\citenamefont {Chen}\ \emph {et~al.}(2023)\citenamefont {Chen}, \citenamefont {Fan},\ and\ \citenamefont {Li}}]{Chen:2023txq}%
  \BibitemOpen
  \bibfield  {author} {\bibinfo {author} {\bibfnamefont {X.}~\bibnamefont {Chen}}, \bibinfo {author} {\bibfnamefont {J.}~\bibnamefont {Fan}}, \ and\ \bibinfo {author} {\bibfnamefont {L.}~\bibnamefont {Li}},\ }\bibfield  {title} {\enquote {\bibinfo {title} {{New inflationary probes of axion dark matter}},}\ }\href {\doibase 10.1007/JHEP12(2023)197} {\bibfield  {journal} {\bibinfo  {journal} {JHEP}\ }\textbf {\bibinfo {volume} {12}},\ \bibinfo {pages} {197} (\bibinfo {year} {2023})},\ \Eprint {http://arxiv.org/abs/2303.03406} {arXiv:2303.03406 [hep-ph]} \BibitemShut {NoStop}%
\bibitem [{\citenamefont {Affleck}\ and\ \citenamefont {Dine}(1985)}]{Affleck:1984fy}%
  \BibitemOpen
  \bibfield  {author} {\bibinfo {author} {\bibfnamefont {I.}~\bibnamefont {Affleck}}\ and\ \bibinfo {author} {\bibfnamefont {M.}~\bibnamefont {Dine}},\ }\bibfield  {title} {\enquote {\bibinfo {title} {{A New Mechanism for Baryogenesis}},}\ }\href {\doibase 10.1016/0550-3213(85)90021-5} {\bibfield  {journal} {\bibinfo  {journal} {Nucl. Phys. B}\ }\textbf {\bibinfo {volume} {249}},\ \bibinfo {pages} {361--380} (\bibinfo {year} {1985})}\BibitemShut {NoStop}%
\bibitem [{\citenamefont {Nilles}(1984)}]{Nilles:1983ge}%
  \BibitemOpen
  \bibfield  {author} {\bibinfo {author} {\bibfnamefont {H.~P.}\ \bibnamefont {Nilles}},\ }\bibfield  {title} {\enquote {\bibinfo {title} {{Supersymmetry, Supergravity and Particle Physics}},}\ }\href {\doibase 10.1016/0370-1573(84)90008-5} {\bibfield  {journal} {\bibinfo  {journal} {Phys. Rept.}\ }\textbf {\bibinfo {volume} {110}},\ \bibinfo {pages} {1--162} (\bibinfo {year} {1984})}\BibitemShut {NoStop}%
\bibitem [{\citenamefont {Mazumdar}\ and\ \citenamefont {Rocher}(2011)}]{Mazumdar:2010sa}%
  \BibitemOpen
  \bibfield  {author} {\bibinfo {author} {\bibfnamefont {A.}~\bibnamefont {Mazumdar}}\ and\ \bibinfo {author} {\bibfnamefont {J.}~\bibnamefont {Rocher}},\ }\bibfield  {title} {\enquote {\bibinfo {title} {{Particle physics models of inflation and curvaton scenarios}},}\ }\href {\doibase 10.1016/j.physrep.2010.08.001} {\bibfield  {journal} {\bibinfo  {journal} {Phys. Rept.}\ }\textbf {\bibinfo {volume} {497}},\ \bibinfo {pages} {85--215} (\bibinfo {year} {2011})},\ \Eprint {http://arxiv.org/abs/1001.0993} {arXiv:1001.0993 [hep-ph]} \BibitemShut {NoStop}%
\bibitem [{\citenamefont {Co}\ \emph {et~al.}(2022)\citenamefont {Co}, \citenamefont {Harigaya},\ and\ \citenamefont {Pierce}}]{Co:2022qpr}%
  \BibitemOpen
  \bibfield  {author} {\bibinfo {author} {\bibfnamefont {R.~T.}\ \bibnamefont {Co}}, \bibinfo {author} {\bibfnamefont {K.}~\bibnamefont {Harigaya}}, \ and\ \bibinfo {author} {\bibfnamefont {A.}~\bibnamefont {Pierce}},\ }\bibfield  {title} {\enquote {\bibinfo {title} {{Cosmic perturbations from a rotating field}},}\ }\href {\doibase 10.1088/1475-7516/2022/10/037} {\bibfield  {journal} {\bibinfo  {journal} {JCAP}\ }\textbf {\bibinfo {volume} {10}},\ \bibinfo {pages} {037} (\bibinfo {year} {2022})},\ \Eprint {http://arxiv.org/abs/2202.01785} {arXiv:2202.01785 [hep-ph]} \BibitemShut {NoStop}%
\bibitem [{\citenamefont {Moxhay}\ and\ \citenamefont {Yamamoto}(1985)}]{Moxhay:1984am}%
  \BibitemOpen
  \bibfield  {author} {\bibinfo {author} {\bibfnamefont {P.}~\bibnamefont {Moxhay}}\ and\ \bibinfo {author} {\bibfnamefont {K.}~\bibnamefont {Yamamoto}},\ }\bibfield  {title} {\enquote {\bibinfo {title} {{{Peccei-Quinn} Symmetry Breaking by Radiative Corrections in Supergravity}},}\ }\href {\doibase 10.1016/0370-2693(85)91655-7} {\bibfield  {journal} {\bibinfo  {journal} {Phys. Lett. B}\ }\textbf {\bibinfo {volume} {151}},\ \bibinfo {pages} {363--366} (\bibinfo {year} {1985})}\BibitemShut {NoStop}%
\bibitem [{\citenamefont {Kasuya}\ \emph {et~al.}(1997)\citenamefont {Kasuya}, \citenamefont {Kawasaki},\ and\ \citenamefont {Yanagida}}]{Kasuya:1996ns}%
  \BibitemOpen
  \bibfield  {author} {\bibinfo {author} {\bibfnamefont {S.}~\bibnamefont {Kasuya}}, \bibinfo {author} {\bibfnamefont {M.}~\bibnamefont {Kawasaki}}, \ and\ \bibinfo {author} {\bibfnamefont {T.}~\bibnamefont {Yanagida}},\ }\bibfield  {title} {\enquote {\bibinfo {title} {{Cosmological axion problem in chaotic inflationary universe}},}\ }\href {\doibase 10.1016/S0370-2693(97)00809-5} {\bibfield  {journal} {\bibinfo  {journal} {Phys. Lett. B}\ }\textbf {\bibinfo {volume} {409}},\ \bibinfo {pages} {94--100} (\bibinfo {year} {1997})},\ \Eprint {http://arxiv.org/abs/hep-ph/9608405} {arXiv:hep-ph/9608405} \BibitemShut {NoStop}%
\bibitem [{\citenamefont {Kawasaki}\ \emph {et~al.}(2009)\citenamefont {Kawasaki}, \citenamefont {Nakayama},\ and\ \citenamefont {Takahashi}}]{Kawasaki:2008mc}%
  \BibitemOpen
  \bibfield  {author} {\bibinfo {author} {\bibfnamefont {M.}~\bibnamefont {Kawasaki}}, \bibinfo {author} {\bibfnamefont {K.}~\bibnamefont {Nakayama}}, \ and\ \bibinfo {author} {\bibfnamefont {F.}~\bibnamefont {Takahashi}},\ }\bibfield  {title} {\enquote {\bibinfo {title} {{Hilltop Non-Gaussianity}},}\ }\href {\doibase 10.1088/1475-7516/2009/01/026} {\bibfield  {journal} {\bibinfo  {journal} {JCAP}\ }\textbf {\bibinfo {volume} {01}},\ \bibinfo {pages} {026} (\bibinfo {year} {2009})},\ \Eprint {http://arxiv.org/abs/0810.1585} {arXiv:0810.1585 [hep-ph]} \BibitemShut {NoStop}%
\bibitem [{\citenamefont {Schwinger}(1960)}]{JS1960}%
  \BibitemOpen
  \bibfield  {author} {\bibinfo {author} {\bibfnamefont {J.}~\bibnamefont {Schwinger}},\ }\bibfield  {title} {\enquote {\bibinfo {title} {The special canonical group},}\ }\href {\doibase 10.1073/pnas.46.10.1401} {\bibfield  {journal} {\bibinfo  {journal} {Proceedings of the National Academy of Sciences}\ }\textbf {\bibinfo {volume} {46}},\ \bibinfo {pages} {1401--1415} (\bibinfo {year} {1960})},\ \Eprint {http://arxiv.org/abs/https://www.pnas.org/doi/pdf/10.1073/pnas.46.10.1401} {https://www.pnas.org/doi/pdf/10.1073/pnas.46.10.1401} \BibitemShut {NoStop}%
\bibitem [{\citenamefont {Bakshi}\ and\ \citenamefont {Mahanthappa}(1963{\natexlab{a}})}]{Bakshi:1962dv}%
  \BibitemOpen
  \bibfield  {author} {\bibinfo {author} {\bibfnamefont {P.~M.}\ \bibnamefont {Bakshi}}\ and\ \bibinfo {author} {\bibfnamefont {K.~T.}\ \bibnamefont {Mahanthappa}},\ }\bibfield  {title} {\enquote {\bibinfo {title} {{Expectation value formalism in quantum field theory. 1.}}}\ }\href {\doibase 10.1063/1.1703883} {\bibfield  {journal} {\bibinfo  {journal} {J. Math. Phys.}\ }\textbf {\bibinfo {volume} {4}},\ \bibinfo {pages} {1--11} (\bibinfo {year} {1963}{\natexlab{a}})}\BibitemShut {NoStop}%
\bibitem [{\citenamefont {Bakshi}\ and\ \citenamefont {Mahanthappa}(1963{\natexlab{b}})}]{Bakshi:1963bn}%
  \BibitemOpen
  \bibfield  {author} {\bibinfo {author} {\bibfnamefont {P.~M.}\ \bibnamefont {Bakshi}}\ and\ \bibinfo {author} {\bibfnamefont {K.~T.}\ \bibnamefont {Mahanthappa}},\ }\bibfield  {title} {\enquote {\bibinfo {title} {{Expectation value formalism in quantum field theory. 2.}}}\ }\href {\doibase 10.1063/1.1703879} {\bibfield  {journal} {\bibinfo  {journal} {J. Math. Phys.}\ }\textbf {\bibinfo {volume} {4}},\ \bibinfo {pages} {12--16} (\bibinfo {year} {1963}{\natexlab{b}})}\BibitemShut {NoStop}%
\bibitem [{\citenamefont {Keldysh}(1964)}]{Keldysh:1964ud}%
  \BibitemOpen
  \bibfield  {author} {\bibinfo {author} {\bibfnamefont {L.~V.}\ \bibnamefont {Keldysh}},\ }\bibfield  {title} {\enquote {\bibinfo {title} {{Diagram technique for nonequilibrium processes}},}\ }\href@noop {} {\bibfield  {journal} {\bibinfo  {journal} {Zh. Eksp. Teor. Fiz.}\ }\textbf {\bibinfo {volume} {47}},\ \bibinfo {pages} {1515--1527} (\bibinfo {year} {1964})}\BibitemShut {NoStop}%
\bibitem [{\citenamefont {Dimopoulos}\ \emph {et~al.}(2003)\citenamefont {Dimopoulos}, \citenamefont {Lyth}, \citenamefont {Notari},\ and\ \citenamefont {Riotto}}]{Dimopoulos:2003az}%
  \BibitemOpen
  \bibfield  {author} {\bibinfo {author} {\bibfnamefont {K.}~\bibnamefont {Dimopoulos}}, \bibinfo {author} {\bibfnamefont {D.~H.}\ \bibnamefont {Lyth}}, \bibinfo {author} {\bibfnamefont {A.}~\bibnamefont {Notari}}, \ and\ \bibinfo {author} {\bibfnamefont {A.}~\bibnamefont {Riotto}},\ }\bibfield  {title} {\enquote {\bibinfo {title} {{The Curvaton as a pseudoNambu-Goldstone boson}},}\ }\href {\doibase 10.1088/1126-6708/2003/07/053} {\bibfield  {journal} {\bibinfo  {journal} {JHEP}\ }\textbf {\bibinfo {volume} {07}},\ \bibinfo {pages} {053} (\bibinfo {year} {2003})},\ \Eprint {http://arxiv.org/abs/hep-ph/0304050} {arXiv:hep-ph/0304050} \BibitemShut {NoStop}%
\bibitem [{\citenamefont {Preskill}\ \emph {et~al.}(1983)\citenamefont {Preskill}, \citenamefont {Wise},\ and\ \citenamefont {Wilczek}}]{Preskill:1982cy}%
  \BibitemOpen
  \bibfield  {author} {\bibinfo {author} {\bibfnamefont {J.}~\bibnamefont {Preskill}}, \bibinfo {author} {\bibfnamefont {M.~B.}\ \bibnamefont {Wise}}, \ and\ \bibinfo {author} {\bibfnamefont {F.}~\bibnamefont {Wilczek}},\ }\bibfield  {title} {\enquote {\bibinfo {title} {{Cosmology of the Invisible Axion}},}\ }\href {\doibase 10.1016/0370-2693(83)90637-8} {\bibfield  {journal} {\bibinfo  {journal} {Phys. Lett. B}\ }\textbf {\bibinfo {volume} {120}},\ \bibinfo {pages} {127--132} (\bibinfo {year} {1983})}\BibitemShut {NoStop}%
\bibitem [{\citenamefont {Dine}\ and\ \citenamefont {Fischler}(1983)}]{Dine:1982ah}%
  \BibitemOpen
  \bibfield  {author} {\bibinfo {author} {\bibfnamefont {M.}~\bibnamefont {Dine}}\ and\ \bibinfo {author} {\bibfnamefont {W.}~\bibnamefont {Fischler}},\ }\bibfield  {title} {\enquote {\bibinfo {title} {{The Not So Harmless Axion}},}\ }\href {\doibase 10.1016/0370-2693(83)90639-1} {\bibfield  {journal} {\bibinfo  {journal} {Phys. Lett. B}\ }\textbf {\bibinfo {volume} {120}},\ \bibinfo {pages} {137--141} (\bibinfo {year} {1983})}\BibitemShut {NoStop}%
\bibitem [{\citenamefont {Abbott}\ and\ \citenamefont {Sikivie}(1983)}]{Abbott:1982af}%
  \BibitemOpen
  \bibfield  {author} {\bibinfo {author} {\bibfnamefont {L.~F.}\ \bibnamefont {Abbott}}\ and\ \bibinfo {author} {\bibfnamefont {P.}~\bibnamefont {Sikivie}},\ }\bibfield  {title} {\enquote {\bibinfo {title} {{A Cosmological Bound on the Invisible Axion}},}\ }\href {\doibase 10.1016/0370-2693(83)90638-X} {\bibfield  {journal} {\bibinfo  {journal} {Phys. Lett. B}\ }\textbf {\bibinfo {volume} {120}},\ \bibinfo {pages} {133--136} (\bibinfo {year} {1983})}\BibitemShut {NoStop}%
\bibitem [{\citenamefont {Co}\ \emph {et~al.}(2020)\citenamefont {Co}, \citenamefont {Hall},\ and\ \citenamefont {Harigaya}}]{Co:2019jts}%
  \BibitemOpen
  \bibfield  {author} {\bibinfo {author} {\bibfnamefont {R.~T.}\ \bibnamefont {Co}}, \bibinfo {author} {\bibfnamefont {L.~J.}\ \bibnamefont {Hall}}, \ and\ \bibinfo {author} {\bibfnamefont {K.}~\bibnamefont {Harigaya}},\ }\bibfield  {title} {\enquote {\bibinfo {title} {{Axion Kinetic Misalignment Mechanism}},}\ }\href {\doibase 10.1103/PhysRevLett.124.251802} {\bibfield  {journal} {\bibinfo  {journal} {Phys. Rev. Lett.}\ }\textbf {\bibinfo {volume} {124}},\ \bibinfo {pages} {251802} (\bibinfo {year} {2020})},\ \Eprint {http://arxiv.org/abs/1910.14152} {arXiv:1910.14152 [hep-ph]} \BibitemShut {NoStop}%
\bibitem [{\citenamefont {Lyth}\ and\ \citenamefont {Wands}(2002)}]{Lyth:2001nq}%
  \BibitemOpen
  \bibfield  {author} {\bibinfo {author} {\bibfnamefont {D.~H.}\ \bibnamefont {Lyth}}\ and\ \bibinfo {author} {\bibfnamefont {D.}~\bibnamefont {Wands}},\ }\bibfield  {title} {\enquote {\bibinfo {title} {{Generating the curvature perturbation without an inflaton}},}\ }\href {\doibase 10.1016/S0370-2693(01)01366-1} {\bibfield  {journal} {\bibinfo  {journal} {Phys. Lett. B}\ }\textbf {\bibinfo {volume} {524}},\ \bibinfo {pages} {5--14} (\bibinfo {year} {2002})},\ \Eprint {http://arxiv.org/abs/hep-ph/0110002} {arXiv:hep-ph/0110002} \BibitemShut {NoStop}%
\bibitem [{\citenamefont {Bartolo}\ \emph {et~al.}(2004{\natexlab{b}})\citenamefont {Bartolo}, \citenamefont {Matarrese},\ and\ \citenamefont {Riotto}}]{Bartolo:2003jx}%
  \BibitemOpen
  \bibfield  {author} {\bibinfo {author} {\bibfnamefont {N.}~\bibnamefont {Bartolo}}, \bibinfo {author} {\bibfnamefont {S.}~\bibnamefont {Matarrese}}, \ and\ \bibinfo {author} {\bibfnamefont {A.}~\bibnamefont {Riotto}},\ }\bibfield  {title} {\enquote {\bibinfo {title} {{On nonGaussianity in the curvaton scenario}},}\ }\href {\doibase 10.1103/PhysRevD.69.043503} {\bibfield  {journal} {\bibinfo  {journal} {Phys. Rev. D}\ }\textbf {\bibinfo {volume} {69}},\ \bibinfo {pages} {043503} (\bibinfo {year} {2004}{\natexlab{b}})},\ \Eprint {http://arxiv.org/abs/hep-ph/0309033} {arXiv:hep-ph/0309033} \BibitemShut {NoStop}%
\bibitem [{\citenamefont {Sasaki}\ \emph {et~al.}(2006)\citenamefont {Sasaki}, \citenamefont {Valiviita},\ and\ \citenamefont {Wands}}]{Sasaki:2006kq}%
  \BibitemOpen
  \bibfield  {author} {\bibinfo {author} {\bibfnamefont {M.}~\bibnamefont {Sasaki}}, \bibinfo {author} {\bibfnamefont {J.}~\bibnamefont {Valiviita}}, \ and\ \bibinfo {author} {\bibfnamefont {D.}~\bibnamefont {Wands}},\ }\bibfield  {title} {\enquote {\bibinfo {title} {{Non-Gaussianity of the primordial perturbation in the curvaton model}},}\ }\href {\doibase 10.1103/PhysRevD.74.103003} {\bibfield  {journal} {\bibinfo  {journal} {Phys. Rev. D}\ }\textbf {\bibinfo {volume} {74}},\ \bibinfo {pages} {103003} (\bibinfo {year} {2006})},\ \Eprint {http://arxiv.org/abs/astro-ph/0607627} {arXiv:astro-ph/0607627} \BibitemShut {NoStop}%
\bibitem [{\citenamefont {Chun}\ \emph {et~al.}(2004)\citenamefont {Chun}, \citenamefont {Dimopoulos},\ and\ \citenamefont {Lyth}}]{Chun:2004gx}%
  \BibitemOpen
  \bibfield  {author} {\bibinfo {author} {\bibfnamefont {E.~J.}\ \bibnamefont {Chun}}, \bibinfo {author} {\bibfnamefont {K.}~\bibnamefont {Dimopoulos}}, \ and\ \bibinfo {author} {\bibfnamefont {D.}~\bibnamefont {Lyth}},\ }\bibfield  {title} {\enquote {\bibinfo {title} {{Curvaton and QCD axion in supersymmetric theories}},}\ }\href {\doibase 10.1103/PhysRevD.70.103510} {\bibfield  {journal} {\bibinfo  {journal} {Phys. Rev. D}\ }\textbf {\bibinfo {volume} {70}},\ \bibinfo {pages} {103510} (\bibinfo {year} {2004})},\ \Eprint {http://arxiv.org/abs/hep-ph/0402059} {arXiv:hep-ph/0402059} \BibitemShut {NoStop}%
\bibitem [{\citenamefont {Dimopoulos}\ and\ \citenamefont {Lazarides}(2006)}]{Dimopoulos:2005bp}%
  \BibitemOpen
  \bibfield  {author} {\bibinfo {author} {\bibfnamefont {K.}~\bibnamefont {Dimopoulos}}\ and\ \bibinfo {author} {\bibfnamefont {G.}~\bibnamefont {Lazarides}},\ }\bibfield  {title} {\enquote {\bibinfo {title} {{Modular inflation and the orthogonal axion as curvaton}},}\ }\href {\doibase 10.1103/PhysRevD.73.023525} {\bibfield  {journal} {\bibinfo  {journal} {Phys. Rev. D}\ }\textbf {\bibinfo {volume} {73}},\ \bibinfo {pages} {023525} (\bibinfo {year} {2006})},\ \Eprint {http://arxiv.org/abs/hep-ph/0511310} {arXiv:hep-ph/0511310} \BibitemShut {NoStop}%
\bibitem [{\citenamefont {Lyth}\ and\ \citenamefont {Rodriguez}(2005)}]{Lyth:2005fi}%
  \BibitemOpen
  \bibfield  {author} {\bibinfo {author} {\bibfnamefont {D.~H.}\ \bibnamefont {Lyth}}\ and\ \bibinfo {author} {\bibfnamefont {Y.}~\bibnamefont {Rodriguez}},\ }\bibfield  {title} {\enquote {\bibinfo {title} {{The Inflationary prediction for primordial non-Gaussianity}},}\ }\href {\doibase 10.1103/PhysRevLett.95.121302} {\bibfield  {journal} {\bibinfo  {journal} {Phys. Rev. Lett.}\ }\textbf {\bibinfo {volume} {95}},\ \bibinfo {pages} {121302} (\bibinfo {year} {2005})},\ \Eprint {http://arxiv.org/abs/astro-ph/0504045} {arXiv:astro-ph/0504045} \BibitemShut {NoStop}%
\bibitem [{\citenamefont {Chingangbam}\ and\ \citenamefont {Huang}(2009)}]{Chingangbam:2009xi}%
  \BibitemOpen
  \bibfield  {author} {\bibinfo {author} {\bibfnamefont {P.}~\bibnamefont {Chingangbam}}\ and\ \bibinfo {author} {\bibfnamefont {Q.-G.}\ \bibnamefont {Huang}},\ }\bibfield  {title} {\enquote {\bibinfo {title} {{The Curvature Perturbation in the Axion-type Curvaton Model}},}\ }\href {\doibase 10.1088/1475-7516/2009/04/031} {\bibfield  {journal} {\bibinfo  {journal} {JCAP}\ }\textbf {\bibinfo {volume} {04}},\ \bibinfo {pages} {031} (\bibinfo {year} {2009})},\ \Eprint {http://arxiv.org/abs/0902.2619} {arXiv:0902.2619 [astro-ph.CO]} \BibitemShut {NoStop}%
\bibitem [{\citenamefont {Lyth}(1997)}]{Lyth:1996im}%
  \BibitemOpen
  \bibfield  {author} {\bibinfo {author} {\bibfnamefont {D.~H.}\ \bibnamefont {Lyth}},\ }\bibfield  {title} {\enquote {\bibinfo {title} {{What would we learn by detecting a gravitational wave signal in the cosmic microwave background anisotropy?}}}\ }\href {\doibase 10.1103/PhysRevLett.78.1861} {\bibfield  {journal} {\bibinfo  {journal} {Phys. Rev. Lett.}\ }\textbf {\bibinfo {volume} {78}},\ \bibinfo {pages} {1861--1863} (\bibinfo {year} {1997})},\ \Eprint {http://arxiv.org/abs/hep-ph/9606387} {arXiv:hep-ph/9606387} \BibitemShut {NoStop}%
\bibitem [{\citenamefont {Creminelli}\ and\ \citenamefont {Zaldarriaga}(2004)}]{Creminelli:2004yq}%
  \BibitemOpen
  \bibfield  {author} {\bibinfo {author} {\bibfnamefont {P.}~\bibnamefont {Creminelli}}\ and\ \bibinfo {author} {\bibfnamefont {M.}~\bibnamefont {Zaldarriaga}},\ }\bibfield  {title} {\enquote {\bibinfo {title} {{Single field consistency relation for the 3-point function}},}\ }\href {\doibase 10.1088/1475-7516/2004/10/006} {\bibfield  {journal} {\bibinfo  {journal} {JCAP}\ }\textbf {\bibinfo {volume} {10}},\ \bibinfo {pages} {006} (\bibinfo {year} {2004})},\ \Eprint {http://arxiv.org/abs/astro-ph/0407059} {arXiv:astro-ph/0407059} \BibitemShut {NoStop}%
\bibitem [{\citenamefont {Lyth}(2004)}]{Lyth:2003dt}%
  \BibitemOpen
  \bibfield  {author} {\bibinfo {author} {\bibfnamefont {D.~H.}\ \bibnamefont {Lyth}},\ }\bibfield  {title} {\enquote {\bibinfo {title} {{Can the curvaton paradigm accommodate a low inflation scale?}}}\ }\href {\doibase 10.1016/j.physletb.2003.11.019} {\bibfield  {journal} {\bibinfo  {journal} {Phys. Lett. B}\ }\textbf {\bibinfo {volume} {579}},\ \bibinfo {pages} {239--244} (\bibinfo {year} {2004})},\ \Eprint {http://arxiv.org/abs/hep-th/0308110} {arXiv:hep-th/0308110} \BibitemShut {NoStop}%
\bibitem [{\citenamefont {Huang}\ and\ \citenamefont {Wang}(2008)}]{Huang:2008bg}%
  \BibitemOpen
  \bibfield  {author} {\bibinfo {author} {\bibfnamefont {Q.-G.}\ \bibnamefont {Huang}}\ and\ \bibinfo {author} {\bibfnamefont {Y.}~\bibnamefont {Wang}},\ }\bibfield  {title} {\enquote {\bibinfo {title} {{Curvaton Dynamics and the Non-Linearity Parameters in Curvaton Model}},}\ }\href {\doibase 10.1088/1475-7516/2008/09/025} {\bibfield  {journal} {\bibinfo  {journal} {JCAP}\ }\textbf {\bibinfo {volume} {09}},\ \bibinfo {pages} {025} (\bibinfo {year} {2008})},\ \Eprint {http://arxiv.org/abs/0808.1168} {arXiv:0808.1168 [hep-th]} \BibitemShut {NoStop}%
\bibitem [{\citenamefont {Chung}\ and\ \citenamefont {Tadepalli}(2024{\natexlab{a}})}]{Chung:2024ctx}%
  \BibitemOpen
  \bibfield  {author} {\bibinfo {author} {\bibfnamefont {D.~J.~H.}\ \bibnamefont {Chung}}\ and\ \bibinfo {author} {\bibfnamefont {S.~C.}\ \bibnamefont {Tadepalli}},\ }\bibfield  {title} {\enquote {\bibinfo {title} {{Large Blue Spectral Index From a Conformal Limit of a Rotating Complex Scalar}},}\ }\href@noop {} {\  (\bibinfo {year} {2024}{\natexlab{a}})},\ \Eprint {http://arxiv.org/abs/2406.12976} {arXiv:2406.12976 [astro-ph.CO]} \BibitemShut {NoStop}%
\bibitem [{\citenamefont {Takahashi}(2014)}]{Takahashi:2014bxa}%
  \BibitemOpen
  \bibfield  {author} {\bibinfo {author} {\bibfnamefont {T.}~\bibnamefont {Takahashi}},\ }\bibfield  {title} {\enquote {\bibinfo {title} {{Primordial non-Gaussianity and the inflationary Universe}},}\ }\href {\doibase 10.1093/ptep/ptu060} {\bibfield  {journal} {\bibinfo  {journal} {PTEP}\ }\textbf {\bibinfo {volume} {2014}},\ \bibinfo {pages} {06B105} (\bibinfo {year} {2014})}\BibitemShut {NoStop}%
\bibitem [{\citenamefont {Chen}\ \emph {et~al.}(2022)\citenamefont {Chen}, \citenamefont {Ebadi},\ and\ \citenamefont {Kumar}}]{Chen:2022vzh}%
  \BibitemOpen
  \bibfield  {author} {\bibinfo {author} {\bibfnamefont {X.}~\bibnamefont {Chen}}, \bibinfo {author} {\bibfnamefont {R.}~\bibnamefont {Ebadi}}, \ and\ \bibinfo {author} {\bibfnamefont {S.}~\bibnamefont {Kumar}},\ }\bibfield  {title} {\enquote {\bibinfo {title} {{Classical cosmological collider physics and primordial features}},}\ }\href {\doibase 10.1088/1475-7516/2022/08/083} {\bibfield  {journal} {\bibinfo  {journal} {JCAP}\ }\textbf {\bibinfo {volume} {08}},\ \bibinfo {pages} {083} (\bibinfo {year} {2022})},\ \Eprint {http://arxiv.org/abs/2205.01107} {arXiv:2205.01107 [hep-ph]} \BibitemShut {NoStop}%
\bibitem [{\citenamefont {Arkani-Hamed}\ and\ \citenamefont {Maldacena}(2015)}]{Arkani-Hamed:2015bza}%
  \BibitemOpen
  \bibfield  {author} {\bibinfo {author} {\bibfnamefont {N.}~\bibnamefont {Arkani-Hamed}}\ and\ \bibinfo {author} {\bibfnamefont {J.}~\bibnamefont {Maldacena}},\ }\bibfield  {title} {\enquote {\bibinfo {title} {{Cosmological Collider Physics}},}\ }\href@noop {} {\  (\bibinfo {year} {2015})},\ \Eprint {http://arxiv.org/abs/1503.08043} {arXiv:1503.08043 [hep-th]} \BibitemShut {NoStop}%
\bibitem [{\citenamefont {Chen}(2010)}]{Chen:2010xka}%
  \BibitemOpen
  \bibfield  {author} {\bibinfo {author} {\bibfnamefont {X.}~\bibnamefont {Chen}},\ }\bibfield  {title} {\enquote {\bibinfo {title} {{Primordial Non-Gaussianities from Inflation Models}},}\ }\href {\doibase 10.1155/2010/638979} {\bibfield  {journal} {\bibinfo  {journal} {Adv. Astron.}\ }\textbf {\bibinfo {volume} {2010}},\ \bibinfo {pages} {638979} (\bibinfo {year} {2010})},\ \Eprint {http://arxiv.org/abs/1002.1416} {arXiv:1002.1416 [astro-ph.CO]} \BibitemShut {NoStop}%
\bibitem [{\citenamefont {Chluba}\ \emph {et~al.}(2015)\citenamefont {Chluba}, \citenamefont {Hamann},\ and\ \citenamefont {Patil}}]{Chluba:2015bqa}%
  \BibitemOpen
  \bibfield  {author} {\bibinfo {author} {\bibfnamefont {J.}~\bibnamefont {Chluba}}, \bibinfo {author} {\bibfnamefont {J.}~\bibnamefont {Hamann}}, \ and\ \bibinfo {author} {\bibfnamefont {S.~P.}\ \bibnamefont {Patil}},\ }\bibfield  {title} {\enquote {\bibinfo {title} {{Features and New Physical Scales in Primordial Observables: Theory and Observation}},}\ }\href {\doibase 10.1142/S0218271815300232} {\bibfield  {journal} {\bibinfo  {journal} {Int. J. Mod. Phys. D}\ }\textbf {\bibinfo {volume} {24}},\ \bibinfo {pages} {1530023} (\bibinfo {year} {2015})},\ \Eprint {http://arxiv.org/abs/1505.01834} {arXiv:1505.01834 [astro-ph.CO]} \BibitemShut {NoStop}%
\bibitem [{\citenamefont {Chen}(2012)}]{Chen:2011zf}%
  \BibitemOpen
  \bibfield  {author} {\bibinfo {author} {\bibfnamefont {X.}~\bibnamefont {Chen}},\ }\bibfield  {title} {\enquote {\bibinfo {title} {{Primordial Features as Evidence for Inflation}},}\ }\href {\doibase 10.1088/1475-7516/2012/01/038} {\bibfield  {journal} {\bibinfo  {journal} {JCAP}\ }\textbf {\bibinfo {volume} {01}},\ \bibinfo {pages} {038} (\bibinfo {year} {2012})},\ \Eprint {http://arxiv.org/abs/1104.1323} {arXiv:1104.1323 [hep-th]} \BibitemShut {NoStop}%
\bibitem [{\citenamefont {Slosar}\ \emph {et~al.}(2019)\citenamefont {Slosar} \emph {et~al.}}]{Slosar:2019gvt}%
  \BibitemOpen
  \bibfield  {author} {\bibinfo {author} {\bibfnamefont {A.}~\bibnamefont {Slosar}} \emph {et~al.},\ }\bibfield  {title} {\enquote {\bibinfo {title} {{Scratches from the Past: Inflationary Archaeology through Features in the Power Spectrum of Primordial Fluctuations}},}\ }\href@noop {} {\bibfield  {journal} {\bibinfo  {journal} {Bull. Am. Astron. Soc.}\ }\textbf {\bibinfo {volume} {51}},\ \bibinfo {pages} {98} (\bibinfo {year} {2019})},\ \Eprint {http://arxiv.org/abs/1903.09883} {arXiv:1903.09883 [astro-ph.CO]} \BibitemShut {NoStop}%
\bibitem [{\citenamefont {Chung}\ and\ \citenamefont {Tadepalli}(2022)}]{Chung:2021lfg}%
  \BibitemOpen
  \bibfield  {author} {\bibinfo {author} {\bibfnamefont {D.~J.~H.}\ \bibnamefont {Chung}}\ and\ \bibinfo {author} {\bibfnamefont {S.~C.}\ \bibnamefont {Tadepalli}},\ }\bibfield  {title} {\enquote {\bibinfo {title} {{Analytic treatment of underdamped axionic blue isocurvature perturbations}},}\ }\href {\doibase 10.1103/PhysRevD.105.123511} {\bibfield  {journal} {\bibinfo  {journal} {Phys. Rev. D}\ }\textbf {\bibinfo {volume} {105}},\ \bibinfo {pages} {123511} (\bibinfo {year} {2022})},\ \Eprint {http://arxiv.org/abs/2110.02272} {arXiv:2110.02272 [astro-ph.CO]} \BibitemShut {NoStop}%
\bibitem [{\citenamefont {Chung}\ and\ \citenamefont {Tadepalli}(2024{\natexlab{b}})}]{Chung:2023xcv}%
  \BibitemOpen
  \bibfield  {author} {\bibinfo {author} {\bibfnamefont {D.~J.~H.}\ \bibnamefont {Chung}}\ and\ \bibinfo {author} {\bibfnamefont {S.~C.}\ \bibnamefont {Tadepalli}},\ }\bibfield  {title} {\enquote {\bibinfo {title} {{Power spectrum in the chaotic regime of axionic blue isocurvature perturbations}},}\ }\href {\doibase 10.1103/PhysRevD.109.023539} {\bibfield  {journal} {\bibinfo  {journal} {Phys. Rev. D}\ }\textbf {\bibinfo {volume} {109}},\ \bibinfo {pages} {023539} (\bibinfo {year} {2024}{\natexlab{b}})},\ \Eprint {http://arxiv.org/abs/2309.17010} {arXiv:2309.17010 [astro-ph.CO]} \BibitemShut {NoStop}%
\bibitem [{\citenamefont {Chen}\ \emph {et~al.}(2008)\citenamefont {Chen}, \citenamefont {Easther},\ and\ \citenamefont {Lim}}]{Chen:2008wn}%
  \BibitemOpen
  \bibfield  {author} {\bibinfo {author} {\bibfnamefont {X.}~\bibnamefont {Chen}}, \bibinfo {author} {\bibfnamefont {R.}~\bibnamefont {Easther}}, \ and\ \bibinfo {author} {\bibfnamefont {E.~A.}\ \bibnamefont {Lim}},\ }\bibfield  {title} {\enquote {\bibinfo {title} {{Generation and Characterization of Large Non-Gaussianities in Single Field Inflation}},}\ }\href {\doibase 10.1088/1475-7516/2008/04/010} {\bibfield  {journal} {\bibinfo  {journal} {JCAP}\ }\textbf {\bibinfo {volume} {04}},\ \bibinfo {pages} {010} (\bibinfo {year} {2008})},\ \Eprint {http://arxiv.org/abs/0801.3295} {arXiv:0801.3295 [astro-ph]} \BibitemShut {NoStop}%
\bibitem [{\citenamefont {Braglia}\ \emph {et~al.}(2022)\citenamefont {Braglia}, \citenamefont {Chen},\ and\ \citenamefont {Hazra}}]{Braglia:2021rej}%
  \BibitemOpen
  \bibfield  {author} {\bibinfo {author} {\bibfnamefont {M.}~\bibnamefont {Braglia}}, \bibinfo {author} {\bibfnamefont {X.}~\bibnamefont {Chen}}, \ and\ \bibinfo {author} {\bibfnamefont {D.~K.}\ \bibnamefont {Hazra}},\ }\bibfield  {title} {\enquote {\bibinfo {title} {{Primordial standard clock models and CMB residual anomalies}},}\ }\href {\doibase 10.1103/PhysRevD.105.103523} {\bibfield  {journal} {\bibinfo  {journal} {Phys. Rev. D}\ }\textbf {\bibinfo {volume} {105}},\ \bibinfo {pages} {103523} (\bibinfo {year} {2022})},\ \Eprint {http://arxiv.org/abs/2108.10110} {arXiv:2108.10110 [astro-ph.CO]} \BibitemShut {NoStop}%
\bibitem [{\citenamefont {Hamann}\ and\ \citenamefont {Wons}(2022)}]{Hamann:2021eyw}%
  \BibitemOpen
  \bibfield  {author} {\bibinfo {author} {\bibfnamefont {J.}~\bibnamefont {Hamann}}\ and\ \bibinfo {author} {\bibfnamefont {J.}~\bibnamefont {Wons}},\ }\bibfield  {title} {\enquote {\bibinfo {title} {{Optimising inflationary features the Bayesian way}},}\ }\href {\doibase 10.1088/1475-7516/2022/03/036} {\bibfield  {journal} {\bibinfo  {journal} {JCAP}\ }\textbf {\bibinfo {volume} {03}},\ \bibinfo {pages} {036} (\bibinfo {year} {2022})},\ \Eprint {http://arxiv.org/abs/2112.08571} {arXiv:2112.08571 [astro-ph.CO]} \BibitemShut {NoStop}%
\bibitem [{\citenamefont {Laureijs}\ \emph {et~al.}(2011)\citenamefont {Laureijs} \emph {et~al.}}]{EUCLID:2011zbd}%
  \BibitemOpen
  \bibfield  {author} {\bibinfo {author} {\bibfnamefont {R.}~\bibnamefont {Laureijs}} \emph {et~al.} (\bibinfo {collaboration} {EUCLID}),\ }\bibfield  {title} {\enquote {\bibinfo {title} {{Euclid Definition Study Report}},}\ }\href@noop {} {\  (\bibinfo {year} {2011})},\ \Eprint {http://arxiv.org/abs/1110.3193} {arXiv:1110.3193 [astro-ph.CO]} \BibitemShut {NoStop}%
\bibitem [{\citenamefont {Ballardini}\ \emph {et~al.}(2024)\citenamefont {Ballardini} \emph {et~al.}}]{Euclid:2023shr}%
  \BibitemOpen
  \bibfield  {author} {\bibinfo {author} {\bibfnamefont {M.}~\bibnamefont {Ballardini}} \emph {et~al.} (\bibinfo {collaboration} {Euclid}),\ }\bibfield  {title} {\enquote {\bibinfo {title} {{Euclid: The search for primordial features}},}\ }\href {\doibase 10.1051/0004-6361/202348162} {\bibfield  {journal} {\bibinfo  {journal} {Astron. Astrophys.}\ }\textbf {\bibinfo {volume} {683}},\ \bibinfo {pages} {A220} (\bibinfo {year} {2024})},\ \Eprint {http://arxiv.org/abs/2309.17287} {arXiv:2309.17287 [astro-ph.CO]} \BibitemShut {NoStop}%
\bibitem [{\citenamefont {Chen}\ and\ \citenamefont {Wang}(2010)}]{Chen:2009zp}%
  \BibitemOpen
  \bibfield  {author} {\bibinfo {author} {\bibfnamefont {X.}~\bibnamefont {Chen}}\ and\ \bibinfo {author} {\bibfnamefont {Y.}~\bibnamefont {Wang}},\ }\bibfield  {title} {\enquote {\bibinfo {title} {{Quasi-Single Field Inflation and Non-Gaussianities}},}\ }\href {\doibase 10.1088/1475-7516/2010/04/027} {\bibfield  {journal} {\bibinfo  {journal} {JCAP}\ }\textbf {\bibinfo {volume} {04}},\ \bibinfo {pages} {027} (\bibinfo {year} {2010})},\ \Eprint {http://arxiv.org/abs/0911.3380} {arXiv:0911.3380 [hep-th]} \BibitemShut {NoStop}%
\bibitem [{\citenamefont {Binetruy}\ and\ \citenamefont {Dvali}(1996)}]{Binetruy:1996xj}%
  \BibitemOpen
  \bibfield  {author} {\bibinfo {author} {\bibfnamefont {P.}~\bibnamefont {Binetruy}}\ and\ \bibinfo {author} {\bibfnamefont {G.~R.}\ \bibnamefont {Dvali}},\ }\bibfield  {title} {\enquote {\bibinfo {title} {{D term inflation}},}\ }\href {\doibase 10.1016/S0370-2693(96)01083-0} {\bibfield  {journal} {\bibinfo  {journal} {Phys. Lett. B}\ }\textbf {\bibinfo {volume} {388}},\ \bibinfo {pages} {241--246} (\bibinfo {year} {1996})},\ \Eprint {http://arxiv.org/abs/hep-ph/9606342} {arXiv:hep-ph/9606342} \BibitemShut {NoStop}%
\bibitem [{\citenamefont {Halyo}(1996)}]{Halyo:1996pp}%
  \BibitemOpen
  \bibfield  {author} {\bibinfo {author} {\bibfnamefont {E.}~\bibnamefont {Halyo}},\ }\bibfield  {title} {\enquote {\bibinfo {title} {{Hybrid inflation from supergravity D terms}},}\ }\href {\doibase 10.1016/0370-2693(96)01001-5} {\bibfield  {journal} {\bibinfo  {journal} {Phys. Lett. B}\ }\textbf {\bibinfo {volume} {387}},\ \bibinfo {pages} {43--47} (\bibinfo {year} {1996})},\ \Eprint {http://arxiv.org/abs/hep-ph/9606423} {arXiv:hep-ph/9606423} \BibitemShut {NoStop}%
\bibitem [{\citenamefont {Gaillard}\ \emph {et~al.}(1995)\citenamefont {Gaillard}, \citenamefont {Murayama},\ and\ \citenamefont {Olive}}]{Gaillard:1995az}%
  \BibitemOpen
  \bibfield  {author} {\bibinfo {author} {\bibfnamefont {M.~K.}\ \bibnamefont {Gaillard}}, \bibinfo {author} {\bibfnamefont {H.}~\bibnamefont {Murayama}}, \ and\ \bibinfo {author} {\bibfnamefont {K.~A.}\ \bibnamefont {Olive}},\ }\bibfield  {title} {\enquote {\bibinfo {title} {{Preserving flat directions during inflation}},}\ }\href {\doibase 10.1016/0370-2693(95)00773-E} {\bibfield  {journal} {\bibinfo  {journal} {Phys. Lett. B}\ }\textbf {\bibinfo {volume} {355}},\ \bibinfo {pages} {71--77} (\bibinfo {year} {1995})},\ \Eprint {http://arxiv.org/abs/hep-ph/9504307} {arXiv:hep-ph/9504307} \BibitemShut {NoStop}%
\bibitem [{\citenamefont {Campbell}\ \emph {et~al.}(1999)\citenamefont {Campbell}, \citenamefont {Gaillard}, \citenamefont {Murayama},\ and\ \citenamefont {Olive}}]{Campbell:1998yi}%
  \BibitemOpen
  \bibfield  {author} {\bibinfo {author} {\bibfnamefont {B.~A.}\ \bibnamefont {Campbell}}, \bibinfo {author} {\bibfnamefont {M.~K.}\ \bibnamefont {Gaillard}}, \bibinfo {author} {\bibfnamefont {H.}~\bibnamefont {Murayama}}, \ and\ \bibinfo {author} {\bibfnamefont {K.~A.}\ \bibnamefont {Olive}},\ }\bibfield  {title} {\enquote {\bibinfo {title} {{Regulating the baryon asymmetry in no scale Affleck-Dine baryogenesis}},}\ }\href {\doibase 10.1016/S0550-3213(98)00574-4} {\bibfield  {journal} {\bibinfo  {journal} {Nucl. Phys. B}\ }\textbf {\bibinfo {volume} {538}},\ \bibinfo {pages} {351--367} (\bibinfo {year} {1999})},\ \Eprint {http://arxiv.org/abs/hep-ph/9805300} {arXiv:hep-ph/9805300} \BibitemShut {NoStop}%
\bibitem [{\citenamefont {Enqvist}\ \emph {et~al.}(2003{\natexlab{a}})\citenamefont {Enqvist}, \citenamefont {Kasuya},\ and\ \citenamefont {Mazumdar}}]{Enqvist:2002rf}%
  \BibitemOpen
  \bibfield  {author} {\bibinfo {author} {\bibfnamefont {K.}~\bibnamefont {Enqvist}}, \bibinfo {author} {\bibfnamefont {S.}~\bibnamefont {Kasuya}}, \ and\ \bibinfo {author} {\bibfnamefont {A.}~\bibnamefont {Mazumdar}},\ }\bibfield  {title} {\enquote {\bibinfo {title} {{Adiabatic density perturbations and matter generation from the MSSM}},}\ }\href {\doibase 10.1103/PhysRevLett.90.091302} {\bibfield  {journal} {\bibinfo  {journal} {Phys. Rev. Lett.}\ }\textbf {\bibinfo {volume} {90}},\ \bibinfo {pages} {091302} (\bibinfo {year} {2003}{\natexlab{a}})},\ \Eprint {http://arxiv.org/abs/hep-ph/0211147} {arXiv:hep-ph/0211147} \BibitemShut {NoStop}%
\bibitem [{\citenamefont {Enqvist}\ \emph {et~al.}(2003{\natexlab{b}})\citenamefont {Enqvist}, \citenamefont {Jokinen}, \citenamefont {Kasuya},\ and\ \citenamefont {Mazumdar}}]{Enqvist:2003mr}%
  \BibitemOpen
  \bibfield  {author} {\bibinfo {author} {\bibfnamefont {K.}~\bibnamefont {Enqvist}}, \bibinfo {author} {\bibfnamefont {A.}~\bibnamefont {Jokinen}}, \bibinfo {author} {\bibfnamefont {S.}~\bibnamefont {Kasuya}}, \ and\ \bibinfo {author} {\bibfnamefont {A.}~\bibnamefont {Mazumdar}},\ }\bibfield  {title} {\enquote {\bibinfo {title} {{MSSM flat direction as a curvaton}},}\ }\href {\doibase 10.1103/PhysRevD.68.103507} {\bibfield  {journal} {\bibinfo  {journal} {Phys. Rev. D}\ }\textbf {\bibinfo {volume} {68}},\ \bibinfo {pages} {103507} (\bibinfo {year} {2003}{\natexlab{b}})},\ \Eprint {http://arxiv.org/abs/hep-ph/0303165} {arXiv:hep-ph/0303165} \BibitemShut {NoStop}%
\bibitem [{\citenamefont {Kasuya}\ \emph {et~al.}(2004)\citenamefont {Kasuya}, \citenamefont {Kawasaki},\ and\ \citenamefont {Takahashi}}]{Kasuya:2003va}%
  \BibitemOpen
  \bibfield  {author} {\bibinfo {author} {\bibfnamefont {S.}~\bibnamefont {Kasuya}}, \bibinfo {author} {\bibfnamefont {M.}~\bibnamefont {Kawasaki}}, \ and\ \bibinfo {author} {\bibfnamefont {F.}~\bibnamefont {Takahashi}},\ }\bibfield  {title} {\enquote {\bibinfo {title} {{MSSM curvaton in the gauge mediated SUSY breaking}},}\ }\href {\doibase 10.1016/j.physletb.2003.10.079} {\bibfield  {journal} {\bibinfo  {journal} {Phys. Lett. B}\ }\textbf {\bibinfo {volume} {578}},\ \bibinfo {pages} {259--268} (\bibinfo {year} {2004})},\ \Eprint {http://arxiv.org/abs/hep-ph/0305134} {arXiv:hep-ph/0305134} \BibitemShut {NoStop}%
\bibitem [{\citenamefont {Hamaguchi}\ \emph {et~al.}(2004)\citenamefont {Hamaguchi}, \citenamefont {Kawasaki}, \citenamefont {Moroi},\ and\ \citenamefont {Takahashi}}]{Hamaguchi:2003dc}%
  \BibitemOpen
  \bibfield  {author} {\bibinfo {author} {\bibfnamefont {K.}~\bibnamefont {Hamaguchi}}, \bibinfo {author} {\bibfnamefont {M.}~\bibnamefont {Kawasaki}}, \bibinfo {author} {\bibfnamefont {T.}~\bibnamefont {Moroi}}, \ and\ \bibinfo {author} {\bibfnamefont {F.}~\bibnamefont {Takahashi}},\ }\bibfield  {title} {\enquote {\bibinfo {title} {{Curvatons in supersymmetric models}},}\ }\href {\doibase 10.1103/PhysRevD.69.063504} {\bibfield  {journal} {\bibinfo  {journal} {Phys. Rev. D}\ }\textbf {\bibinfo {volume} {69}},\ \bibinfo {pages} {063504} (\bibinfo {year} {2004})},\ \Eprint {http://arxiv.org/abs/hep-ph/0308174} {arXiv:hep-ph/0308174} \BibitemShut {NoStop}%
\bibitem [{\citenamefont {Co}\ and\ \citenamefont {Yamada}(2024)}]{Co:2023mhe}%
  \BibitemOpen
  \bibfield  {author} {\bibinfo {author} {\bibfnamefont {R.~T.}\ \bibnamefont {Co}}\ and\ \bibinfo {author} {\bibfnamefont {M.}~\bibnamefont {Yamada}},\ }\bibfield  {title} {\enquote {\bibinfo {title} {{Axion cogenesis without isocurvature perturbations}},}\ }\href {\doibase 10.1103/PhysRevD.110.055009} {\bibfield  {journal} {\bibinfo  {journal} {Phys. Rev. D}\ }\textbf {\bibinfo {volume} {110}},\ \bibinfo {pages} {055009} (\bibinfo {year} {2024})},\ \Eprint {http://arxiv.org/abs/2312.17730} {arXiv:2312.17730 [hep-ph]} \BibitemShut {NoStop}%
\bibitem [{\citenamefont {Akrami}\ \emph {et~al.}(2020{\natexlab{b}})\citenamefont {Akrami} \emph {et~al.}}]{Planck:2018jri}%
  \BibitemOpen
  \bibfield  {author} {\bibinfo {author} {\bibfnamefont {Y.}~\bibnamefont {Akrami}} \emph {et~al.} (\bibinfo {collaboration} {Planck}),\ }\bibfield  {title} {\enquote {\bibinfo {title} {{Planck 2018 results. X. Constraints on inflation}},}\ }\href {\doibase 10.1051/0004-6361/201833887} {\bibfield  {journal} {\bibinfo  {journal} {Astron. Astrophys.}\ }\textbf {\bibinfo {volume} {641}},\ \bibinfo {pages} {A10} (\bibinfo {year} {2020}{\natexlab{b}})},\ \Eprint {http://arxiv.org/abs/1807.06211} {arXiv:1807.06211 [astro-ph.CO]} \BibitemShut {NoStop}%
\bibitem [{\citenamefont {Gordon}\ \emph {et~al.}(2000)\citenamefont {Gordon}, \citenamefont {Wands}, \citenamefont {Bassett},\ and\ \citenamefont {Maartens}}]{Gordon:2000hv}%
  \BibitemOpen
  \bibfield  {author} {\bibinfo {author} {\bibfnamefont {C.}~\bibnamefont {Gordon}}, \bibinfo {author} {\bibfnamefont {D.}~\bibnamefont {Wands}}, \bibinfo {author} {\bibfnamefont {B.~A.}\ \bibnamefont {Bassett}}, \ and\ \bibinfo {author} {\bibfnamefont {R.}~\bibnamefont {Maartens}},\ }\bibfield  {title} {\enquote {\bibinfo {title} {{Adiabatic and entropy perturbations from inflation}},}\ }\href {\doibase 10.1103/PhysRevD.63.023506} {\bibfield  {journal} {\bibinfo  {journal} {Phys. Rev. D}\ }\textbf {\bibinfo {volume} {63}},\ \bibinfo {pages} {023506} (\bibinfo {year} {2000})},\ \Eprint {http://arxiv.org/abs/astro-ph/0009131} {arXiv:astro-ph/0009131} \BibitemShut {NoStop}%
\bibitem [{\citenamefont {Langlois}\ and\ \citenamefont {van Tent}(2011)}]{Langlois:2011hn}%
  \BibitemOpen
  \bibfield  {author} {\bibinfo {author} {\bibfnamefont {D.}~\bibnamefont {Langlois}}\ and\ \bibinfo {author} {\bibfnamefont {B.}~\bibnamefont {van Tent}},\ }\bibfield  {title} {\enquote {\bibinfo {title} {{Hunting for Isocurvature Modes in the CMB non-Gaussianities}},}\ }\href {\doibase 10.1088/0264-9381/28/22/222001} {\bibfield  {journal} {\bibinfo  {journal} {Class. Quant. Grav.}\ }\textbf {\bibinfo {volume} {28}},\ \bibinfo {pages} {222001} (\bibinfo {year} {2011})},\ \Eprint {http://arxiv.org/abs/1104.2567} {arXiv:1104.2567 [astro-ph.CO]} \BibitemShut {NoStop}%
\bibitem [{\citenamefont {Langlois}\ and\ \citenamefont {van Tent}(2012)}]{Langlois:2012tm}%
  \BibitemOpen
  \bibfield  {author} {\bibinfo {author} {\bibfnamefont {D.}~\bibnamefont {Langlois}}\ and\ \bibinfo {author} {\bibfnamefont {B.}~\bibnamefont {van Tent}},\ }\bibfield  {title} {\enquote {\bibinfo {title} {{Isocurvature modes in the CMB bispectrum}},}\ }\href {\doibase 10.1088/1475-7516/2012/07/040} {\bibfield  {journal} {\bibinfo  {journal} {JCAP}\ }\textbf {\bibinfo {volume} {07}},\ \bibinfo {pages} {040} (\bibinfo {year} {2012})},\ \Eprint {http://arxiv.org/abs/1204.5042} {arXiv:1204.5042 [astro-ph.CO]} \BibitemShut {NoStop}%
\bibitem [{\citenamefont {Hazumi}\ \emph {et~al.}(2012)\citenamefont {Hazumi} \emph {et~al.}}]{Hazumi:2012gjy}%
  \BibitemOpen
  \bibfield  {author} {\bibinfo {author} {\bibfnamefont {M.}~\bibnamefont {Hazumi}} \emph {et~al.},\ }\bibfield  {title} {\enquote {\bibinfo {title} {{LiteBIRD: a small satellite for the study of B-mode polarization and inflation from cosmic background radiation detection}},}\ }\href {\doibase 10.1117/12.926743} {\bibfield  {journal} {\bibinfo  {journal} {Proc. SPIE Int. Soc. Opt. Eng.}\ }\textbf {\bibinfo {volume} {8442}},\ \bibinfo {pages} {844219} (\bibinfo {year} {2012})}\BibitemShut {NoStop}%
\bibitem [{\citenamefont {Sekimoto}\ \emph {et~al.}(2020)\citenamefont {Sekimoto} \emph {et~al.}}]{LiteBIRD:2020tzb}%
  \BibitemOpen
  \bibfield  {author} {\bibinfo {author} {\bibfnamefont {Y.}~\bibnamefont {Sekimoto}} \emph {et~al.} (\bibinfo {collaboration} {LiteBIRD}),\ }\bibfield  {title} {\enquote {\bibinfo {title} {{Concept Design of Low Frequency Telescope for CMB B-mode Polarization satellite LiteBIRD}},}\ }\href {\doibase 10.1117/12.2561841} {\bibfield  {journal} {\bibinfo  {journal} {Proc. SPIE Int. Soc. Opt. Eng.}\ }\textbf {\bibinfo {volume} {11453}},\ \bibinfo {pages} {1145310} (\bibinfo {year} {2020})},\ \Eprint {http://arxiv.org/abs/2101.06342} {arXiv:2101.06342 [astro-ph.IM]} \BibitemShut {NoStop}%
\bibitem [{\citenamefont {Lonappan}\ \emph {et~al.}(2024)\citenamefont {Lonappan} \emph {et~al.}}]{LiteBIRD:2023iiy}%
  \BibitemOpen
  \bibfield  {author} {\bibinfo {author} {\bibfnamefont {A.~I.}\ \bibnamefont {Lonappan}} \emph {et~al.} (\bibinfo {collaboration} {LiteBIRD}),\ }\bibfield  {title} {\enquote {\bibinfo {title} {{LiteBIRD science goals and forecasts: a full-sky measurement of gravitational lensing of the CMB}},}\ }\href {\doibase 10.1088/1475-7516/2024/06/009} {\bibfield  {journal} {\bibinfo  {journal} {JCAP}\ }\textbf {\bibinfo {volume} {06}},\ \bibinfo {pages} {009} (\bibinfo {year} {2024})},\ \Eprint {http://arxiv.org/abs/2312.05184} {arXiv:2312.05184 [astro-ph.CO]} \BibitemShut {NoStop}%
\bibitem [{\citenamefont {Montandon}\ \emph {et~al.}(2021)\citenamefont {Montandon}, \citenamefont {Patanchon},\ and\ \citenamefont {van Tent}}]{Montandon:2020kuk}%
  \BibitemOpen
  \bibfield  {author} {\bibinfo {author} {\bibfnamefont {T.}~\bibnamefont {Montandon}}, \bibinfo {author} {\bibfnamefont {G.}~\bibnamefont {Patanchon}}, \ and\ \bibinfo {author} {\bibfnamefont {B.}~\bibnamefont {van Tent}},\ }\bibfield  {title} {\enquote {\bibinfo {title} {{Isocurvature modes: joint analysis of the CMB power spectrum and bispectrum}},}\ }\href {\doibase 10.1088/1475-7516/2021/01/004} {\bibfield  {journal} {\bibinfo  {journal} {JCAP}\ }\textbf {\bibinfo {volume} {01}},\ \bibinfo {pages} {004} (\bibinfo {year} {2021})},\ \Eprint {http://arxiv.org/abs/2007.05457} {arXiv:2007.05457 [astro-ph.CO]} \BibitemShut {NoStop}%
\bibitem [{\citenamefont {Kawasaki}\ \emph {et~al.}(2000)\citenamefont {Kawasaki}, \citenamefont {Yamaguchi},\ and\ \citenamefont {Yanagida}}]{Kawasaki:2000yn}%
  \BibitemOpen
  \bibfield  {author} {\bibinfo {author} {\bibfnamefont {M.}~\bibnamefont {Kawasaki}}, \bibinfo {author} {\bibfnamefont {M.}~\bibnamefont {Yamaguchi}}, \ and\ \bibinfo {author} {\bibfnamefont {T.}~\bibnamefont {Yanagida}},\ }\bibfield  {title} {\enquote {\bibinfo {title} {{Natural chaotic inflation in supergravity}},}\ }\href {\doibase 10.1103/PhysRevLett.85.3572} {\bibfield  {journal} {\bibinfo  {journal} {Phys. Rev. Lett.}\ }\textbf {\bibinfo {volume} {85}},\ \bibinfo {pages} {3572--3575} (\bibinfo {year} {2000})},\ \Eprint {http://arxiv.org/abs/hep-ph/0004243} {arXiv:hep-ph/0004243} \BibitemShut {NoStop}%
\bibitem [{\citenamefont {Kallosh}\ and\ \citenamefont {Linde}(2010)}]{Kallosh:2010ug}%
  \BibitemOpen
  \bibfield  {author} {\bibinfo {author} {\bibfnamefont {R.}~\bibnamefont {Kallosh}}\ and\ \bibinfo {author} {\bibfnamefont {A.}~\bibnamefont {Linde}},\ }\bibfield  {title} {\enquote {\bibinfo {title} {{New models of chaotic inflation in supergravity}},}\ }\href {\doibase 10.1088/1475-7516/2010/11/011} {\bibfield  {journal} {\bibinfo  {journal} {JCAP}\ }\textbf {\bibinfo {volume} {11}},\ \bibinfo {pages} {011} (\bibinfo {year} {2010})},\ \Eprint {http://arxiv.org/abs/1008.3375} {arXiv:1008.3375 [hep-th]} \BibitemShut {NoStop}%
\bibitem [{\citenamefont {Chung}\ \emph {et~al.}(2023)\citenamefont {Chung}, \citenamefont {M\"unchmeyer},\ and\ \citenamefont {Tadepalli}}]{Chung:2023syw}%
  \BibitemOpen
  \bibfield  {author} {\bibinfo {author} {\bibfnamefont {D.~J.~H.}\ \bibnamefont {Chung}}, \bibinfo {author} {\bibfnamefont {M.}~\bibnamefont {M\"unchmeyer}}, \ and\ \bibinfo {author} {\bibfnamefont {S.~C.}\ \bibnamefont {Tadepalli}},\ }\bibfield  {title} {\enquote {\bibinfo {title} {{Search for isocurvature with large-scale structure: A forecast for Euclid and MegaMapper using EFTofLSS}},}\ }\href {\doibase 10.1103/PhysRevD.108.103542} {\bibfield  {journal} {\bibinfo  {journal} {Phys. Rev. D}\ }\textbf {\bibinfo {volume} {108}},\ \bibinfo {pages} {103542} (\bibinfo {year} {2023})},\ \Eprint {http://arxiv.org/abs/2306.09456} {arXiv:2306.09456 [astro-ph.CO]} \BibitemShut {NoStop}%
\bibitem [{\citenamefont {Mukaida}\ and\ \citenamefont {Nakayama}(2013)}]{Mukaida:2012qn}%
  \BibitemOpen
  \bibfield  {author} {\bibinfo {author} {\bibfnamefont {K.}~\bibnamefont {Mukaida}}\ and\ \bibinfo {author} {\bibfnamefont {K.}~\bibnamefont {Nakayama}},\ }\bibfield  {title} {\enquote {\bibinfo {title} {{Dynamics of oscillating scalar field in thermal environment}},}\ }\href {\doibase 10.1088/1475-7516/2013/01/017} {\bibfield  {journal} {\bibinfo  {journal} {JCAP}\ }\textbf {\bibinfo {volume} {01}},\ \bibinfo {pages} {017} (\bibinfo {year} {2013})},\ \Eprint {http://arxiv.org/abs/1208.3399} {arXiv:1208.3399 [hep-ph]} \BibitemShut {NoStop}%
\bibitem [{\citenamefont {Baumann}(2018)}]{Baumann:2018muz}%
  \BibitemOpen
  \bibfield  {author} {\bibinfo {author} {\bibfnamefont {D.}~\bibnamefont {Baumann}},\ }\bibfield  {title} {\enquote {\bibinfo {title} {{Primordial Cosmology}},}\ }\href {\doibase 10.22323/1.305.0009} {\bibfield  {journal} {\bibinfo  {journal} {PoS}\ }\textbf {\bibinfo {volume} {TASI2017}},\ \bibinfo {pages} {009} (\bibinfo {year} {2018})},\ \Eprint {http://arxiv.org/abs/1807.03098} {arXiv:1807.03098 [hep-th]} \BibitemShut {NoStop}%
\bibitem [{\citenamefont {Chen}\ \emph {et~al.}(2015)\citenamefont {Chen}, \citenamefont {Namjoo},\ and\ \citenamefont {Wang}}]{Chen:2014cwa}%
  \BibitemOpen
  \bibfield  {author} {\bibinfo {author} {\bibfnamefont {X.}~\bibnamefont {Chen}}, \bibinfo {author} {\bibfnamefont {M.~H.}\ \bibnamefont {Namjoo}}, \ and\ \bibinfo {author} {\bibfnamefont {Y.}~\bibnamefont {Wang}},\ }\bibfield  {title} {\enquote {\bibinfo {title} {{Models of the Primordial Standard Clock}},}\ }\href {\doibase 10.1088/1475-7516/2015/02/027} {\bibfield  {journal} {\bibinfo  {journal} {JCAP}\ }\textbf {\bibinfo {volume} {02}},\ \bibinfo {pages} {027} (\bibinfo {year} {2015})},\ \Eprint {http://arxiv.org/abs/1411.2349} {arXiv:1411.2349 [astro-ph.CO]} \BibitemShut {NoStop}%
\end{thebibliography}%

\end{document}